\documentclass{elsarticle}

\usepackage[margin=1in]{geometry}

\usepackage{bbm}
\usepackage{graphicx}
\usepackage{pstool}
\usepackage{wrapfig}
\usepackage{caption}
\usepackage{subcaption}
\usepackage[section]{placeins} 

\usepackage{fancyhdr} 
\usepackage{lastpage} 
\usepackage{extramarks} 

\usepackage{xcolor} 
\usepackage{enumerate} 
\usepackage{paralist} 
\usepackage{amsmath, amsthm, amssymb, mathtools}
\usepackage{mathabx, pifont, stmaryrd} 
\usepackage[explicit]{titlesec} 
\usepackage{etoolbox} 
\usepackage{bibentry} 
\makeatletter\let\saved@bibitem\@bibitem\makeatother 
\usepackage[colorlinks, bookmarksopen, bookmarksnumbered,
            citecolor=red,urlcolor=red]{hyperref} 
\makeatletter\let\@bibitem\saved@bibitem\makeatother 
\usepackage{rotating}

\usepackage{algorithm}
\usepackage{algorithmic}

\newtheorem{remark}{Remark}

\theoremstyle{definition}

\usepackage{bm} 

\DeclareMathOperator*{\argmin}{arg\,min}

\newcommand{\func}[3]{\ensuremath{#1 : #2 \rightarrow #3}}

\newcommand{\norm}[1]{\ensuremath{\left\| #1 \right\|}}

\newcommand{\suchthat}{\mathrel{}\middle|\mathrel{}}

\newcommand{\pder}[2]{\ensuremath{\frac{\partial #1}{\partial #2}}}

\newcommand{\Ecal}{\ensuremath{\mathcal{E}}}
\newcommand{\Fcal}{\ensuremath{\mathcal{F}}}
\newcommand{\Gcal}{\ensuremath{\mathcal{G}}}
\newcommand{\Hcal}{\ensuremath{\mathcal{H}}}

\newcommand{\Ocal}{\ensuremath{\mathcal{O}}}
\newcommand{\Pcal}{\ensuremath{\mathcal{P}}}

\newcommand{\Scal}{\ensuremath{\mathcal{S}}}
\newcommand{\Tcal}{\ensuremath{\mathcal{T}}}

\newcommand{\Vcal}{\ensuremath{\mathcal{V}}}
\newcommand{\Wcal}{\ensuremath{\mathcal{W}}}

\newcommand{\Gbb}{\ensuremath{\mathbb{G}}}

\newcommand{\Rbb}{\ensuremath{\mathbb{R} }}
\newcommand{\Sbb}{\ensuremath{\mathbb{S} }}

\newcommand\Rbm{{\ensuremath{\bm{R}}}}

\newcommand\rbm{{\ensuremath{\bm{r}}}}

\newcommand\ubm{{\ensuremath{\bm{u}}}}

\newcommand\xbm{{\ensuremath{\bm{x}}}}
\newcommand\ybm{{\ensuremath{\bm{y}}}}

\newcommand\phibold{{\ensuremath{\boldsymbol{\phi}}}}

\newcommand\zerobold{\ensuremath{\mathbf{0}}}

\usepackage{tikz}
\usepackage{pgfplots}
\usepackage{pgfplotstable, filecontents, booktabs}
\pgfplotsset{compat=1.9}

\usetikzlibrary{pgfplots.groupplots}
\usepgfplotslibrary{fillbetween}
\usetikzlibrary{calc,fit,matrix,arrows,automata,positioning,shapes}
\usetikzlibrary{arrows.meta}

\pgfplotsset{select coords between index/.style 2 args={
    x filter/.code={
        \ifnum\coordindex<#1\fi
        \ifnum\coordindex>#2\fi
    }
}}

\tikzset{
 invisible/.style={opacity=0},
 visible on/.style={alt={#1{}{invisible}}},
 alt/.code args={<#1>#2#3}{%
   \alt<#1>{\pgfkeysalso{#2}}{\pgfkeysalso{#3}}
 },
}

\newcommand{\colorbarMatlabParula}[5]{
\begin{tikzpicture}
\begin{axis}[
   hide axis, scale only axis,
   height=0pt, width=0pt,
   colormap={parula}{rgb255=(62,38,168) rgb255=(62,39,172) rgb255=(63,40,175) rgb255=(63,41,178) rgb255=(64,42,180) rgb255=(64,43,183) rgb255=(65,44,186) rgb255=(65,45,189) rgb255=(66,46,191) rgb255=(66,47,194) rgb255=(67,48,197) rgb255=(67,49,200) rgb255=(67,50,202) rgb255=(68,51,205) rgb255=(68,52,208) rgb255=(69,53,210) rgb255=(69,55,213) rgb255=(69,56,215) rgb255=(70,57,217) rgb255=(70,58,220) rgb255=(70,59,222) rgb255=(70,61,224) rgb255=(71,62,225) rgb255=(71,63,227) rgb255=(71,65,229) rgb255=(71,66,230) rgb255=(71,68,232) rgb255=(71,69,233) rgb255=(71,70,235) rgb255=(72,72,236) rgb255=(72,73,237) rgb255=(72,75,238) rgb255=(72,76,240) rgb255=(72,78,241) rgb255=(72,79,242) rgb255=(72,80,243) rgb255=(72,82,244) rgb255=(72,83,245) rgb255=(72,84,246) rgb255=(71,86,247) rgb255=(71,87,247) rgb255=(71,89,248) rgb255=(71,90,249) rgb255=(71,91,250) rgb255=(71,93,250) rgb255=(70,94,251) rgb255=(70,96,251) rgb255=(70,97,252) rgb255=(69,98,252) rgb255=(69,100,253) rgb255=(68,101,253) rgb255=(67,103,253) rgb255=(67,104,254) rgb255=(66,106,254) rgb255=(65,107,254) rgb255=(64,109,254) rgb255=(63,110,255) rgb255=(62,112,255) rgb255=(60,113,255) rgb255=(59,115,255) rgb255=(57,116,255) rgb255=(56,118,254) rgb255=(54,119,254) rgb255=(53,121,253) rgb255=(51,122,253) rgb255=(50,124,252) rgb255=(49,125,252) rgb255=(48,127,251) rgb255=(47,128,250) rgb255=(47,130,250) rgb255=(46,131,249) rgb255=(46,132,248) rgb255=(46,134,248) rgb255=(46,135,247) rgb255=(45,136,246) rgb255=(45,138,245) rgb255=(45,139,244) rgb255=(45,140,243) rgb255=(45,142,242) rgb255=(44,143,241) rgb255=(44,144,240) rgb255=(43,145,239) rgb255=(42,147,238) rgb255=(41,148,237) rgb255=(40,149,236) rgb255=(39,151,235) rgb255=(39,152,234) rgb255=(38,153,233) rgb255=(38,154,232) rgb255=(37,155,232) rgb255=(37,156,231) rgb255=(36,158,230) rgb255=(36,159,229) rgb255=(35,160,229) rgb255=(35,161,228) rgb255=(34,162,228) rgb255=(33,163,227) rgb255=(32,165,227) rgb255=(31,166,226) rgb255=(30,167,225) rgb255=(29,168,225) rgb255=(29,169,224) rgb255=(28,170,223) rgb255=(27,171,222) rgb255=(26,172,221) rgb255=(25,173,220) rgb255=(23,174,218) rgb255=(22,175,217) rgb255=(20,176,216) rgb255=(18,177,214) rgb255=(16,178,213) rgb255=(14,179,212) rgb255=(11,179,210) rgb255=(8,180,209) rgb255=(6,181,207) rgb255=(4,182,206) rgb255=(2,183,204) rgb255=(1,183,202) rgb255=(0,184,201) rgb255=(0,185,199) rgb255=(0,186,198) rgb255=(1,186,196) rgb255=(2,187,194) rgb255=(4,187,193) rgb255=(6,188,191) rgb255=(9,189,189) rgb255=(13,189,188) rgb255=(16,190,186) rgb255=(20,190,184) rgb255=(23,191,182) rgb255=(26,192,181) rgb255=(29,192,179) rgb255=(32,193,177) rgb255=(35,193,175) rgb255=(37,194,174) rgb255=(39,194,172) rgb255=(41,195,170) rgb255=(43,195,168) rgb255=(44,196,166) rgb255=(46,196,165) rgb255=(47,197,163) rgb255=(49,197,161) rgb255=(50,198,159) rgb255=(51,199,157) rgb255=(53,199,155) rgb255=(54,200,153) rgb255=(56,200,150) rgb255=(57,201,148) rgb255=(59,201,146) rgb255=(61,202,144) rgb255=(64,202,141) rgb255=(66,202,139) rgb255=(69,203,137) rgb255=(72,203,134) rgb255=(75,203,132) rgb255=(78,204,129) rgb255=(81,204,127) rgb255=(84,204,124) rgb255=(87,204,122) rgb255=(90,204,119) rgb255=(94,205,116) rgb255=(97,205,114) rgb255=(100,205,111) rgb255=(103,205,108) rgb255=(107,205,105) rgb255=(110,205,102) rgb255=(114,205,100) rgb255=(118,204,97) rgb255=(121,204,94) rgb255=(125,204,91) rgb255=(129,204,89) rgb255=(132,204,86) rgb255=(136,203,83) rgb255=(139,203,81) rgb255=(143,203,78) rgb255=(147,202,75) rgb255=(150,202,72) rgb255=(154,201,70) rgb255=(157,201,67) rgb255=(161,200,64) rgb255=(164,200,62) rgb255=(167,199,59) rgb255=(171,199,57) rgb255=(174,198,55) rgb255=(178,198,53) rgb255=(181,197,51) rgb255=(184,196,49) rgb255=(187,196,47) rgb255=(190,195,45) rgb255=(194,195,44) rgb255=(197,194,42) rgb255=(200,193,41) rgb255=(203,193,40) rgb255=(206,192,39) rgb255=(208,191,39) rgb255=(211,191,39) rgb255=(214,190,39) rgb255=(217,190,40) rgb255=(219,189,40) rgb255=(222,188,41) rgb255=(225,188,42) rgb255=(227,188,43) rgb255=(230,187,45) rgb255=(232,187,46) rgb255=(234,186,48) rgb255=(236,186,50) rgb255=(239,186,53) rgb255=(241,186,55) rgb255=(243,186,57) rgb255=(245,186,59) rgb255=(247,186,61) rgb255=(249,186,62) rgb255=(251,187,62) rgb255=(252,188,62) rgb255=(254,189,61) rgb255=(254,190,60) rgb255=(254,192,59) rgb255=(254,193,58) rgb255=(254,194,57) rgb255=(254,196,56) rgb255=(254,197,55) rgb255=(254,199,53) rgb255=(254,200,52) rgb255=(254,202,51) rgb255=(253,203,50) rgb255=(253,205,49) rgb255=(253,206,49) rgb255=(252,208,48) rgb255=(251,210,47) rgb255=(251,211,46) rgb255=(250,213,46) rgb255=(249,214,45) rgb255=(249,216,44) rgb255=(248,217,43) rgb255=(247,219,42) rgb255=(247,221,42) rgb255=(246,222,41) rgb255=(246,224,40) rgb255=(245,225,40) rgb255=(245,227,39) rgb255=(245,229,38) rgb255=(245,230,38) rgb255=(245,232,37) rgb255=(245,233,36) rgb255=(245,235,35) rgb255=(245,236,34) rgb255=(245,238,33) rgb255=(246,239,32) rgb255=(246,241,31) rgb255=(246,242,30) rgb255=(247,244,28) rgb255=(247,245,27) rgb255=(248,247,26) rgb255=(248,248,24) rgb255=(249,249,22) rgb255=(249,251,21) },
   colorbar horizontal,
   point meta min=#1, point meta max=#5,
   colorbar style={width=10cm, xtick={#1,#2,#3,#4,#5}}
]
\addplot [draw=none] coordinates {(0,0)};
\end{axis}
\end{tikzpicture}
}

\usepackage{amsmath}

\usepackage{enumitem}

\newcommand{\physF}{\mathcal{F}}
\newcommand{\physS}{\mathcal{S}}
\newcommand{\physU}{U}

\newcommand{\physUnitN}{\eta}

\newcommand{\transF}{\bar{\physF}} 
\newcommand{\transS}{\bar{\physS}}
\newcommand{\transU}{\bar{\physU}}

\newcommand{\transUnitN}{\bar{\physUnitN}}

\newbool{fastcompile}
\setbool{fastcompile}{false}

\title{A sharp-interface discontinuous Galerkin method for simulation of two-phase flow of real gases based on implicit shock tracking}
\author[rvt1]{Charles Naudet\fnref{fn1}}
\ead{cnaudet@nd.edu}

\author[rvt2]{Brian Taylor\fnref{fn3}}

\author[rvt1]{Matthew J. Zahr\fnref{fn2}\corref{cor1}}
\ead{mzahr@nd.edu}

\address[rvt1]{Department of Aerospace and Mechanical Engineering, University
               of Notre Dame, Notre Dame, IN 46556, United States}
\address[rvt2]{Air Force Research Laboratory, Eglin AFB, FL 32542}
\cortext[cor1]{Corresponding author}

\fntext[fn1]{Graduate Student, Department of Aerospace and Mechanical
             Engineering, University of Notre Dame}
 \fntext[fn3]{Senior Research Engineer, Air Force Research Laboratory}
\fntext[fn2]{Assistant Professor, Department of Aerospace and Mechanical
             Engineering, University of Notre Dame}

\begin{document}

\begin{abstract}
	We present a high-order, sharp-interface method for simulation of
	two-phase flow of real gases using implicit shock tracking. The
	method is based on a phase-field formulation of two-phase, compressible,
	inviscid flow with a trivial mixture model. Implicit shock tracking is
	a high-order, optimization-based discontinuous Galerkin method that 
	automatically aligns mesh faces with non-smooth flow features to
	represent them perfectly with inter-element jumps. It is used to
	accurately approximate shocks and rarefactions without stabilization
	and converge the phase-field solution to a sharp interface one by
	aligning mesh faces with the material interface. Time-dependent problems
	are formulated as steady problems in a space-time domain where complex
	wave interactions (e.g., intersections and reflections) manifest as
	space-time triplet points. The space-time formulation avoids complex
	re-meshing and solution transfer that would be required to track moving
	waves with mesh faces using the method of lines. The approach is
	applied to several two-phase flow Riemann problems involving gases with
	ideal, stiffened gas, and Becker-Kistiakowsky-Wilson (BKW) equations of
	state, including a spherically symmetric underwater explosion problem. In all
	cases, the method aligns element faces with all shocks (including secondary
	shocks that form at time $t > 0$), rarefactions, and material interfaces, and
	accurately resolves the flow field on coarse space-time grids.
\end{abstract}

\maketitle

\section{Introduction}
\label{sec:intro}
\label{sec:intro}
Compressible, two-phase flows arise in the study of bubbles and interfaces
\cite{nicklin1962two,saye2013multiscale},
mixing processes \cite{witte1969mixing},
bubbly flows \cite{khan2020two,ma2015using},
granular solids \cite{Powers2004, Powers1990}, and
explosive mixtures \cite{banks2007high, Gonthier2000},
to name a few. Such flows are challenging to simulate
because treatment of the material interface plays a large role in the stability and
accuracy of the method. We propose a novel simulation methodology for compressible,
two-phase flow that circumvents several of the challenges faced by modern tools.

\subsection{Review of two-phase flow simulation technology}
\label{sec:intro:litrev}
Numerical methods to simulate two-phase flow either approximate the material
interface as a smooth transition between materials (\textit{diffuse interface})
or treat the interface as a true discontinuity in the material (\textit{sharp
interface}).

\subsubsection{Diffuse-interface approaches}
\label{sec:intro:litrev:diffuse}
Shocks and material interfaces arise in compressible, two-phase flow simulations and
lead to non-physical oscillations in the solution when greater-than-first-order
discretization methods are used. In single-phase flow problems, these oscillations are
suppressed using shock capturing methods such as
limiting \cite{1979_vanleer_consdiff, Baumann1999},
non-oscillatory reconstruction methods \cite{1987_harten_eno,Liu1994},
and artificial viscosity methods \cite{2006_persson_shkcapt,Barter2010}, for example.
For two-phase flows, conservative interface-capturing schemes diffuse the material
interface, which leads to spurious oscillations from pressure calculations in the
mixture region \cite{banks2007high, Abgrall2001}. Non-conservative
\cite{Karni1992, Karni1994} and conservative \cite{Quirk1996, Shyue1998}
interface-capturing methods based on the primitive form of the governing
equations have been proposed to improve the accuracy of pressure computations
near the interface. Other approaches to suppress non-physical oscillations
near the material interface include specialized discretizations of the
species mass fraction equation \cite{Abgrall1996, saurel1999simple},
inclusion of an energy correction equation \cite{Jenny1997, banks2007high},
and introduction of a pressure non-equilibrium model that reduces to
a single-velocity, single-pressure model \cite{Saurel2009,Karni2004, Ha2015}.
Many of these approaches have been used in combination with high-order
discretizations \cite{Kawai2011} such as discontinuous Galerkin (DG)
\cite{Ha2015, Frahan2015, Wang2012} and weighted essentially non-oscillatory (WENO)
reconstructions \cite{Johnsen2006, Gonthier2000} to increase accuracy per degree
of freedom. However, these diffuse-interface methods will be, at best, first-order
accurate near shocks or material interfaces because a smooth profile is being used
to approximate a discontinuity. This is usually offset with significant adaptive
mesh refinement near shocks and interfaces \cite{banks2007high}.

\subsubsection{Sharp-interface approaches}
\label{sec:intro:litrev:sharp}
In contrast to diffuse-interface methods, sharp-interface methods discretize the
governing equations in each fluid domain separately, and interface conditions
are used to evolve the material interface in time. Mesh-based Lagrangian methods
\cite{ball2000shock,kamran2013compressible} deform the grid at each time step to
align with the material interface,
which ensures a sharp interface but leads to a distorted or tangled mesh for
large interface deformations. Meshless Lagrangian methods
\cite{liu2010smoothed,sun2015modified,shobeyri2010simulating} have been
developed to address the mesh distortion problem; however, they suffer from
their own shortcomings such as large computational cost, instabilities due
to particle clustering, difficulty enforcing boundary conditions, and overall
lower accuracy. Marker-and-cell (MAC) methods \cite{harlow1965numerical,McKee2008}
advect particles in a Lagrangian manner with the flow velocity to reconstruct a
sharp interface. While accurate, this can be computationally expensive in three
dimensions due to the large number of particles that must be tracked. Arbitrary
Lagrangian-Eulerian \cite{donea1982arbitrary,cruchaga2001moving}
and front-tracking \cite{chern1986front} methods reduce the cost of MAC methods by
only adding a set of connected marker particles to the material interface. These
approaches tend to allow for larger interface deformations than pure Lagrangian
approaches; however, they will still lead to distorted meshes for large deformations
and cannot handle topological changes to the interface.

An alternative to Lagrangian or mixed Eulerian-Lagragian approaches are pure Eulerian
methods that use a fixed grid. These methods can be classified as either interface
capturing (Section~\ref{sec:intro:litrev:diffuse}) or interface tracking.
Interface-tracking methods combine the flow equations with a mathematical model
to reconstruct the interface, e.g., such as
volume of fluid (VOF) \cite{Rudman1997, Rudman1998, Nguyen2016, Gueyffier1999, Welch2000, PilliodJr2004} or
level sets \cite{Osher1988, Sussman1994, Sussman1999, VanderPijl2005}. The VOF method
advects the volume fraction of each fluid using an interface advection equation
and ensures mass conservation; however, geometric reconstruction of the interface
is difficult and leads to accuracy degradation \cite{Nguyen2016}.
Level set methods \cite{VanderPijl2005,Sussman1999,Sussman1994,saye2017implicit}
implicitly define the interface through a signed distance function that is advected
with the fluid velocity. These methods can reconstruct complex interface topologies
and have proven to lead to highly accurate two-phase flow simulations, especially when
coupled with high-order cut-cell methods \cite{saye2017implicit,kummer2017extended}.
However, they violate mass conservation \cite{Sussman1999}, require sufficiently
smooth interfaces (to be represented by a level set function), incur additional cost
from the level set advection equation, and are quite intricate.

Overall, sharp-interface methods avoid the non-physical numerical oscillations that
arise at the material interface using diffuse-interface methods. However, these methods
have their own shortcomings. They suffer from an accumulation of error in the
location of the material interface as it evolves. They are strictly
used to obtain a sharp material interface, whereas stabilization techniques must be
used to resolve shock waves and other contact discontinuities, which degrades global
solution accuracy to first-order. This can lead to further loss of accuracy if shock
waves interact with the material interface. 

\subsection{Proposed approach: Sharp-interface method based on space-time implicit shock tracking}
In this work, we propose a high-order DG discretization of
two-phase flow with sharp interface treatment based on implicit shock tracking. Our
approach uses the High-Order Implicit Shock Tracking (HOIST) method
\cite{Zahr2018,2020_zahr_HOIST,2021_huang_HOIST} to align
the computational grid with all non-smooth flow features (shocks, material interfaces,
head/tail of rarefactions) to represent them perfectly with inter-element jumps in
the DG basis, leaving the intra-element polynomial basis to represent smooth regions
of the flow with high-order accuracy. Alignment of the mesh faces with non-smooth
features is achieved without explicit shock detection or re-meshing; rather, it is
the solution of an optimization problem that minimizes the enriched DG residual
(based on a test space with higher polynomial degree than the trial space) and
therefore alignment is achieved implicitly. The optimization problem is initialized
from a shock-agnostic grid.

Two variants of implicit shock tracking have been developed: (1) a method-of-lines
approach \cite{Shi2022} that solves an optimization problem at each time
step to deform the grid as time evolves to ensure the same element faces track the
non-smooth features as they evolve and (2) a space-time approach
\cite{corrigan2019moving,Corrigan2019a,naudet2024space,nourgaliev2021implicit}
that recast the time-dependent problem in $d'$ dimensions as a steady problem
in $(d'+1)$ dimensions and aligns element faces of the space-time mesh with the
non-smooth features in space-time. The time domain in the latter approach is usually
split into a number of slabs to avoid coupling the entire temporal domain. The
method-of-lines approach
is more efficient because it works directly with the unsteady conservation law without
increasing the dimensionality of the problem; however, it will inevitably lead to mesh
entanglement if the non-smooth features move substantially throughout the domain or
intersect. On the other hand, these complex features manifest as curves and triple
points in space-time, both of which are handled robustly by implicit shock tracking
\cite{corrigan2019moving,2021_huang_HOIST,naudet2024space}. Thus, we build our method
on space-time implicit shock tracking to robustly handle complex wave dynamics.

The space-time formulation of implicit shock tracking introduces a special complication
for two-phase flow. Because the shock-aligned grid is computed as the solution of an
optimization problem and initialized from a non-aligned grid, the material interface
is not present until the end of the simulation making it difficult to assign an equation
of state to each space-time location. We overcome this problem by introducing an additional
equation governing the advection of the material phase, $\phi(x,t) \in [0,1]$, where
$\phi(x,t) = 1$ means material 1 occupies the space-time point $(x,t)$,
$\phi(x,t) = 0$ means material 2 occupies the point, and $0 < \phi(x,t) < 1$
represent a non-physical mixture of the phases. Because we are using implicit shock
tracking to represent the interface as a perfect discontinuity, we expect
$\phi(x,t) \in \{0,1\}$ at convergence. Thus, non-trivial mixtures are only used
to help discover the true interface location. As a result, the mixture equation of
state must only be theromdynamically consistent with the individual materials in the
limit as $\phi \rightarrow 0$ and $\phi \rightarrow 1$. Other approaches that
employ a phase variable have stringent requirements on the mixture equation
of state for $0 < \phi < 1$ \cite{sun2007sharp}.

\begin{figure}
	\centering
 	\begin{tikzpicture}
\begin{groupplot} [
	group style={
		group size = 1 by 3,
		horizontal sep = 0.05cm,
		vertical sep = 0.4cm
	},
	title style={at={(current bounding box.north west)}, anchor=west}
]

\nextgroupplot[axis equal image, width=1.0\textwidth, xtick={0, 0.5, 1}, ytick={0, 0.1, 0.2}, xticklabels={}, yticklabels={}, xlabel={}, ylabel={time}, xmin=0, xmax=1, ymin=0, ymax=0.1]
\addplot [] graphics [xmin=0,xmax=1,ymin=0,ymax=0.1] { 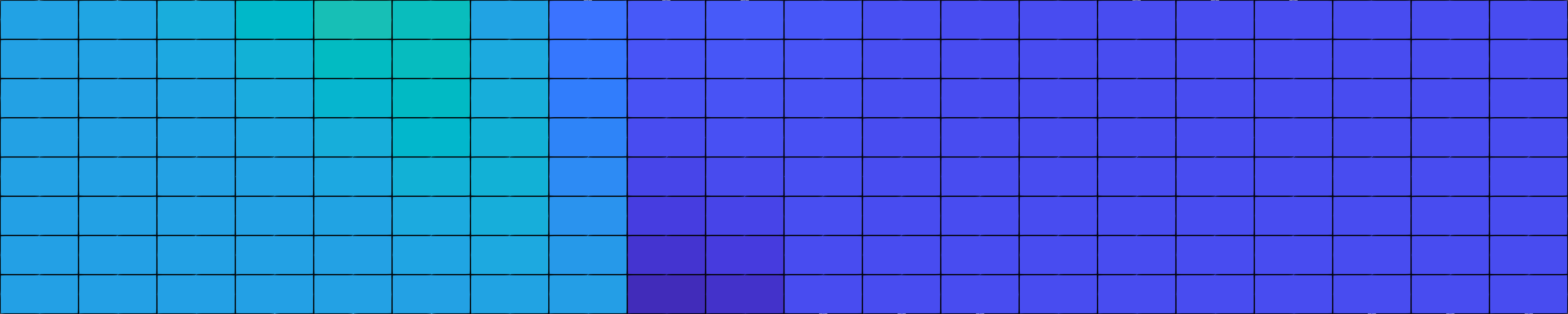};

\nextgroupplot[axis equal image, width=1.0\textwidth, xtick={0, 0.5, 1}, ytick={0, 0.1, 0.2}, xticklabels={}, yticklabels={}, xlabel={}, ylabel={time}, xmin=0, xmax=1, ymin=0, ymax=0.1]
\addplot [] graphics [xmin=0,xmax=1,ymin=0,ymax=0.1] { 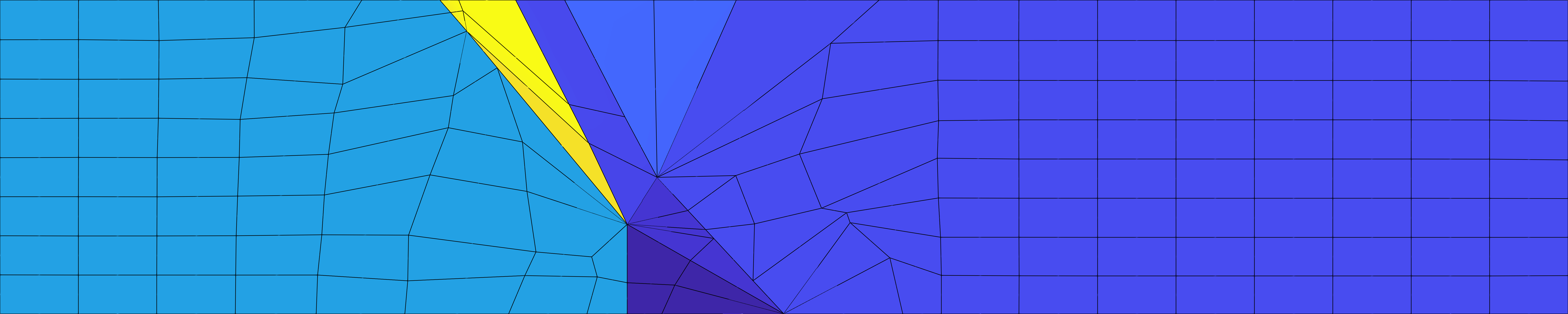};

\nextgroupplot[axis equal image, width=1.0\textwidth, xtick={}, ytick={}, xticklabels={}, yticklabels={}, xlabel={space}, ylabel={time}, xmin=0, xmax=1, ymin=0, ymax=0.1]
\addplot [] graphics [xmin=0,xmax=1,ymin=0,ymax=0.1] { 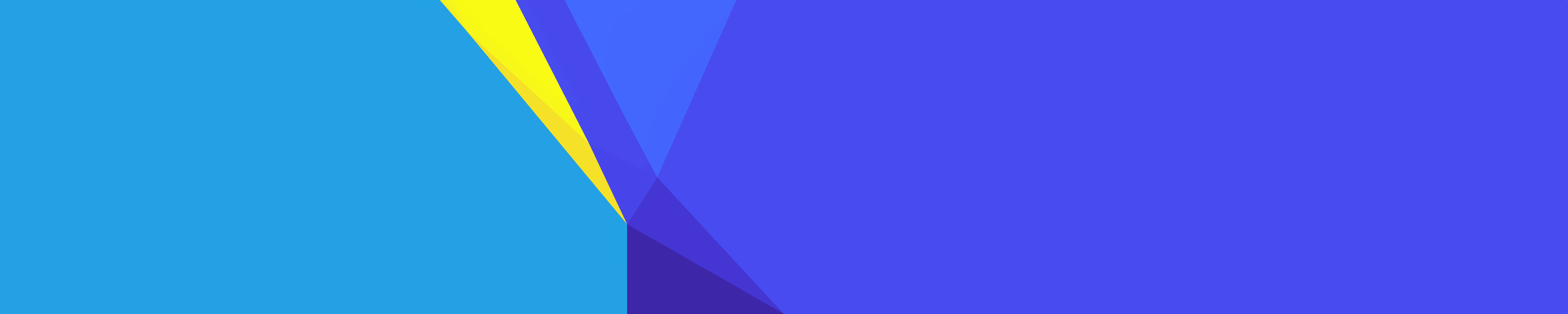};

\end{groupplot}
\end{tikzpicture}
 	\colorbarMatlabParula{0.125}{1}{1.5}{2}{2.5}        
	\caption{Shock-agnostic space-time mesh and first-order solution (density) used to initialize HOIST solver (\textit{top}), and converged HOIST solution (density) with (\textit{middle}) and without (\textit{bottom}) mesh edges.}
 	\label{fig:two_shock_example_dens}
\end{figure}

\begin{figure}[htbp] 
	\centering
 	\begin{tikzpicture}
\begin{groupplot} [
        group style={
                group size = 1 by 3,
                horizontal sep = 0.05cm,
                vertical sep = 0.4cm
        },
        title style={at={(current bounding box.north west)}, anchor=west}
]       

\nextgroupplot[axis equal image, width=1.0\textwidth, xtick={0, 0.5, 1}, ytick={0, 0.1, 0.2}, xticklabels={}, yticklabels={}, xlabel={}, ylabel={time}, xmin=0, xmax=1, ymin=0, ymax=0.1]
\addplot [] graphics [xmin=0,xmax=1,ymin=0,ymax=0.1] { 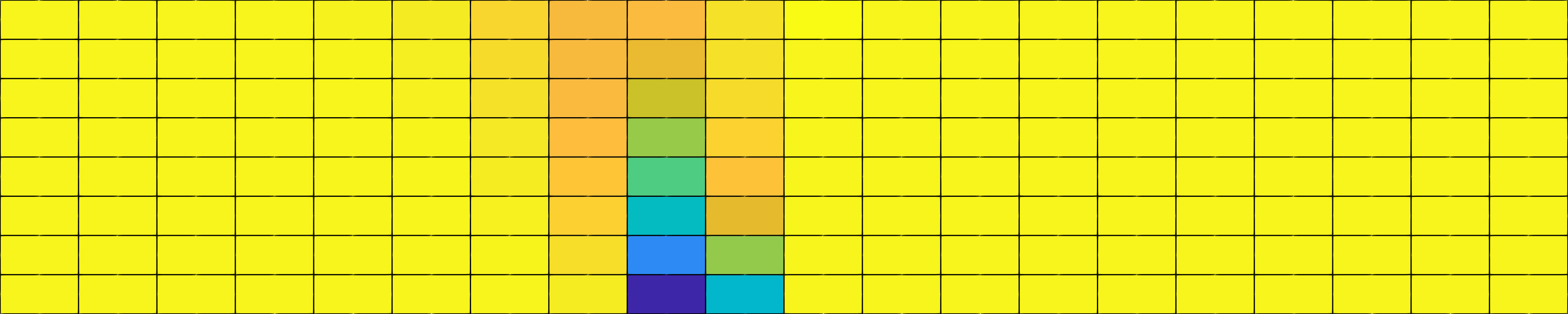};

\nextgroupplot[axis equal image, width=1.0\textwidth, xtick={0, 0.5, 1}, ytick={0, 0.1, 0.2}, xticklabels={}, yticklabels={}, xlabel={}, ylabel={time}, xmin=0, xmax=1, ymin=0, ymax=0.1]
\addplot [] graphics [xmin=0,xmax=1,ymin=0,ymax=0.1] { 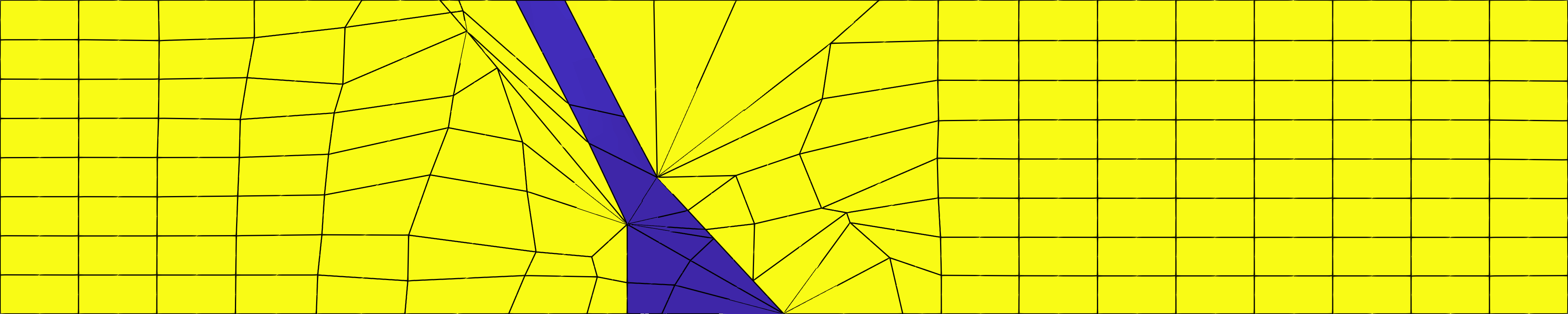};

\nextgroupplot[axis equal image, width=1.0\textwidth, xtick={}, ytick={}, xticklabels={}, yticklabels={}, xlabel={space}, ylabel={time}, xmin=0, xmax=1, ymin=0, ymax=0.1]
\addplot [] graphics [xmin=0,xmax=1,ymin=0,ymax=0.1] { 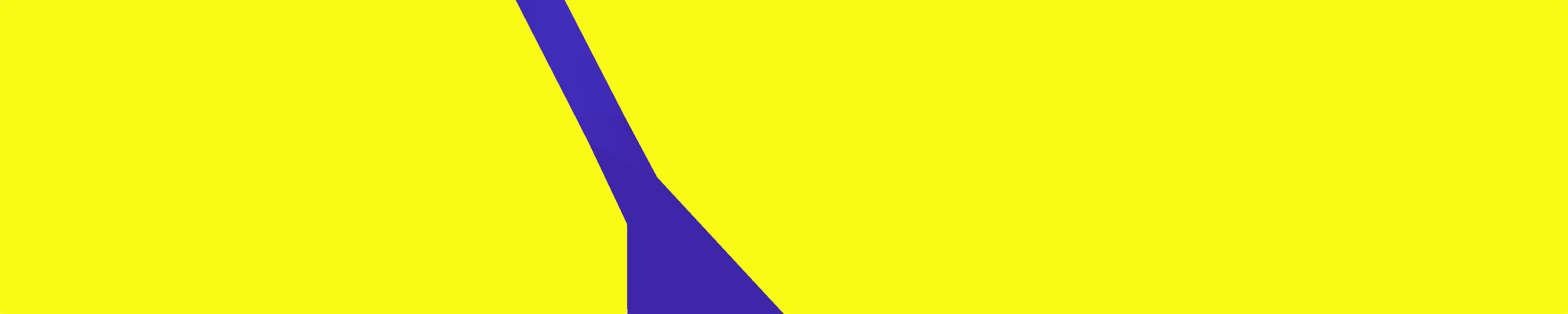};

\end{groupplot}
\end{tikzpicture}
 	\colorbarMatlabParula{0}{0.25}{0.5}{0.75}{1}        
	\caption{Shock-agnostic space-time mesh and first-order solution (phase) used to initialize HOIST solver (\textit{top}), and converged HOIST solution (phase) with (\textit{middle}) and without (\textit{bottom}) mesh edges.}
 	\label{fig:two_shock_example_phase}
\end{figure}

To demonstrate the proposed approach, we consider a one-dimensional Riemann problem
of the Euler equations (\ref{eqn:multiphase_gov_eqn}) involving two ideal gases with
initial condition
 \begin{equation} \label{eqn:motiv_eq}
 \rho(x,0) = \begin{cases} 1 & x<0.4 \\ \alpha & x \geq 0.5 \\  0.125 & \text{else} \end{cases}, \quad
  v(x,0) = \begin{cases} 0 & x<0.4 \\  \beta & x \geq 0.5 \\0 & \text{else} \end{cases} , \quad
  P(x,0) = \begin{cases} 0.1 & x<0.4 \\ \omega  & x \geq 0.5 \\ 0.1 & \text{else} \end{cases}, \quad
          \gamma(x, 0) = \begin{cases} 5/3 & x<0.4 \\ 5/3  & x \geq 0.5 \\ 7/5 & \text{else} \end{cases}
\end{equation}
where $\rho(x,t) \in \Rbb_{> 0}$ is the density of the fluid at
$x \in \Omega_x \coloneqq (0, 1)$ and $t \in \Tcal \coloneqq (0, 0.2]$,
$v(x, t) \in \Rbb$ is the velocity of the fluid at $(x,t) \in \Omega_x\times\Tcal$,
$P(x,t) \in \Rbb_{\ge 0}$ is the pressure of the fluid at $(x,t) \in \Omega_x\times\Tcal$,
and $\gamma(x,t) \in \Rbb_{\ge 0}$ is the ratio of specific heats at $(x,t) \in \Omega_x\times\Tcal$, and $\alpha = 0.4263194281784951$, $\beta = -0.92745262004894879$, $\omega = 0.30313017805064679$.
This problem features a stationary material interface at $x = 0.4$ and a left-moving
shock and material interface at $x = 0.5$. Once the shock impinges on the stationary
interface, the material interface begins to move and both a reflected and pass-through
shock are created. The proposed approach is initialized from a first-order solution on
a shock-agnostic mesh, which includes material mixtures because the mesh is not aligned
with the interface yet. However, at convergence, mesh faces are aligned with the interface,
which allows the contact discontinuity to be represented perfectly and a solution completely
devoid of mixtures is obtained
(Figure~\ref{fig:two_shock_example_dens}-\ref{fig:two_shock_example_phase}).
The shock-contact interaction and subsequent shock reflection in the space-time
solution would require re-meshing and solution transfer in the method-of-lines setting.
However, these simply manifest as triple points in space-time, which are handled robustly
by implicit shock tracking.

\subsection{Contributions}
The contributions of this work are three-fold. First and foremost, the proposed
high-order sharp-interface methodology for real gas two-phase flow simulations
based on implicit shock tracking is novel. The method overcomes many of the traditional
challenges of sharp-interface approaches to two-phase flow at the cost of increased
complexity and computational expense from the space-time formulation. A similar
approach \cite{corrigan2019moving,Corrigan2019a,nourgaliev2021implicit}
uses the Moving Discontinuous Galerkin Method with Interface
Condition Enforcement (MDG-ICE) version of implicit shock fitting instead of
HOIST for ideal gases. This work also presents the flux Jacobian and its
complete eigenvalue decomposition for both a single- and two-phase real gas
in multiple spatial dimensions. These eigenvalue decompositions are used to
construct upwinded numerical flux functions (e.g., Roe's flux \cite{roe1981approximate});
however, they are usually only presented in one dimension \cite{saurel1999simple}.

\subsection{Outline}
\label{sec:intro:outline}
The remainder of this paper is organized as follows. Section~\ref{sec:gov_eqns_total}
systematically builds up a space-time formulation of a time-dependent conservation
law from a traditional formulation that separates the temporal and spatial independent
variables and transforms the space-time conservation law to a reference domain such that
domain deformations appear explicitly in the conservation law. This section specializes
this general formulation to the single- and two-phase, compressible Euler equations,
including a complete description the flux Jacobians and their eigenvalue decompositions
(commonly use to build numerical flux functions). The section closes with an overview
of the three equations of state considered in this work (ideal gas, stiffened gas, and
Becker-Kistiakowsky-Wilson model). Section~\ref{sec:govern:disc} details a high-order
DG discretization of the transformed space-time conservation law.
Section~\ref{sec:ist} uses this discretization to formulate the implicit
shock tracking method and review relevant details in the space-time setting from
\cite{naudet2024space}. Section~\ref{sec:numexp} applies the proposed methods to a
sequence of increasingly difficult problems beginning with a simple single-phase
validation case and closes with a spherically symmetric underwater blast problem.
Finally, Section~\ref{sec:conclude} offers conclusions.

\section{Governing equations}
\label{sec:gov_eqns_total}
In this section we introduce the governing equations considered in this manuscript, namely, inviscid, compressible single- and two-phase flow involving real gases. We begin with
a general form of a time-dependent, inviscid conservation law
(Section~\ref{sec:govern:spatial}), its formulation as a space-time
conservation law (Section~\ref{sec:govern:sptm}), and
a reformulation of the space-time conservation law on a
fixed reference domain (Section~\ref{sec:govern:transf}) because the proposed
shock tracking method will deform the space-time domain to align element faces
with discontinuities. Next, we introduce the Euler equations of gasdynamics for
single-phase (Section~\ref{sec:gov_singlephase}) and two-phase
(Section~\ref{sec:gov_multiphase}) of real gases with a generic equation of state.
The inviscid projected Jacobian and its analytical eigenvalue decomposition are
provided as these play a key role in the construction of numerical fluxes 
schemes \cite{glaister1988approximate}. We use a phase variable to
track the two phases with the mixture model described in
Section~\ref{sec:multiphase_p_and_derivs}. We close this section with the specific
real gases considered in the numerical experiements (Section~\ref{sec:gov_eos}) to
demonstrate the versatility of the framework in
Sections~\ref{sec:gov_singlephase}-\ref{sec:gov_multiphase} for any equation of state.

\subsection{Time-dependent system of conservation laws}
\label{sec:govern:spatial}
Consider a general system of $m$ hyperbolic partial differential equations
posed in the spatial domain $\Omega_x \subset \Rbb^{d'}$ over the time interval
$\mathcal{T} \coloneqq (t_0, t_1) \subset \Rbb_{\ge 0}$
\begin{equation} \label{eqn:gen_cons_law}
 \pder{\physU_x}{t}+ \nabla_x \cdot \physF_x (\physU_x) = \physS_x(\physU_x),
\end{equation}
where $t \in \mathcal{T}$ is the temporal coordinate,
$x = (x_1, ..., x_{d'}) \in \Omega_x$ is the spatial coordinate,
$\physU_x(\cdot,t) : \Omega_x \rightarrow \Rbb^m$ is the conservative state
implicitly defined as the solution to (\ref{eqn:gen_cons_law}),
$\physF_x : \Rbb^m \rightarrow \Rbb^{m \times d'}$
with $\physF_x : W_x \mapsto \physF_x(W_x)$ is the physical flux function,
$\physS_x : \Rbb^m \rightarrow \Rbb^m$ is the physical source term,
$(\nabla_x \cdot)$ is the divergence operator on the domain $\Omega_x$ defined as
$(\nabla_x\cdot\psi)_i := \partial_{x_j} \psi_{ij}$ (summation implied on repeated
index), and $\partial \Omega_x$ is the boundary of the spatial domain (with appropriate boundary conditions prescribed). In general, the solution $U(x,t)$ may contain
discontinuities, in which case, the conservation law (\ref{eqn:gen_cons_law}) holds
away from the discontinuities and the Rankine-Hugoniot conditions hold at discontinuities.

The Jacobian of the projected inviscid flux is 
\begin{equation} \label{eqn:space_inv_proj_jacob}
  B_x : \Rbb^m \times \Sbb_{d'} \rightarrow \Rbb^{m\times m}, \qquad
	B_x : (W_x, \eta_x) \mapsto \frac{\partial [\physF_x(W_x) \eta_x]}{\partial W_x},
\end{equation}
where $\Sbb_{d'} \coloneqq \{ \eta\in\Rbb^{d'} \mid \norm{\eta} = 1\}$. The eigenvalue
decomposition of the Jacobian is
\begin{equation} \label{eqn:spat_eigval_decomp}
  B_x(W_x, \eta_x) = V_x(W_x, \eta_x) \Lambda_x(W_x, \eta_x) V_x(W_x, \eta_x)^{-1},
\end{equation}
where $\Lambda_x : \Rbb^m \times \Sbb_{d'} \rightarrow \Rbb^{m\times m}$
is a diagonal matrix containing the real eigenvalues of $B_x$ and the columns of
$V_x : \Rbb^m \times \Sbb_{d'} \rightarrow \Rbb^{m\times m}$
contain the corresponding right eigenvectors.

\subsection{Space-time formulation} 
\label{sec:govern:sptm} 
The conservation law in (\ref{eqn:gen_cons_law}) describes a general time-dependent
system of conservation laws in a $d'$-dimensional spatial domain. Because the proposed
method tracks discontinuities over space-time slabs, we reformulate
(\ref{eqn:gen_cons_law}) as a steady conservation law in a space-time domain
\cite{lowrie1998space,sudirham2006space, klaij2006space}.
To this end, we define the space-time domain as
$\Omega \coloneqq \Omega_x \times \Tcal \subset \Rbb^d$ ($d = d'+1$) with boundary
$\partial\Omega$, and let $z=(x,t)\in\Omega$ denote the space-time coordinate.
Because the space-time domain is a Cartesian product of the spatial domain with
a time interval, we will refer to it as a \textit{space-time slab}. The temporal
domain $\Tcal$ may be the entire time window of interest, or a portion of it.
The boundary of a space-time slab $\partial\Omega$ consists of three pieces: 1) the
spatial boundary, $\partial\Omega_x \times \Tcal$, 2) the bottom of the slab,
$\Omega_x \times \{t_0\}$, and 3) the top of the slab, $\Omega_x \times \{t_1\}$.
Without loss of generality, we formulate the space-time conservation law, as well
as its transformation (Section~\ref{sec:govern:sptm}) and discretization
(Section~\ref{sec:govern:disc}) over a single arbitrary slab corresponding
to the time interval $\Tcal$; in practice, we use a sequence of space-time slabs
to cover the temporal domain of interest.

The conservation law in (\ref{eqn:gen_cons_law}) can be written as a
steady conservation law over the space-time slab as
\begin{equation} \label{eqn:sptm_claw}
\nabla \cdot \physF (\physU) = \physS(\physU),
\end{equation}
where $U : \Omega \rightarrow \Rbb^m$ is the space-time conservative vector implicitly
defined as the solution of (\ref{eqn:sptm_claw}) and related to the solution of the
spatial conservation law as $\physU : z \mapsto  \physU_x(x,t)$,
$\physF : \Rbb^m \rightarrow \Rbb^{m \times d}$ and
$\physS : \Rbb^m \rightarrow \Rbb^{m \times d}$ are the
space-time flux function and source term, respectively, and related
to the spatial conservation law terms as
\begin{equation} \label{eqn:sptm_flux_src}
 \physF : W \mapsto \begin{bmatrix} \physF_x(W) & W \end{bmatrix}, \qquad
 \physS : W \mapsto \physS_x(W),
\end{equation}
and $(\nabla\cdot)$ is the space-time divergence operator defined as
$\nabla\cdot\begin{bmatrix}\psi&\phi\end{bmatrix} = \nabla_x\cdot\psi + \partial_t \phi$.

The Jacobian of the space-time projected inviscid flux
$B : \Rbb^m \times \Sbb_d \rightarrow \Rbb^{m\times m}$
is
\begin{equation} \label{eqn:space_time_proj_jac}
  B : (W, \eta) \mapsto \frac{\partial [\physF (W) \eta]}{\partial W}.
\end{equation}
Any $\eta\in\Sbb_d$ can be written as $\eta = (n_x, n_t)$ where
$n_x \in \Rbb^{d'}$ and $n_t \in \Rbb$, and $n_x$ can be expanded
$n_x = \eta_x \norm{n_x}$ with $\eta_x \in \Sbb_{d'}$ ($\eta_x$
is uniquely defined as $n_x/\norm{n_x}$ in the case where $n_x \neq 0$,
otherwise it is arbitrary). From this expansion and the form of the inviscid
flux in (\ref{eqn:sptm_flux_src}), the space-time Jacobian and be related to
the original Jacobian
\begin{equation} \label{eqn:jac_sp_sptm}
 B(W, \eta) = B_x(W,\eta_x) \norm{n_x} + n_t I_m.
\end{equation}
The eigenvalue decomposition of the projected Jacobian is denoted
\begin{equation}
 B(W, \eta) = V(W,\eta) \Lambda(W,\eta) V(W,\eta)^{-1},
\end{equation}
where $\Lambda : \Rbb^m \times \Sbb_d$ is a diagonal matrix containing the
real eigenvalues of $B$ and the columns of
$V : \Rbb^m \times \Sbb_d \rightarrow \Rbb^{m\times m}$
contain the corresponding right eigenvectors. Owing to the relationship
between the projected Jacobians of the spatial and space-time inviscid
fluxes (\ref{eqn:jac_sp_sptm}), their eigenvalue decompositions are related as
\begin{equation} \label{eqn:sptm_eigval_decomp}
 \Lambda(W, \eta) = \Lambda_x(W,\eta_x)\norm{n_x} + n_t I_m, \qquad
 V(W, \eta) = V_x(W,\eta_x)
\end{equation}

\subsection{Transformed space-time conservation law on a fixed reference domain}
\label{sec:govern:transf}
Because the proposed numerical method is based on deforming the space-time domain
to track discontinuities with the computational grid, it is convenient to recast
the space-time conservation law such that domain deformations appear explicitly.
To this end, we define $\bar\Omega \subset \Rbb^d$ as a fixed space-time reference
domain, which we require to take the form of a space-time slab, i.e.,
$\bar\Omega \coloneqq \Omega_x \times \Tcal$. Let $\Gbb$ be the collection of
diffeomorphisms from the reference domain to the physical domain, i.e., for any
$\Gcal \in \Gbb$, $\Gcal : \bar\Omega \rightarrow \Omega$ with
$\Gcal : \bar{z} \mapsto \Gcal(\bar{z})$, that preserve the space-time slab
structure of the physical domain. The space-time conservation law in
(\ref{eqn:sptm_claw}) can be reformulated as a conservation law over the
reference domain as \cite{Zahr2018,naudet2024space}
\begin{equation} \label{eqn:trans1}
  \bar{\nabla} \cdot \transF(\transU; G) = \transS (\transU; g),
\end{equation}
where $\bar{U} : \bar\Omega \rightarrow \Rbb^m$ is the solution of the transformed
conservation law, $\bar\Fcal : \Rbb^m \times \Rbb^{d\times d} \rightarrow \Rbb^{m\times d}$
and $\bar\Scal : \Rbb^m \times \Rbb \rightarrow \Rbb^m$ are the transformed flux function
and source term, respectively, $\bar\nabla\cdot$ is the divergence operator in the
reference domain $\bar\Omega$ defined as
$(\bar\nabla \cdot \psi)_i = \partial_{\bar{z}_j}\psi_{ij}$, 
and the deformation gradient $G : \bar\Omega \rightarrow \Rbb^{d\times d}$
and mapping Jacobian $g : \bar\Omega \rightarrow \Rbb$ are
\begin{equation}
 G : \bar{z} \mapsto \partial_{\bar{z}}\Gcal(\bar{z}), \qquad
 g: \bar{z} \mapsto \det G(\bar{z}).
\end{equation}
The reference domain quantities are defined in terms of the corresponding
physical domain quantities as \cite{Zahr2018}
\begin{equation} \label{eqn:trans2}
 \transU(\bar{z}) = U(\Gcal(\bar{z})),
 \qquad
 \transF : (\bar{W}; \Theta) \mapsto (\det\Theta) \Fcal(\bar{W}) \Theta^{-T},
 \qquad
 \transS : (\bar{W}; q) \mapsto q S(\bar{W}).
\end{equation}

\subsection{Inviscid, compressible single-phase flow of real gas}
\label{sec:gov_singlephase}
Consider flow of an inviscid, compressible fluid through a domain
$\Omega_x \subset \mathbb{R}^{d'}$, where $d' \geq 1$ is the spatial dimension,
governed by the Euler equations (\ref{eqn:gen_cons_law}) with
\begin{eqnarray} 
	U_x = \begin{bmatrix} \rho \\
	\rho v \\
	\rho E \end{bmatrix}, \quad & \Fcal_x(U_x) = \begin{bmatrix} \rho v^T \\
	\rho vv^T + P(\rho, e) I_{d'} \\
	[ \rho E + P(\rho, e) ] v^T \end{bmatrix}, \quad
	\Scal_x(U_x) = 0,
\end{eqnarray}
where $\rho : \Omega_x \times \Tcal \rightarrow \Rbb_{\ge 0}$ is the density
of the fluid, $v : \Omega_x \times \Tcal \rightarrow \Rbb^{d'}$ is the velocity
of the fluid, $E : \Omega_x \times \Tcal \rightarrow \Rbb$ is the total energy
of the fluid, and $e : \Omega_x \times \Tcal \rightarrow \Rbb_{\ge 0}$ is the
specific internal energy of the fluid. The \textit{equation of state} defines
the pressure of the fluid, $P : \Rbb_{\ge 0} \times \Rbb_{\ge 0} \rightarrow \Rbb_{\ge 0}$,
as a function of its density and specific internal energy, i.e.,
$P : (\rho, e) \mapsto P(\rho, e)$. The partial derivatives of this function play a
central role in the flux Jacobian and its eigenvalue decomposition so we introduce
the following short-hand notation:
$P_\rho : \Rbb_{\ge 0} \times \Rbb_{\ge 0} \rightarrow \Rbb$ and 
$P_e : \Rbb_{\ge 0} \times \Rbb_{\ge 0} \rightarrow \Rbb$, defined as
\begin{equation}
  P_\rho : (\rho, e) \mapsto \frac{\partial P}{ \partial \rho} (\rho, e) \bigg|_{e}, \qquad
  P_e : (\rho, e) \mapsto \frac{\partial P}{ \partial e} (\rho, e) \bigg|_{\rho}
\end{equation}
The internal energy relates to the conservative variables as
\begin{equation} \label{eqn:int_ener_to_total_ener}
e(\rho, \rho v, \rho E) = \frac{\rho E}{\rho} - \frac{\rho v \cdot \rho v}{2 \rho^2}
\end{equation}
and enthalpy of the fluid is defined as $H = (\rho E + P)/ \rho$. For convenience, we
introduce a new function, $\bar{P}$, that represents the pressure \textit{as a function
of the conservative variables}
\begin{equation}
  \bar{P} : \Rbb_{\ge 0} \times \Rbb^{d'} \times \Rbb \rightarrow \Rbb_{\ge 0}, \qquad
  \bar{P} : (\rho, \rho v, \rho E) \mapsto P(\rho, e(\rho, \rho v, \rho E)).
\end{equation}
This new definition will allows us to easily distinguish between derivatives of pressure
under constant internal energy versus the derivatives of pressure under constant momentum
or total energy without cumbersome notation. We use the chain rule to obtain the following
expressions for the derivatives of pressure with the conservative variables held constant
\begin{equation} \label{eqn:pressure_derivs}
\begin{aligned}
	\frac{\partial \bar{P}}{\partial \rho} (\rho, \rho v, \rho E)\bigg|_{\rho v, \rho E} &= \frac{\partial P}{ \partial \rho} (\rho, e) \bigg|_{e}  +  \frac{\partial P}{ \partial e} (\rho, e) \bigg|_{\rho} \frac{\partial e}{\partial \rho} (\rho, \rho v, \rho E)  \bigg|_{\rho v, \rho E} &&= P_\rho(\rho,e) - \frac{P_e(\rho,e)}{\rho}(E-\norm{v}^2) \\
	\frac{\partial \bar{P}}{\partial \rho v}  (\rho,  \rho v, \rho E ) \bigg|_{\rho, \rho E} &= \frac{\partial P}{ \partial e} (\rho, e) \bigg|_{\rho} \frac{\partial e}{\partial \rho v} (\rho, \rho v, \rho E)  \bigg|_{\rho, \rho E} &&= -\frac{P_e(\rho, e)}{\rho}v \\
	\frac{\partial \bar{P}}{\partial \rho E}  (\rho, \rho v, \rho E )\bigg|_{\rho, \rho v}  &= \frac{\partial P}{ \partial e} (\rho, e) \bigg|_{\rho} \frac{\partial e}{\partial \rho E} (\rho, \rho v, \rho E)  \bigg|_{\rho, \rho v} &&=  \frac{P_e(\rho,e)}{\rho}.
\end{aligned}
\end{equation}
For brevity, we introduce the following short-hand notation: 
$\bar{P}_\rho : \Rbb_{\ge 0} \times \Rbb^{d'} \times \Rbb \rightarrow \Rbb$,
$\bar{P}_{\rho v} : \Rbb_{\ge 0} \times \Rbb^{d'} \times \Rbb \rightarrow \Rbb^{d'}$,
$\bar{P}_{\rho E} : \Rbb_{\ge 0} \times \Rbb^{d'} \times \Rbb \rightarrow \Rbb$, defined as
\begin{equation}
\begin{aligned} \label{eqn:pressure_derivs2}
	\bar{P}_\rho : &(\rho, \rho v, \rho E) \mapsto \pder{\bar{P}}{\rho}(\rho,\rho v,\rho E)\bigg|_{\rho v, \rho E}, \\
	\bar{P}_{\rho v} : &(\rho, \rho v, \rho E) \mapsto \pder{\bar{P}}{\rho v}(\rho,\rho v, \rho E)\bigg|_{\rho, \rho E}, \\
	\bar{P}_{\rho E} : &(\rho, \rho v, \rho E) \mapsto \pder{\bar{P}}{\rho E}(\rho, \rho v, \rho E)\bigg|_{\rho, \rho v}.
\end{aligned}
\end{equation}

From these definitions, the spatial projected inviscid Jacobian for the Euler equations
with a general equation of state is
\begin{equation} \label{eqn:proj_jac_non_ideal}
	B_x : (U_x, \eta_x) \mapsto
	\begin{bmatrix}
		0 & \eta_x^T & 0\\
		-v_n v + \bar{P}_\rho \eta_x & v_n I_{d'} + v \eta_x^T +  \eta_x \bar{P}_{\rho v}^T & \bar{P}_{\rho E} \eta_x \\
		\left(\bar{P}_\rho - H \right) v_n & H \eta_x^T + v_n \bar{P}_{\rho v}^T & \left(1 + \bar{P}_{\rho E} \right) v_n
	\end{bmatrix}, 
\end{equation}
which, using (\ref{eqn:pressure_derivs}), simplifies to
\begin{eqnarray} \label{eqn:proj_jac_non_ideal_simp}
	B_x : (U_x, \eta_x) \mapsto
	\begin{bmatrix}
		0 & \eta_x^T & 0\\
		-v_n v + \bar{P}_\rho \eta_x & v_n I_{d'} + v \eta_x^T - \frac{P_e}{\rho}  \eta_x v^T & \frac{P_e}{\rho} \eta_x \\
		\left(\bar{P}_\rho - H \right) v_n & H \eta_x^T - v_n \frac{P_e}{\rho} v^T  & \left(1 + \frac{P_e}{\rho}  \right) v_n,
	\end{bmatrix},
\end{eqnarray}
where $\eta_x \in \Rbb^{d'}$ is the unit normal, $v_n \coloneqq v \cdot \eta_x$, and
$I_{d'}$ is the $d'\times d'$ identity matrix. The matrix of right eigenvectors is given
by 
\begin{equation} \label{eqn:right_evec_non_ideal}
	V_x : (U_x, \eta_x) \mapsto
	\begin{bmatrix} 
		1 & \eta_x^T & 1 \\
		v - c \eta_x & (v-\eta_x)\eta_x^T  + I_{d'} & v + c \eta_x \\
		H - v_n c & v^T + (\theta - v_n) \eta_x^T & H + v_n c 
	\end{bmatrix}, 
\end{equation}
where $\theta$ is defined as
\begin{equation} \label{eqn:eqn_for_theta}
 \theta = ||v||^2 - \bar{P}_\rho   \frac{\rho}{P_e}.
\end{equation}
The sound speed, $c$, is defined as
\begin{equation} \label{eqn:sound_speed_nonideal_single}
c^2 = \bar{P}_\rho + (H - ||v||^2) \frac{P_e}{\rho};
\end{equation}
see \ref{sec:speed_of_sound_deriv_single} for derivation. The diagonal matrix of
eigenvalues corresponding to these right eigenvectors is
\begin{equation} \label{eqn:evals_single}
	\Lambda_x : (U_x, \eta_x) \mapsto
	\begin{bmatrix} 
		v_n-c & 0& 0  \\
		0 & v_n I_{d'} & 0  \\
		0 & 0 & v_n+c
	\end{bmatrix}  
\end{equation}
Finally, the matrix of left eigenvectors is given by 
\begin{equation} \label{eqn:left_evec_non_ideal}
	V_x^{-1} : (U_x,\eta_x) \mapsto
	\frac{1}{2 c^2 \rho}
	\begin{bmatrix} 
		\rho \bar{P}_\rho+ c \rho v_n & -c \rho \eta_x^T - P_e v^T & P_e  \\
		2 c^2 \rho (\eta_x v_n - v) + 2 \eta_x \omega & 2 c^2 \rho (I_{d'} - \eta_x \eta_x^T) + 2 P_e \eta_x v^T & -2 \eta_x P_e   \\
		\rho \bar{P}_\rho - c \rho v_n & c \rho \eta_x^T - P_e v^T & P_e
	\end{bmatrix}, 
\end{equation}
where $\omega  = P_e (H - ||v||^2)$.

\subsection{Inviscid, compressible two-phase flow of real gases}
\label{sec:gov_multiphase}
Next, consider flow of two inviscid, compressible fluids through a domain
$\Omega_x \subset \mathbb{R}^{d'}$, where $d' \geq 1$ is the spatial dimension,
governed by the Euler equations using a phase variable with interface advection
equation \cite{sun2007sharp}
(\ref{eqn:gen_cons_law}) with
\begin{eqnarray} \label{eqn:multiphase_gov_eqn}
	U_x = \begin{bmatrix} \rho \\
	\rho v \\
	\rho E \\
	\rho \phi \end{bmatrix}, \quad  & \Fcal_x(U_x) = \begin{bmatrix} \rho v^T \\
	\rho vv^T + P(\rho, e, \rho \phi) I_{d'} \\
	[ \rho E + P(\rho, e, \rho \phi) ]v^T \\
	\rho \phi v^T \end{bmatrix}, \quad
	\Scal_x(U_x) = 0,
\end{eqnarray}
where $\rho$, $v$, and $E$ are defined in Section~\ref{sec:gov_singlephase}, and
$\phi : \Omega_x \times \Tcal \rightarrow [0,1]$ tracks the phase of the fluid,
i.e., if $\phi(x,t) = 1$ then the fluid 1 occupies $x\in\Omega_x$ at time $t$, if
$\phi(x,t) = 0$ then fluid 2 occupies $x\in\Omega_x$ at time $t$, otherwise some
mixture of the two fluids (Section~\ref{sec:multiphase_p_and_derivs}) occupies the
space-time location. In the two-phase setting, the equation of state defines the
pressure of the fluid,
$P : \Rbb_{\ge 0} \times \Rbb_{\ge 0} \times \Rbb_{\ge 0} \rightarrow \Rbb_{\ge 0}$,
as a function of its density, specific internal energy, and phase, i.e.,
$P : (\rho, e, \rho\phi) \mapsto P(\rho, e, \rho\phi)$. We use the following short-hand
notation for the partial derivatives of this function:
$P_\rho : \Rbb_{\ge 0} \times \Rbb_{\ge 0} \times \Rbb_{\ge 0} \rightarrow \Rbb$ and
$P_{e} : \Rbb_{\ge 0} \times \Rbb_{\ge 0} \times \Rbb_{\ge 0} \rightarrow \Rbb$,
defined as
\begin{equation}
  P_\rho : (\rho, e, \rho\phi) \mapsto \frac{\partial P}{ \partial \rho} (\rho, e, \rho\phi) \bigg|_{e,\rho\phi}, \qquad
  P_e : (\rho, e, \rho\phi) \mapsto \frac{\partial P}{ \partial \rho} (\rho, e, \rho\phi) \bigg|_{\rho,\rho\phi}.
\end{equation}
Following Section~\ref{sec:gov_singlephase} we introduce a new function, $\bar{P}$, that
represents the pressure as a function of the conservative variables
\begin{equation}
  \bar{P} : \Rbb_{\ge 0} \times \Rbb^{d'} \times \Rbb \times \Rbb_{\ge 0} \rightarrow \Rbb_{\ge 0}, \qquad
  \bar{P} : (\rho, \rho v, \rho E, \rho \phi) \mapsto P(\rho, e(\rho, \rho v, \rho E), \rho \phi).
\end{equation}
We use the chain rule to obtain the following expressions for the derivatives of
pressure with the conservative variables held constant
\begin{equation} \label{eqn:pressure_derivs_multi}
\begin{aligned}
        \pder{\bar{P}}{\rho}(\rho, \rho v, \rho E,\rho\phi)\bigg|_{\rho v, \rho E,\rho\phi} &= \pder{P}{\rho} (\rho, e, \rho\phi) \bigg|_{e,\rho\phi} + \pder{P}{e} (\rho, e, \rho\phi) \bigg|_{\rho,\rho\phi} \pder{e}{\rho} (\rho,\rho v,\rho E)  \bigg|_{\rho v, \rho E} \\
        \pder{\bar{P}}{\rho v}(\rho,  \rho v, \rho E, \rho\phi) \bigg|_{\rho, \rho E, \rho\phi} &= \pder{P}{e} (\rho, e, \rho\phi) \bigg|_{\rho,\rho\phi} \pder{e}{\rho v} (\rho,\rho v, \rho E, \rho\phi)  \bigg|_{\rho, \rho E, \rho\phi} = -\frac{P_e(\rho, e, \rho\phi)}{\rho}v \\
	\pder{\bar{P}}{\rho E}  (\rho, \rho v, \rho E, \rho\phi)\bigg|_{\rho, \rho v, \rho\phi}  &= \pder{P}{e} (\rho, e, \rho\phi) \bigg|_{\rho,\rho\phi} \pder{e}{\rho E} (\rho, \rho v, \rho E, \rho\phi)  \bigg|_{\rho, \rho v, \rho\phi} =  \frac{P_e(\rho,e,\rho\phi)}{\rho}\\
	\pder{\bar{P}}{\rho\phi} (\rho, \rho v, \rho E, \rho \phi )\bigg|_{\rho, \rho v, \rho E} &= \pder{P}{\rho\phi} (\rho, e, \rho \phi) \bigg|_{\rho, e}.
\end{aligned}
\end{equation}
For brevity, we introduce the following short-hand notation:
$\bar{P}_\rho : \Rbb_{\ge 0} \times \Rbb^{d'} \times \Rbb \times \Rbb_{\ge 0} \rightarrow \Rbb$,
$\bar{P}_{\rho v} : \Rbb_{\ge 0} \times \Rbb^{d'} \times \Rbb \times \Rbb_{\ge 0} \rightarrow \Rbb^{d'}$,
$\bar{P}_{\rho E} : \Rbb_{\ge 0} \times \Rbb^{d'} \times \Rbb \times \Rbb_{\ge 0} \rightarrow \Rbb$, defined as
\begin{equation}
\begin{aligned}
        \bar{P}_\rho : &(\rho, \rho v, \rho E, \rho\phi) \mapsto \pder{\bar{P}}{\rho}(\rho,\rho v,\rho E, \rho\phi)\bigg|_{\rho v, \rho E,\rho\phi}, \\
        \bar{P}_{\rho v} : &(\rho, \rho v, \rho E, \rho\phi) \mapsto \pder{\bar{P}}{\rho v}(\rho,\rho v, \rho E,\rho\phi)\bigg|_{\rho, \rho E, \rho\phi}, \\
        \bar{P}_{\rho E} : &(\rho, \rho v, \rho E, \rho\phi) \mapsto \pder{\bar{P}}{\rho E}(\rho, \rho v, \rho E, \rho\phi)\bigg|_{\rho, \rho v, \rho\phi}, \\
        \bar{P}_{\rho \phi} : &(\rho, \rho v, \rho E, \rho\phi) \mapsto \pder{\bar{P}}{\rho \phi}(\rho, \rho v, \rho E, \rho\phi)\bigg|_{\rho, \rho v, \rho E}.
\end{aligned}
\end{equation}

From these definitions, the spatial projected inviscid Jacobian for the two-phase
Euler equations with general equations of state is
\begin{equation} \label{eqn:proj_invis_job_multi_orig}
	B_x : (U_x, \eta_x) \mapsto
	\begin{bmatrix}
		0 &  \eta_x^T & 0 & 0 \\
		-v_n v + \bar{P}_\rho \eta_x & v_n I_{d'} + v \eta_x^T +  \eta_x \bar{P}_{\rho v}^T &  \bar{P}_{\rho E} \eta_x & \bar{P}_{\rho \phi} \eta_x \\
		\left(\bar{P}_\rho- H\right) v_n & H \eta_x^T + v_n \bar{P}_{\rho v}^T  & (1 + \bar{P}_{\rho E})v_n & \bar{P}_{\rho \phi} v_n  \\
	- \phi v_n & \phi \eta_x^T & 0 & v_n
	\end{bmatrix}, 
\end{equation}
which, using (\ref{eqn:pressure_derivs_multi}), simplifies to
\begin{equation} \label{eqn:proj_invis_job_multi_simp}
	B_x : (U_x, \eta_x) \mapsto
	\begin{bmatrix}
		0 & \eta_x^T & 0 & 0 \\
		-v_n v + \bar{P}_\rho \eta_x & v_n I_{d'} + v \eta_x^T - \frac{P_e}{\rho} \eta_x v^T &  \frac{P_e}{\rho}  \eta_x & \bar{P}_{\rho \phi}\eta_x \\
		\left(\bar{P}_\rho- H\right) v_n & H \eta_x^T - v_n \frac{P_e}{\rho} v^T  & (1 + \frac{P_e}{\rho} )v_n & \bar{P}_{\rho \phi} v_n  \\
		-\phi v_n & \phi \eta_x^T & 0 & v_n
	\end{bmatrix}  
\end{equation}
where $\eta_x \in \Rbb^{d'}$ is the unit normal, $v_n \coloneqq v \cdot \eta_x$, and
$I_{d'}$ is the $d'\times d'$ identity matrix. The matrix of right eigenvectors is given
by
\begin{equation} \label{eqn:evecs_multi}
	V_x : (U_x, \eta_x) \mapsto
	\begin{bmatrix} 
		1 & \eta_x^T & 0 & 1 \\
		v - c \eta_x & (v-\eta_x)\eta_x^T  + I_{d'} & 0 & v + c \eta_x \\
		H - v_n c & v^T + (\theta - v_n) \eta_x^T & - \bar{P}_{\rho \phi} & H + v_n c  \\
		\phi & \vec{0}^T & \frac{P_e}{\rho} & \phi
	\end{bmatrix}, 
\end{equation}
where $\theta$ is defined in (\ref{eqn:eqn_for_theta}) and the sound speed, $c$, is 
defined as
\begin{equation} \label{eqn:sndsp_gen_mix}
c^2 = \bar{P}_\rho + (H - ||v||^2)  \frac{P_e}{\rho}  + \phi P_{\rho \phi}.
\end{equation}
Derivation of the (\ref{eqn:sndsp_gen_mix}) is provided in
\ref{sec:speed_of_sound_deriv_multi}.
The diagonal matrix of eigenvalues corresponding to these right eigenvectors is
\begin{equation} \label{eqn:evals_multi}
	\Lambda_x : (U_x, \eta_x) \mapsto
	\begin{bmatrix} 
		v_n-c & 0& 0 & 0 \\
		0 & v_n I_{d'} & 0 & 0 \\
		0 & 0 & v_n & 0  \\
		0 & 0 & 0& v_n+c
	\end{bmatrix}  
\end{equation}
Finally, the matrix of left eigenvectors is given by
\begin{equation}
	V^{-1} : (U_x,\eta_x) \mapsto
	\frac{1}{2 c^2 \rho}
	\begin{bmatrix} 
		\rho \bar{P}_\rho+ c \rho v_n & -c \rho \eta_x^T - P_e v^T & P_e & \bar{P}_{\rho \phi} \rho \\
		2 c^2 \rho (\eta_x v_n - v) + 2 \eta_x \omega & 2 c^2 \rho (I_{d'} - \eta_x \eta_x^T) + 2 P_e \eta_x v^T & -2 \eta_x P_e & -2 \rho \bar{P}_{\rho \phi} \eta_x  \\
	\	\frac{-2 \phi \bar{P}_\rho \rho^2}{P_e} & 2 \phi \rho v^T & 2 \phi \rho & \frac{2 \rho \omega}{P_e} \\
		\rho \bar{P}_\rho - c \rho v_n & c \rho \eta_x^T - P_e v^T & P_e & \bar{P}_{\rho \phi}  \rho
	\end{bmatrix}  
\end{equation}
where $\omega  = \left(H P_e - P_e ||v||^2 + \rho \phi \bar{P}_{\rho \phi} \right)$.

\subsection{Two-phase mixture equation of state}
\label{sec:multiphase_p_and_derivs}
Recall from Section~\ref{sec:intro} that the method proposed in this manuscript
is a \textit{sharp-interface} model for two-phase flow; the phase-field formulation is
used as a means to converge to the sharp interface, i.e., $\phi(x, t) \in \{0,1\}$ using
implicit shock tracking. As such, the requirements on our mixture model are much weaker
than most phase-field models where $\phi(x,t) \in [0,1]$ and require thermodynamic
consistency in mixture regions, i.e., $(x,t)\in\Omega_x\times\Tcal$ such that 
$0 < \phi(x,t) < 1$ \cite{sun2007sharp}. In our setting, we only require the mixture
model is consistent in the sense that all theromdynamic properties (e.g., pressure and
sound speed) reduce to that of the individual fluids as $\phi \rightarrow {0,1}$,
e.g., in the case of pressure,
\begin{equation} \label{eqn:gen_pres_mixture}
	\lim_{\phi\rightarrow 1} P(\rho,e,\rho\phi) = P_1(\rho, e), \qquad
	\lim_{\phi\rightarrow 0} P(\rho,e,\rho\phi) = P_2(\rho, e),
\end{equation}
where $P_1(\rho,e)$ and $P_2(\rho,e)$ are the equations of state of the two fluids
under consideration. No requirements are imposed on mixtures $0 < \phi < 1$. Thus,
we choose the simplest mixture model
\begin{equation} \label{eqn:mixture_eos}
	P(\rho, e, \rho \phi) = \frac{\rho \phi}{\rho} P_1(\rho, e) + \left(1 - \frac{\rho \phi}{\rho}\right) P_2(\rho, e)
\end{equation}
which clearly satisfies (\ref{eqn:gen_pres_mixture}) and the sound speed also approaches
the material truths as $\phi \rightarrow {0,1}$ (derivation in
\ref{sec:multiphase_sound_speed_to_material}).

The partial derivatives of $P$ are needed to define the inviscid projected Jacobian and
eigenvalue decompositions of the two-phase Euler equations
(Section \ref{sec:gov_multiphase}). First, the partial derivative with respect to density,
$P_\rho(\rho, e, \rho\phi)$, is
\begin{equation} \label{eqn:mixture_deriv_rho}
	\begin{aligned}
		P_\rho(\rho,e,\rho\phi) &= \pder{}{\rho}\left(\frac{\rho \phi}{\rho} P_1(\rho, e) + \left(1 - \frac{\rho \phi}{\rho}\right) P_2(\rho, e)\right) \\
		&= \frac{\phi}{\rho} \left(P_2(\rho,e) - P_1(\rho,e) \right) + \phi \left(\pder{P_1}{\rho}(\rho,e)\bigg|_{e} - \pder{P_2}{\rho}(\rho,e)\bigg|_{e}\right) + \pder{P_2}{\rho}(\rho,e)\bigg|_{e}.
	\end{aligned}
\end{equation}
The partial derivative with respect to internal energy, $P_e(\rho,e,\rho\phi)$, is
\begin{equation}\label{eqn:mixture_deriv_e}
	\begin{aligned}
		P_e(\rho,e,\rho\phi) &= \pder{}{e}\left(\frac{\rho \phi}{\rho} P_1(\rho, e) + \left(1 - \frac{\rho \phi}{\rho}\right) P_2(\rho, e)\right) \\
		&= \phi\pder{P_1}{e}(\rho,e)\bigg|_{\rho} + \left(1 - \phi\right) \pder{P_2}{e}(\rho,e)\bigg|_{\rho}
	\end{aligned}
\end{equation}
Finally, the partial derivative with respect to the phase,
$P_{\rho\phi}(\rho, e, \rho\phi)$, is
\begin{equation} \label{eqn:mixture_deriv_rhophi}
	P_{\rho\phi}(\rho,e,\rho\phi) = \pder{}{\rho\phi}\left(\frac{\rho \phi}{\rho} P_1(\rho, e) + \left(1 - \frac{\rho \phi}{\rho}\right) P_2(\rho, e)\right) = \frac{ P_1 - P_2 }{\rho}
\end{equation}
From these three expressions, all terms in Section~\ref{sec:gov_multiphase} are
well-defined once an equation of state is selected for each fluid
(Section~\ref{sec:gov_eos}).

\begin{remark} \label{rem:mixture1}
	In the special case where both fluids are identical, i.e.,
	$P_1(\rho, e) = P_2(\rho, e)$, the mixture equation of state
	(\ref{eqn:mixture_eos}) and its partial derivatives
	(\ref{eqn:mixture_eos})-(\ref{eqn:mixture_deriv_rhophi})
	are independent of $\phi$.
\end{remark}

\subsection{Equations of state considered} \label{sec:gov_eos}
In this section we outline the equations of state considered (Section~\ref{sec:numexp})
in the notation of Section~\ref{sec:gov_singlephase}. In particular, we
consider an ideal gas (Section~\ref{sec:gov_eos:eos_ig}), a stiffened gas model of
water (Section~\ref{sec:gov_eos:eos_sg}), and the Becker-Kistiakowsky-Wilson (BKW)
equation of state \cite{becker1922stosswelle, kistiakowsky1941hydrodynamic, mader1963detonation, suceska2023bkw, neron2024revisiting} used to model gaseous byproducts in detonations
(Section~\ref{sec:gov_eos:eos_BKW}). These single-phase real gas equations of state are
used to construct the two-phase mixture equation of state (\ref{eqn:mixture_eos}) in
Section~\ref{sec:multiphase_p_and_derivs}.

\subsubsection{Ideal gas} \label{sec:gov_eos:eos_ig}
First, we introduce an ideal gas using the formalism of Section~\ref{sec:gov_singlephase}
as a simple demonstration and validation case in Section~\ref{sec:numexp}. An ideal gas
is calorically perfect with equation of state
\begin{equation} \label{eqn:P_IG}
	P^{IG}(\rho, e) = (\gamma -1) \rho e,
\end{equation}
where $\gamma > 1$ is the ratio of specific heats. The ideal gas can also be written in
terms of temperature, $T^{IG} : \Rbb_{\ge 0} \rightarrow \Rbb_{\ge 0}$, as
\begin{equation}\label{eqn:P_IG_2}
	P^{IG}(\rho, e) = \rho R T^{IG}(e),
\end{equation}
where $R \in \Rbb_{>0}$ is the universal gas constant and
$T^{IG} : e \mapsto (\gamma-1) e/R$.
The partial derivatives of this equation of state are
\begin{equation} \label{eqn:P_derivs_IG}
	P_\rho^{IG}(\rho, e) = (\gamma-1)e, \qquad
	P_e^{IG}(\rho, e) = (\gamma-1)\rho.
\end{equation}
The equations for the projected flux Jacobian (\ref{eqn:proj_jac_non_ideal_simp}) and its
eigenvalue decomposition (\ref{eqn:right_evec_non_ideal})-(\ref{eqn:left_evec_non_ideal})
reduce to their usual definitions for an ideal gas \cite{naudet2024space} if
(\ref{eqn:P_IG})-(\ref{eqn:P_derivs_IG}) are substituted for $P$, $P_\rho$, and $P_e$
(\ref{sec:idealgas_lim}).

\subsubsection{Stiffened gas} \label{sec:gov_eos:eos_sg}
Next, we introduce stiffened gas model of water \cite{allaire2002five}
with equation of state
\begin{equation}
	P^{SG}(\rho, e) = (\gamma-1)\rho e - \gamma P^\star,
\end{equation}
where $P^\star$ is a constant representing the molecular attraction between water
molecules. In terms of temperature,
$T^{SG} : \Rbb_{\ge 0}\times\Rbb_{\ge 0} \rightarrow \Rbb_{\ge 0}$,
the stiffened gas equation of state is
\begin{equation}
	P^{SG}(\rho, e) = \rho R T^{SG}(\rho, e) Z^{SG}(\rho, e),
\end{equation}
where $T^{SG} : (\rho, e) \mapsto (e - P^\star/\rho) / C_V$,
$Z^{SG} : (\rho, e) \mapsto 1 - P^\star/(\rho R T^{SG}(\rho, e))$ is the
compressibility factor, and $C_V \in \Rbb_{\ge 0}$ is specific heat at constant volume.
Because this stiffened gas model is just an ideal gas law with an additional constant
term, its partial derivatives are the identical to those of the ideal gas equation of
state, i.e.,
\begin{equation} \label{eqn:P_derivs_SG}
        P_\rho^{SG}(\rho, e) = (\gamma-1)e, \qquad
        P_e^{SG}(\rho, e) = (\gamma-1)\rho.
\end{equation}

\subsubsection{Becker-Kistiakowsky-Wilson (BKW) equation of state}
\label{sec:gov_eos:eos_BKW}
The Becker-Kistiakowsky-Wilson (BKW) equation of state is often used as
a simplified model of the gaseous byproduct from a detonation \cite{becker1922stosswelle, kistiakowsky1941hydrodynamic, mader1963detonation, suceska2023bkw, neron2024revisiting} and takes the form
\begin{equation}
	P^{BKW}(\rho, e) = \hat{P}^{BKW}(\rho, T^{BKW}(\rho, e)), \qquad
	\hat{P}^{BKW}(\rho, T) = \rho R T \hat{Z}^{BKW}(\rho, T),
\end{equation}
where the compressibility factor,
$\hat{Z}^{BKW} : \Rbb_{\ge 0} \times \Rbb_{\ge 0} \rightarrow \Rbb$, is defined as
\begin{equation}
	\hat{Z}^{BKW} : (\rho, T) \mapsto
	(1 + \hat{X}(\rho, T) \mathrm{exp}(\beta \hat{X}(\rho, T))),
\end{equation}
the function $\hat{X} : \Rbb_{\ge 0} \times \Rbb_{\ge 0} \rightarrow \Rbb$
is defined as 
\begin{equation}
	\hat{X} : (\rho, T) \mapsto \frac{\kappa \rho}{(T + \theta)^\alpha},
\end{equation}
and $\beta, \kappa, \theta, \alpha$ are all scalar constants.
The temperature, $T^{BKW} : \Rbb_{\ge 0} \times \Rbb_{\ge 0} \rightarrow \Rbb_{\ge 0}$,
$T^{BKW} : (\rho, e) \mapsto T^{BKW}(\rho, e)$, is implicitly defined as the solution
($T$) of the nonlinear scalar equation (given $\rho$ and $e$)
\begin{equation} \label{eqn:bkw_nleqn}
	e^{BKW}(\rho, T) = e,
\end{equation}
which must be solved using nonlinear iterations. For this equation of state,
the internal energy,
$e^{BKW} : \Rbb_{\ge 0} \times \Rbb_{\ge 0} \rightarrow \Rbb_{\ge 0}$,
is defined in terms of the density and temperature as
\begin{equation} \label{eqn:bkw_e}
	e^{BKW}: (\rho, T) \mapsto \frac{\alpha RT^2 X(\rho,T) \mathrm{exp}(\beta X(\rho, T))}{T + \theta} + aT^2 + bT + c,
\end{equation}
where $a, b, c$ are constant coefficients calibrated to the gas under consideration.

Recall, the partial derivatives of the equation of state are crucial to define
the conservation law flux and its Jacobian (Section~\ref{sec:gov_singlephase}).
The partial derivatives of the BKW equation of state are
\begin{equation}
	\begin{aligned}
		P_\rho^{BKW}(\rho, e) &= \pder{\hat{P}^{BKW}}{\rho}(\rho, T^{BKW}(\rho, e)) \bigg|_T + \pder{\hat{P}^{BKW}}{T}(\rho, T^{BKW}(\rho, e)) \bigg|_\rho \pder{T^{BKW}}{\rho}(\rho,e)\bigg|_e \\
		P_e^{BKW}(\rho, e) &= \pder{\hat{P}^{BKW}}{T}(\rho, T^{BKW}(\rho, e)) \bigg|_\rho \pder{T^{BKW}}{e}(\rho,e)\bigg|_\rho.
	\end{aligned}
\end{equation}
Both of the derivatives depend on derivatives of $T^{BKW}$, which is implicitly defined as
the solution of the nonlinear equation in (\ref{eqn:bkw_nleqn}). To derive these
derivatives, introduce a new function,
$F : \Rbb_{\ge 0} \times \Rbb_{\ge 0} \rightarrow \Rbb$, defined as
\begin{equation} \label{eqn:bkw_nleqn2}
	F(\rho, e) = e^{BKW}(\rho, T^{BKW}(\rho, e)) - e.
\end{equation}
From (\ref{eqn:bkw_nleqn}), $F$ is the zero function, which means
$\pder{F}{\rho}(\rho, e) = \pder{F}{e}(\rho, e) = 0$ for any
$\rho, e \in \Rbb_{\ge 0}$. Equation (\ref{eqn:bkw_nleqn2}) and the fact
that the derivatives of $F$ are zero leads to the following equations
\begin{equation}
	\begin{aligned}
		\pder{F}{\rho}(\rho, e) &= \pder{e^{BKW}}{\rho}(\rho,T^{BKW}(\rho,e))\bigg|_T + \pder{e^{BKW}}{T}(\rho,T^{BKW}(\rho,e))\bigg|_\rho \pder{T^{BKW}}{\rho}(\rho,e)\bigg|_e = 0 \\
		\pder{F}{e}(\rho, e) &= \pder{e^{BKW}}{T}(\rho,T^{BKW}(\rho,e))\bigg|_\rho \pder{T^{BKW}}{e}(\rho,e)\bigg|_\rho - 1 = 0
	\end{aligned}
\end{equation}
These equations can easily be solved for the unknown partial derivatives to yield
\begin{equation}
	\begin{aligned}
		\pder{T^{BKW}}{\rho}(\rho,e)\bigg|_e &= -\left[\pder{e^{BKW}}{T}(\rho,T^{BKW}(\rho,e))\bigg|_\rho\right]^{-1}\pder{e^{BKW}}{\rho}(\rho,T^{BKW}(\rho,e))\bigg|_T \\
		\pder{T^{BKW}}{e}(\rho,e)\bigg|_\rho &= \left[\pder{e^{BKW}}{T}(\rho,T^{BKW}(\rho,e))\bigg|_\rho\right]^{-1}
	\end{aligned}
\end{equation}
after $T^{BKW}(\rho, e)$ has been obtained from the solution of (\ref{eqn:bkw_nleqn}).
All partial derivatives of $e^{BKW}$ can be derived analytically from (\ref{eqn:bkw_e}).

\begin{remark}
In practice, the nonlinear equation (\ref{eqn:bkw_nleqn}) must be solved many times, e.g., at every quadrature node of every element of the finite element mesh (Section~\ref{sec:govern:disc}), for a single discrete residual or Jacobian evaluation, so efficiency of the nonlinear solver is paramount. We use Newton's method due to its quadratic convergence with a linear model to determine the initial guess. In particular, given $\rho, e \in \Rbb_{\ge 0}$ (input to the nonlinear system), we construct a linear model of the form $\tilde{T}(\tilde{e}) = a + b \tilde{e}$ with constants $a, b$ defined such that $\tilde{T}(e^{BKW}(\rho, T_1)) = T_1$ and $\tilde{T}(e^{BKW}(\rho, T_2)) = T_2$, where $T_1, T_2 \in \Rbb_{\ge 0}$ are estimates of the extreme values of temperature encountered during the problem. Then, the nonlinear iterations are initialized with $T^{(0)} = \tilde{T}(e)$. For all problems considered in this work (Section~\ref{sec:numexp}), we use $T_1 = 10 K$ and $T_2 = 3000K$.
\end{remark}




\section{Discontinuous Galerkin discretization of the transformed space-time conservation law}
\label{sec:govern:disc}
We discretize the transformed conservation law (\ref{eqn:trans1}) with a discontinuous
Galerkin method \cite{Hesthaven2007} with high-order piecewise polynomials spaces used
to approximate the state $\transU$ and domain mapping $\Gcal$.  To this end, let
$\bar\Ecal_h$ represent a discretization of the reference domain $\Omega_0$ into
non-overlapping, potentially curved, computational elements. To establish the
finite-dimensional DG formulation, we introduce the DG approximation (trial) space
of discontinuous piecewise polynomials associated with the mesh $\bar\Ecal_h$
\begin{equation}
 \Vcal_h^p \coloneqq
 \left\{
  v \in [L^2(\bar\Omega)]^m \suchthat \left. v\right|_{\bar{K}} \in [\Pcal_p(\bar{K})]^m,~\forall \bar{K}\in\bar\Ecal_h
 \right\},
\end{equation}
where $\Pcal_p(\bar{K})$ is the space of polynomial functions of degree at most
$p \geq 0$ over the domain $\bar{K}$. Furthermore, we define the space of globally
continuous piecewise polynomials of degree $q$ associated with the mesh $\bar\Ecal_h$ as
\begin{equation}
 \Wcal_h \coloneqq
 \left\{
  v \in C^0(\bar\Omega) \suchthat \left. v\right|_{\bar{K}} \in \Pcal_q(\bar{K}),~\forall \bar{K}\in\bar\Ecal_h
 \right\}
\end{equation}
and discretize the domain mapping with the corresponding vector-valued space
$\left[\Wcal_h\right]^d$. With these definitions, the DG variational problem
is: given $\Gcal_h \in \left[\Wcal_h\right]^d$, find $\bar{U}_h\in\Vcal_h^p$
such that for all $\bar\psi_h \in \Vcal_h^{p'}$, we have
\begin{equation} \label{eqn:weak1}
 r_h^{p',p}(\bar\psi_h, \transU_h; \bar\nabla\Gcal_h) = 0
\end{equation}
where $p' \geq p$ and the global residual function
$r_h^{p',p} : \Vcal_h^{p'}\times\Vcal_h^p\times[\Wcal_h]^d \rightarrow \Rbb$
is defined in terms of elemental residuals
$r_{\bar{K}}^{p',p} : \Vcal_h^{p'}\times\Vcal_h^p\times[\Wcal_h]^d \rightarrow \Rbb$
as
\begin{equation} \label{eqn:weak2}
r_h^{p', p} : (\bar\psi_h, \bar{W}_h; \Theta_h) \mapsto
 \sum_{\bar{K}\in\bar\Ecal_h} r_{\bar{K}}^{p', p} (\bar\psi_h, \bar{W}_h; \Theta_h).
\end{equation}
The elemental residuals come directly from a standard DG formulation applied to
the transformed space-time conservation law in (\ref{eqn:trans1})
\begin{equation} \label{eqn:weak3}
\begin{aligned}
 r_{\bar{K}}^{p',p} : (\bar\psi_h, \bar{W}_h; \Theta_h) \mapsto
 &\int_{\partial \bar{K}}\bar\psi_h \cdot \bar{\Hcal} ( W^+_h,  W^-_h, \transUnitN_h; \Theta_h  ) \, dS  \\
- &\int_{\bar{K}} \transF (W_h; \Theta_h) : \bar{\nabla} \bar\psi_h \, dV \\
- &\int_{\bar{K}} \bar\psi_h \cdot \transS (W_h; \det \Theta_h)) \, dV,
\end{aligned}
\end{equation}
where $\bar\eta_h : \partial\bar{K} \mapsto \Rbb^d$ is the outward unit normal to
an element $\bar{K} \in \bar\Ecal_h$, $\bar{W}_h^+$ ($\bar{W}_h^-$) denotes the interior
(exterior) trace of $\bar{W}_h \in \Vcal_h^p$ to the element, and $\bar{\Hcal}$ is the transformed space-time numerical (Roe) flux function \cite{roe1981approximate}.

Following the approach in \cite{2020_zahr_HOIST,naudet2024space}, we define the transformed
numerical flux function,
$\bar\Hcal : \Rbb^m\times\Rbb^m\times\Sbb_d\times\Rbb^{d\times d}\mapsto\Rbb^m$,
as
\begin{equation} \label{eqn:trans3}
 \bar\Hcal : (\bar{W}^+, \bar{W}^-, \bar\eta; \Theta) \mapsto
    \sigma \Hcal(\bar{W}^+, \bar{W}^-, \sigma^{-1}(\det\Theta)\Theta^{-T}\bar\eta )
\end{equation}
where $\sigma = \norm{(\det\Theta) \Theta^{-T}\bar\eta}$ is defined for brevity
and $\Hcal : \Rbb^m \times \Rbb^m \times \Sbb_d \rightarrow \Rbb^m$ is the
space-time numerical flux, defined as \cite{naudet2024space}
\begin{equation} \label{eqn:sptm_numflux}
 \Hcal : (W^+,W^-,\eta) \mapsto \frac{1}{2}\left(\physF(W^+)\eta + \physF(W^-)\eta\right) + \frac{1}{2}\tilde{B}(W^+,W^-,\eta)(W^+-W^-).
\end{equation}
The Jacobian of the linearized Riemann problem
$\tilde{B} : \Rbb^m\times\Rbb^m\times\Sbb_d \rightarrow \Rbb^{m\times m}$ is
\begin{equation} \label{eqn:sptm_linriem}
  \tilde{B}(W^+,W^-,\eta) =  \left|B(\hat{U}(W^+,W^-),\eta)\right|,
\end{equation}
which can be directly written in terms of spatial quantities using
(\ref{eqn:sptm_eigval_decomp})
\begin{equation} \label{eqn:sptm_linriem2}
 \tilde{B}(W^+,W^-,\eta) =
  V_x(\hat{W},\eta_x)
  \left|\Lambda_x(\hat{W},\eta_x)\norm{n_x} + n_t I_m\right|
  V_x(\hat{W},\eta_x)^{-1},
\end{equation}
where $\hat{W} = \hat{U}(W^+,W^-)$ was introduced for brevity and
$\hat{U} : \Rbb^m \times \Rbb^m \rightarrow \Rbb^m$ is
the equation-specific Roe average (\ref{sec:roeavg}).

Finally, we introduce a basis for the test space ($\Vcal_h^{p'}$), trial space
($\Vcal_h^p$), and domain mapping space ($[\Wcal_h]^d$) to reduce the weak
formulation in (\ref{eqn:weak1})-(\ref{eqn:weak3}) to a system of nonlinear
algebraic equations. In the case where $p = p'$, we have
\begin{equation}
 \rbm : \Rbb^{N_\ubm} \times \Rbb^{N_\xbm} \rightarrow \Rbb^{N_\ubm}, \qquad
 \rbm : (\ubm, \xbm) \mapsto \rbm(\ubm,\xbm),
\end{equation}
where $N_\ubm = \mathrm{dim}(\Vcal_h^p)$ and $N_\xbm = \mathrm{dim}([\Wcal_h]^d)$,
which is the residual of a standard space-time DG method. Furthermore, we define
the algebraic enriched residual associated with a test space of degree $p' > p$
($p' = p+1$ in this work) as
\begin{equation}
 \Rbm : \Rbb^{N_\ubm} \times \Rbb^{N_\xbm} \rightarrow \Rbb^{N_\ubm'}, \qquad
 \Rbm : (\ubm, \xbm) \mapsto \Rbm(\ubm,\xbm),
\end{equation}
where $N_\ubm' = \mathrm{dim}(\Vcal_h^{p'})$, which will be used to construct
the implicit shock tracking objective function.

\section{Implicit shock tracking for two-phase flow}
\label{sec:ist}
In this section we formulate the implicit shock tracking optimization problem
for two-phase flow (Section~\ref{sec:ist_form}) using the machinery established
in  Sections~\ref{sec:gov_eqns_total}-\ref{sec:govern:disc} and review important details
of implicit shock tracking for space-time problems (Section~\ref{sec:ist_review}) from
\cite{naudet2024space}.

\subsection{Formulation}
\label{sec:ist_form}
Let $\phibold : \Rbb^{N_\ybm} \rightarrow \Rbb^{N_\xbm}$,
$\phibold : \ybm \mapsto \phibold(\ybm)$ be a boundary-preserving parametrization
of the mesh nodal coordinate $\xbm \in \Rbb^{N_\xbm}$, where $\ybm\in\Rbb^{N_\ybm}$
is a vector of unconstrained degrees of freedom, i.e., for any $\ybm\in\Rbb^{N_\ybm}$,
then $\xbm = \phibold(\ybm)$ are nodal coordinates that preserve all boundaries of the
computational domain \cite{2020_zahr_HOIST,2021_huang_HOIST,naudet2024space}. The HOIST method
is formulated as an optimization problem over the DG solution coefficients and the
unconstrained mesh degrees of freedom
\begin{equation} \label{eqn:pde-opt}
	(\ubm ^\star,\ybm ^\star) \coloneqq \argmin_{\ubm \in\Rbb^{N_{\ubm}},\ybm \in\Rbb^{N_{\ybm}}} f(\ubm,\phibold(\ybm)) \quad \text{subject to:} \quad \rbm (\ubm ,\phibold(\ybm)) = \zerobold,
\end{equation}
where $\func{f }{\Rbb^{N_{\ubm}}\times\Rbb^{N_{\xbm}}}{\Rbb}$ is the
objective function defined in \cite{2020_zahr_HOIST,2021_huang_HOIST}.
The objective function is composed of two terms as
\begin{equation}\label{eqn:obj0}
 f  : (\ubm ,\xbm ) \mapsto f_{\text{err}}(\ubm ,\xbm ) + \kappa^2 f_{\text{msh}}(\xbm ),
\end{equation}
which balances alignment of the mesh with non-smooth features and the quality 
of the elements, and $\kappa$ is the mesh penalty parameter. The mesh alignment 
term ($f_{\text{err}}$) is taken to be the norm of the enriched DG residual
\begin{equation} \label{eqn:ist_obj_fcn}
 f_{\mathrm{err}} : (\ubm ,\xbm ) \rightarrow \frac{1}{2}\norm{\Rbm (\ubm ,\xbm )}_2^2,
\end{equation}
which promotes mesh  alignment by penalizing non-physical oscillations that
arise on meshes that are not discontinuity-aligned. The mesh quality term
($f_{\text{msh}}$) is a measure of the element-wise mesh distortion \cite{2021_huang_HOIST}.
A sequential quadratic programming (SQP) method, globalized with a line search on the
$\ell_1$ merit function, is used to solve the optimization problem in (\ref{eqn:pde-opt})
by simultaneously converging $(\ubm,\xbm)$ to their optimal values
$(\ubm^\star,\xbm^\star)$. The parameter $\kappa\in\Rbb$, which balances the
contributions of the shock tracking and mesh quality terms, is adapted during
the SQP iterations using the algorithm described in \cite{2021_huang_HOIST}.
For a complete description of the HOIST method and SQP solver, the reader is
referred to \cite{2020_zahr_HOIST,2021_huang_HOIST}. 

\subsection{Space-time implicit shock tracking}
\label{sec:ist_review}
In this work, we solve time-dependent two-phase flow using the space-time
HOIST method \cite{naudet2024space}. The transformed space-time conservation law
(\ref{eqn:trans1}) is discretized in space-time resulting in a $(d'+1)$-dimensional
conservation law (Sections~\ref{sec:govern:spatial}-\ref{sec:govern:sptm}). To avoid
coupling the entire temporal domain, the time domain is partitioned, which creates
smaller, more manageable space-time \textit{slabs}. The HOIST method
(Section~\ref{sec:ist_form}) is applied sequentially over individual slabs to
align the space-time slab mesh with discontinuities and compute the corresponding
flow solution. After the aligned mesh and flow solution are available on a given
slab, a space-time mesh of the next slab is formed by (1) extracting the upper temporal
boundary, a $d'$-dimensional mesh, from the current slab, (2) extruding the
$d'$-dimensional to form space-time prisms, and (3) splitting the prisms into
space-time simplices \cite{naudet2024space}. This process ensures the lower
temporal boundary (initial time) of the mesh of any slab conforms to the upper
temporal boundary (final time) of the mesh of the previous slab. As a result,
the lower temporal boundary of the mesh of any slab is aligned with discontinuities
so we choose to freeze the nodes on the lower temporal boundary using the mesh
parametrization $\phibold$. The initial condition of each slab (boundary condition
on the lower temporal boundary) is obtained by transferring the solution on the
upper temporal boundary of previous slab to the quadrature nodes on the lower
temporal boundary of the current slab mesh. This procedure repeats until the
union of all the slabs covers the entire temporal domain of interest. A complete
description of the space-time HOIST method can be found in \cite{naudet2024space}.

Many two-phase flows, particularly the blast problems considered in this work,
have huge disparities in the spatial and temporal scales. This can lead to space-time
meshes with excessive resolution in either the spatial or temporal dimension or highly
skewed meshes. To avoid this, we nondimensionalize the time-dependent conservation law
(\ref{eqn:gen_cons_law}) by introducing a spatial $L^\star$, temporal $t^\star$, and
mass $m^\star$ scaling, and reformulating (\ref{eqn:gen_cons_law}) in terms of
nondimensional quantities \cite{naudet2024space}. These scales are chosen such that:
1) the components of the solution vector ($U_x$) have similar magnitudes to improve
the conditioning/scaling of the optimization problem in (\ref{eqn:pde-opt}) and 2) the
spatial and temporal dimensions have similar magnitudes, which ensures the space-time
meshes are well-conditioned using the extract-extrude-split approach above. The specific
scales for different problems in the numerical experiments (Section \ref{sec:numexp})
are noted in the problem description.

\section{Numerical experiments}
\label{sec:numexp}
In this section, we apply the slab-based HOIST method to a range of two-phase flow
Riemann problems. We begin with a single-phase, ideal gas validation of the real gas
two-phase framework of Section~\ref{sec:gov_eqns_total} to demonstrate it recovers this
limiting case (Section~\ref{sec:numexp:euler:IG}). Next, we present a two-phase validation
where both fluids are ideal gases with different $\gamma$ values
(Section~\ref{sec:numexp:euler:IGIG}) and an ideal-stiffened gas validation
(Section~\ref{sec:numexp:euler:IGSG}). Finally, demonstrate the framework on a spherically
symmetric underwater explosion where the gas is modeled with the BKW equation of state
and water is modeled as a stiffened gas (Section~\ref{sec:numexp:euler:IGBKW}).

\newcommand{\colorbarMatlabParulaMod}[5]{
\begin{tikzpicture}
\begin{axis}[
   hide axis, scale only axis,
   height=0pt, width=0pt,
   colormap={parula}{rgb255=(62,38,168) rgb255=(62,39,172) rgb255=(63,40,175) rgb255=(63,41,178) rgb255=(64,42,180) rgb255=(64,43,183) rgb255=(65,44,186) rgb255=(65,45,189) rgb255=(66,46,191) rgb255=(66,47,194) rgb255=(67,48,197) rgb255=(67,49,200) rgb255=(67,50,202) rgb255=(68,51,205) rgb255=(68,52,208) rgb255=(69,53,210) rgb255=(69,55,213) rgb255=(69,56,215) rgb255=(70,57,217) rgb255=(70,58,220) rgb255=(70,59,222) rgb255=(70,61,224) rgb255=(71,62,225) rgb255=(71,63,227) rgb255=(71,65,229) rgb255=(71,66,230) rgb255=(71,68,232) rgb255=(71,69,233) rgb255=(71,70,235) rgb255=(72,72,236) rgb255=(72,73,237) rgb255=(72,75,238) rgb255=(72,76,240) rgb255=(72,78,241) rgb255=(72,79,242) rgb255=(72,80,243) rgb255=(72,82,244) rgb255=(72,83,245) rgb255=(72,84,246) rgb255=(71,86,247) rgb255=(71,87,247) rgb255=(71,89,248) rgb255=(71,90,249) rgb255=(71,91,250) rgb255=(71,93,250) rgb255=(70,94,251) rgb255=(70,96,251) rgb255=(70,97,252) rgb255=(69,98,252) rgb255=(69,100,253) rgb255=(68,101,253) rgb255=(67,103,253) rgb255=(67,104,254) rgb255=(66,106,254) rgb255=(65,107,254) rgb255=(64,109,254) rgb255=(63,110,255) rgb255=(62,112,255) rgb255=(60,113,255) rgb255=(59,115,255) rgb255=(57,116,255) rgb255=(56,118,254) rgb255=(54,119,254) rgb255=(53,121,253) rgb255=(51,122,253) rgb255=(50,124,252) rgb255=(49,125,252) rgb255=(48,127,251) rgb255=(47,128,250) rgb255=(47,130,250) rgb255=(46,131,249) rgb255=(46,132,248) rgb255=(46,134,248) rgb255=(46,135,247) rgb255=(45,136,246) rgb255=(45,138,245) rgb255=(45,139,244) rgb255=(45,140,243) rgb255=(45,142,242) rgb255=(44,143,241) rgb255=(44,144,240) rgb255=(43,145,239) rgb255=(42,147,238) rgb255=(41,148,237) rgb255=(40,149,236) rgb255=(39,151,235) rgb255=(39,152,234) rgb255=(38,153,233) rgb255=(38,154,232) rgb255=(37,155,232) rgb255=(37,156,231) rgb255=(36,158,230) rgb255=(36,159,229) rgb255=(35,160,229) rgb255=(35,161,228) rgb255=(34,162,228) rgb255=(33,163,227) rgb255=(32,165,227) rgb255=(31,166,226) rgb255=(30,167,225) rgb255=(29,168,225) rgb255=(29,169,224) rgb255=(28,170,223) rgb255=(27,171,222) rgb255=(26,172,221) rgb255=(25,173,220) rgb255=(23,174,218) rgb255=(22,175,217) rgb255=(20,176,216) rgb255=(18,177,214) rgb255=(16,178,213) rgb255=(14,179,212) rgb255=(11,179,210) rgb255=(8,180,209) rgb255=(6,181,207) rgb255=(4,182,206) rgb255=(2,183,204) rgb255=(1,183,202) rgb255=(0,184,201) rgb255=(0,185,199) rgb255=(0,186,198) rgb255=(1,186,196) rgb255=(2,187,194) rgb255=(4,187,193) rgb255=(6,188,191) rgb255=(9,189,189) rgb255=(13,189,188) rgb255=(16,190,186) rgb255=(20,190,184) rgb255=(23,191,182) rgb255=(26,192,181) rgb255=(29,192,179) rgb255=(32,193,177) rgb255=(35,193,175) rgb255=(37,194,174) rgb255=(39,194,172) rgb255=(41,195,170) rgb255=(43,195,168) rgb255=(44,196,166) rgb255=(46,196,165) rgb255=(47,197,163) rgb255=(49,197,161) rgb255=(50,198,159) rgb255=(51,199,157) rgb255=(53,199,155) rgb255=(54,200,153) rgb255=(56,200,150) rgb255=(57,201,148) rgb255=(59,201,146) rgb255=(61,202,144) rgb255=(64,202,141) rgb255=(66,202,139) rgb255=(69,203,137) rgb255=(72,203,134) rgb255=(75,203,132) rgb255=(78,204,129) rgb255=(81,204,127) rgb255=(84,204,124) rgb255=(87,204,122) rgb255=(90,204,119) rgb255=(94,205,116) rgb255=(97,205,114) rgb255=(100,205,111) rgb255=(103,205,108) rgb255=(107,205,105) rgb255=(110,205,102) rgb255=(114,205,100) rgb255=(118,204,97) rgb255=(121,204,94) rgb255=(125,204,91) rgb255=(129,204,89) rgb255=(132,204,86) rgb255=(136,203,83) rgb255=(139,203,81) rgb255=(143,203,78) rgb255=(147,202,75) rgb255=(150,202,72) rgb255=(154,201,70) rgb255=(157,201,67) rgb255=(161,200,64) rgb255=(164,200,62) rgb255=(167,199,59) rgb255=(171,199,57) rgb255=(174,198,55) rgb255=(178,198,53) rgb255=(181,197,51) rgb255=(184,196,49) rgb255=(187,196,47) rgb255=(190,195,45) rgb255=(194,195,44) rgb255=(197,194,42) rgb255=(200,193,41) rgb255=(203,193,40) rgb255=(206,192,39) rgb255=(208,191,39) rgb255=(211,191,39) rgb255=(214,190,39) rgb255=(217,190,40) rgb255=(219,189,40) rgb255=(222,188,41) rgb255=(225,188,42) rgb255=(227,188,43) rgb255=(230,187,45) rgb255=(232,187,46) rgb255=(234,186,48) rgb255=(236,186,50) rgb255=(239,186,53) rgb255=(241,186,55) rgb255=(243,186,57) rgb255=(245,186,59) rgb255=(247,186,61) rgb255=(249,186,62) rgb255=(251,187,62) rgb255=(252,188,62) rgb255=(254,189,61) rgb255=(254,190,60) rgb255=(254,192,59) rgb255=(254,193,58) rgb255=(254,194,57) rgb255=(254,196,56) rgb255=(254,197,55) rgb255=(254,199,53) rgb255=(254,200,52) rgb255=(254,202,51) rgb255=(253,203,50) rgb255=(253,205,49) rgb255=(253,206,49) rgb255=(252,208,48) rgb255=(251,210,47) rgb255=(251,211,46) rgb255=(250,213,46) rgb255=(249,214,45) rgb255=(249,216,44) rgb255=(248,217,43) rgb255=(247,219,42) rgb255=(247,221,42) rgb255=(246,222,41) rgb255=(246,224,40) rgb255=(245,225,40) rgb255=(245,227,39) rgb255=(245,229,38) rgb255=(245,230,38) rgb255=(245,232,37) rgb255=(245,233,36) rgb255=(245,235,35) rgb255=(245,236,34) rgb255=(245,238,33) rgb255=(246,239,32) rgb255=(246,241,31) rgb255=(246,242,30) rgb255=(247,244,28) rgb255=(247,245,27) rgb255=(248,247,26) rgb255=(248,248,24) rgb255=(249,249,22) rgb255=(249,251,21) },
   colorbar horizontal,
   point meta min=#1, point meta max=#5,
   colorbar style={width=0.85\textwidth, xtick={#1,#2,#3,#4,#5},
   scaled ticks=false, tick label style={/pgf/number format/fixed}}
]
\addplot [draw=none] coordinates {(0,0)};
\end{axis}
\end{tikzpicture}
}

\subsection{Sod's shock tube: single-phase, ideal gas validation}
\label{sec:numexp:euler:IG}
Sod's shock tube is a Riemann problem for the Euler equations that
models an idealized shock tube where the membrane separating a high pressure
region from a low pressure one is instantaneously removed. This is a commonly
used validation problem because it has an analytical solution that features a shock
wave, rarefaction wave, and contact discontinuity. This problem serves as a simple
validation of the two-phase real gas HOIST method in the single-phase, ideal gas limit.

We model Sod's shock tube using the two-phase real gas framework of
Section~\ref{sec:gov_eqns_total}, i.e., the two-phase Euler equations
(\ref{eqn:gen_cons_law}) and (\ref{eqn:multiphase_gov_eqn}) with mixture equation
of state (\ref{eqn:mixture_eos}). Both fluids are taken to be ideal gases with the same
ratio of specific heats $\gamma_1 = \gamma_2 = 7/5$, thus reducing to a single-phase,
ideal gas flow. We consider the space-time domain $\Omega_x \coloneqq (0, 1)$ and
$\Tcal = (0, 0.1]$ with initial condition
\begin{equation}
 \rho(x,0) = \begin{cases} 1 & x<0.5 \\ 0.125 & x \geq 0.5 \end{cases}, \quad
  v(x,0) = 0, \quad
  P(x,0) = \begin{cases} 1 & x<0.5 \\ 0.1 & x \geq 0.5 \end{cases} , \quad
  \phi(x,0) = \begin{cases} 1 & x<0.5 \\ 0 & x \geq 0.5.
   \end{cases}
\end{equation}
Because both fluids are identical, the mixture equation of state (\ref{eqn:mixture_eos})
is independent of $\phi$ (Remark~\ref{rem:mixture1}), which makes $\phi(x,t)$ arbitrary.
The solution of this problem contains three waves (shock, contact, rarefaction) that
emanate from $x=0.5$ at $t=0$.

\begin{figure}
	\centering
 	\begin{tikzpicture}
\begin{groupplot} [
group style={group size = 1 by 2, horizontal sep = 0.05cm, vertical sep = 0.8cm},
title style={at={(current bounding box.north west)}, anchor=west}]
\nextgroupplot[axis equal image, width=1.0\textwidth, xtick={0, 0.5, 1}, ytick={0, 0.1, 0.2}, xticklabels={}, yticklabels={}, xlabel={}, ylabel={time}, xmin=0, xmax=1, ymin=0, ymax=0.1]
\addplot []
graphics [xmin=0,xmax=1,ymin=0,ymax=0.1] { 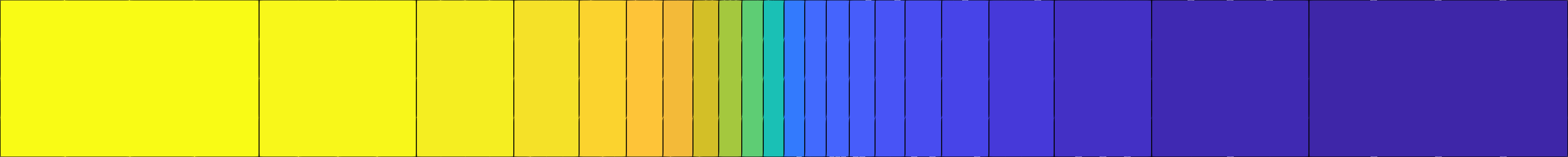};

\nextgroupplot[axis equal image, width=1.0\textwidth, xtick={}, ytick={}, xticklabels={}, yticklabels={}, xlabel={space}, ylabel={time}, xmin=0, xmax=1, ymin=0, ymax=0.1]
\addplot []
graphics [xmin=0,xmax=1,ymin=0,ymax=0.1] { 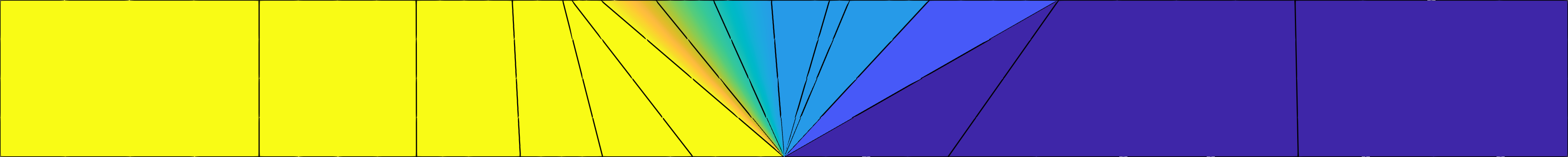};

\end{groupplot}\end{tikzpicture}
 	\colorbarMatlabParula{0.125}{0.25}{0.5}{0.75}{1}
	\caption{Two-phase HOIST solution (density) for single-phase, ideal gas Sod shock tube problem using one slab (\textit{bottom}) and the mesh and solution used to initialize the HOIST solver (\textit{top}).}
 	\label{fig:real_sod_dens}
\end{figure}

\begin{figure}
 \centering
 \raisebox{-0.5\height}{\input{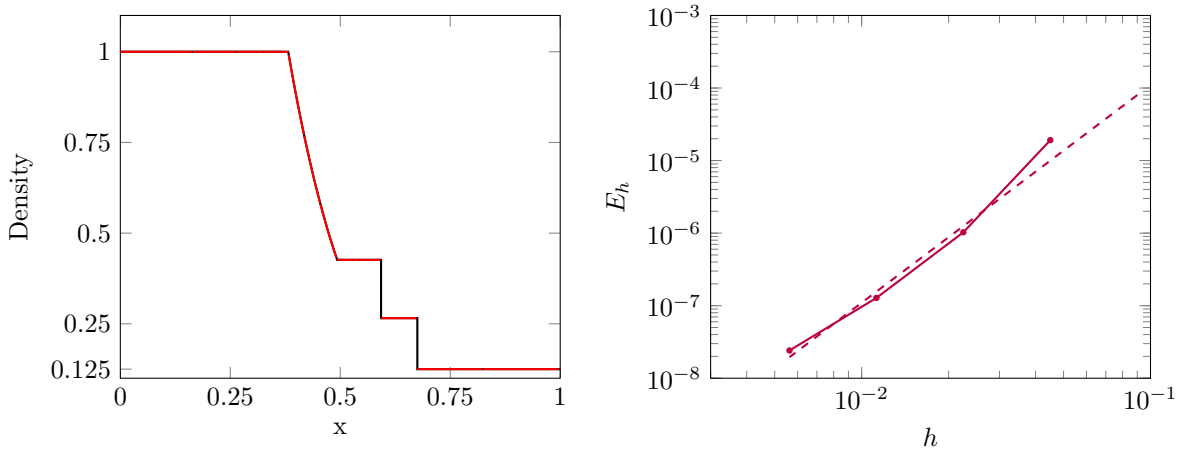}}
 \caption{\textit{Left}: Slice of two-phase HOIST solution (density) (\ref{line:density_sod_multi}) to the single-phase, ideal gas Sod shock tube problem at the final time, and the corresponding analytical solution (\ref{line:density_sod_multi_anyl}). \textit{Right}: Mesh convergence of the two-phase HOIST method ($p=2$) applied to the single-phase Sod shock tube measured using $L^1$ error at time $t=0.1$. Legend (\textit{right}): HOIST solution error (\ref{line:err_multi_sod_conv}) and optimal convergence rate ($p+1$) (\ref{line:err_multi_sod_conv_ideal}).}
	\label{fig:real_sod_slice_anyl}
\end{figure}

A single time slab is used to cover the entire time domain with an initial (non-aligned)
space-time mesh consisting of $22$ linear ($q=1$) quadrilateral elements with quadratic
($p=2$) solution approximation over each. The problem is nondimensionalized using the
scalings $L^\star = t^\star = m^\star = 1$ because the spatial and temporal scales are
similar magnitudes. The HOIST solver is initialized from the $p=0$
DG solution on the initial space-time mesh (Figure~\ref{fig:real_sod_dens}). The final
HOIST solution has tracked the lead shock, contact discontinuity, and the head and tail
of the rarefaction (Figure~\ref{fig:real_sod_dens}). Furthermore, the HOIST solution shows
near perfect agreement with the analytical solution to this problem at $t = 0.1$
(Figure~\ref{fig:real_sod_slice_anyl} (\textit{left})).

To close this section, we perform a mesh convergence study of the HOIST method for this
problem (Figure~\ref{fig:real_sod_slice_anyl} (\textit{right})), which confirms the method achieves the optimal
convergence, $\Ocal(h^{p+1})$, for $p=2$. The error is measured using the $L^1$-norm of
density at the final time $t = 0.1$ against the analytical solution, i.e.,
\begin{equation}
        E_h : h \mapsto \int_0^1 | \rho(x, 0.1) - \rho_h(x, 0.1) | dx,
\end{equation}
where $\rho(x,t)$ is the analytical solution and $\rho_h(x,t)$ is the HOIST
solution on the mesh with grid measure (longest edge in mesh) $h$.  We use
the $L^1$ error because it is expected to converge at the optimal rate for
discontinuous solutions provided the smooth solution and position of the
discontinuity converge at that rate \cite{2021_huang_HOIST, 2020_zahr_HOIST}.

%

\subsection{Sod's shock tube: two-phase, ideal gas validation}
\label{sec:numexp:euler:IGIG}
Next, we apply our method to a variation of Sod's shock tube involving two different
ideal gases to validate the two-phase HOIST method in this limiting case. In
particular, the high-pressure region is an ideal gas with $\gamma_1 = 7/5$ and
the low-pressure region is an ideal gas with $\gamma_2 = 7/3$. We consider the
space-time domain $\Omega_x \coloneqq (0, 1)$ and $\Tcal = (0, 0.1]$ with initial
condition
\begin{equation}
 \rho(x,0) = \begin{cases} 1 & x<0.5 \\ 0.125 & x \geq 0.5 \end{cases}, \quad
  v(x,0) = 0, \quad
  P(x,0) = \begin{cases} 1 & x<0.5 \\ 0.1 & x \geq 0.5 \end{cases} , \quad
  \phi(x,0) = \begin{cases} 1 & x<0.5 \\ 0 & x \geq 0.5. \end{cases}
\end{equation}
We use the convention defined in (\ref{eqn:mixture_eos}) that $\phi(x,t) = 1$
indicates material 1 is occupies $(x,t)\in\Omega_x\times\Tcal$, and material 2
occupies any $(x,t)\in\Omega_x\times\Tcal$ where $\phi(x,t) = 0$. The solution
of this problem contains three waves (shock, contact, rarefaction) that emanate
from $x=0.5$ at $t=0$.

\begin{figure}
	\centering
 	\raisebox{-0.5\height}{\begin{tikzpicture}
\begin{groupplot} [
group style={group size = 1 by 5, horizontal sep = 0.05cm, vertical sep = 0.8cm},
title style={at={(current bounding box.north)}}]
\nextgroupplot[axis equal image, width=1.0\textwidth, xtick={0, 0.5, 1}, ytick={0, 0.1, 0.2}, xticklabels={}, yticklabels={}, xlabel={}, ylabel={time}, xmin=0, xmax=1, ymin=0, ymax=0.1]
\addplot []
graphics [xmin=0,xmax=1,ymin=0,ymax=0.1] { 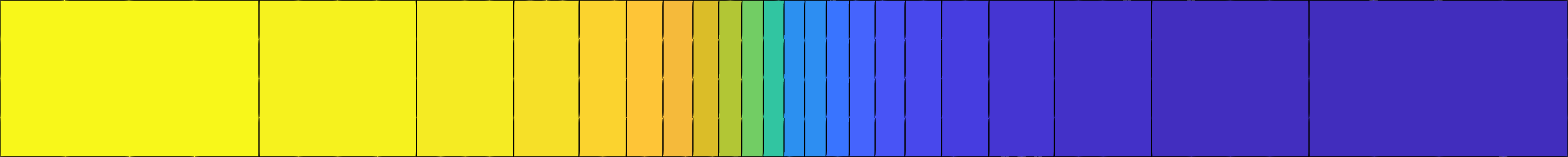};

\nextgroupplot[axis equal image, width=1.0\textwidth, xtick={}, ytick={}, xticklabels={}, yticklabels={}, xlabel={}, ylabel={time}, xmin=0, xmax=1, ymin=0, ymax=0.1]
\addplot []
graphics [xmin=0,xmax=1,ymin=0,ymax=0.1] { 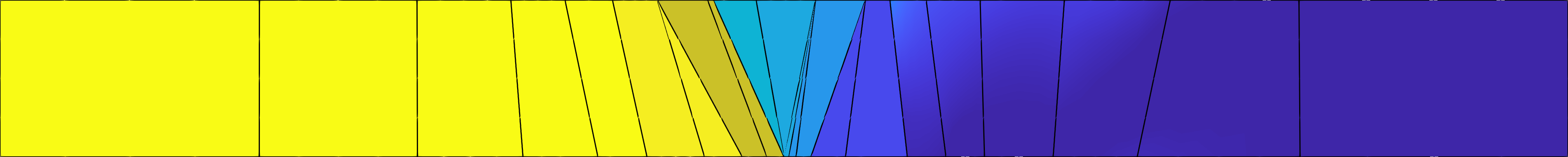};

\nextgroupplot[axis equal image, width=1.0\textwidth, xtick={0, 0.5, 1}, ytick={0, 0.1, 0.2}, xticklabels={}, yticklabels={}, xlabel={}, ylabel={time}, xmin=0, xmax=1, ymin=0, ymax=0.1]
\addplot []
graphics [xmin=0,xmax=1,ymin=0,ymax=0.1] { 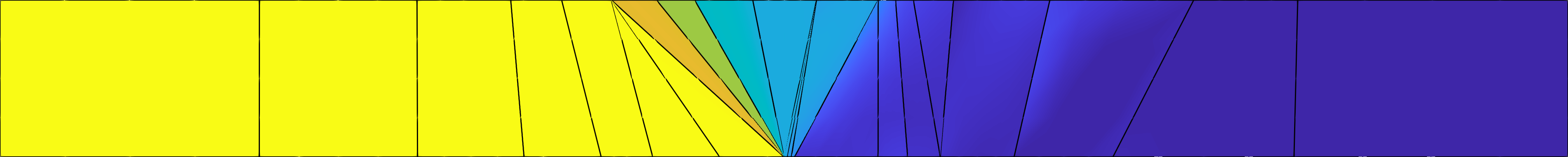};

\nextgroupplot[axis equal image, width=1.0\textwidth, xtick={}, ytick={}, xticklabels={}, yticklabels={}, xlabel={}, ylabel={time}, xmin=0, xmax=1, ymin=0, ymax=0.1]
\addplot []
graphics [xmin=0,xmax=1,ymin=0,ymax=0.1] { 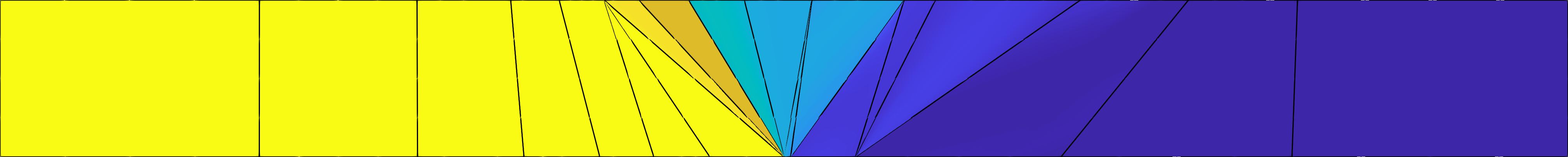};

\nextgroupplot[axis equal image, width=1.0\textwidth, xtick={0, 0.5, 1}, ytick={0, 0.1, 0.2}, xticklabels={}, yticklabels={}, xlabel={space}, ylabel={time}, xmin=0, xmax=1, ymin=0, ymax=0.1]
\addplot []
graphics [xmin=0,xmax=1,ymin=0,ymax=0.1] { 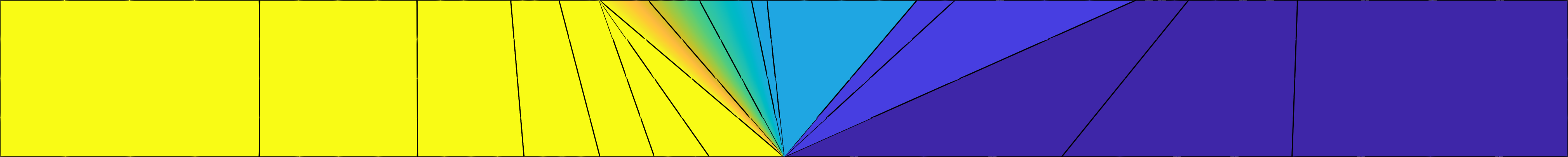};

\end{groupplot}\end{tikzpicture}}
 	\colorbarMatlabParula{0.125}{0.25}{0.5}{0.75}{1}        
	\caption{Two-phase HOIST SQP iterations (density) for two-phase, ideal gas Sod shock tube problem using one slab. SQP iterations $n = 0, 25, 50, 75, 100$ (\textit{top-to-bottom}).}
 	\label{fig:sod_prob_sqp_iters_dens}
\end{figure}

\begin{figure}
	\centering
 	\raisebox{-0.5\height}{\begin{tikzpicture}
\begin{groupplot} [
group style={group size = 1 by 5, horizontal sep = 0.05cm, vertical sep = 0.8cm},
title style={at={(current bounding box.north)}}]
\nextgroupplot[axis equal image, width=1.0\textwidth, xtick={0, 0.5, 1}, ytick={0, 0.1, 0.2}, xticklabels={}, yticklabels={}, xlabel={}, ylabel={time}, xmin=0, xmax=1, ymin=0, ymax=0.1]
\addplot []
graphics [xmin=0,xmax=1,ymin=0,ymax=0.1] { 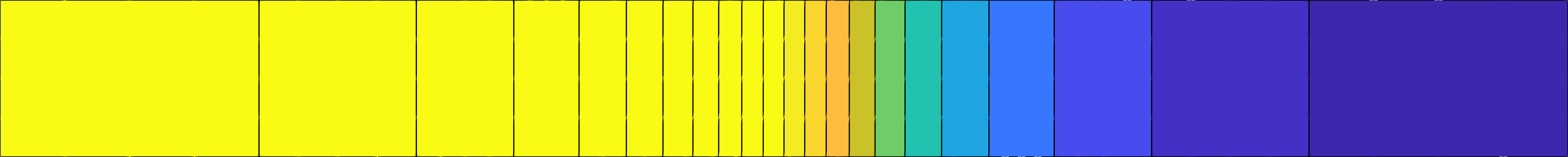};

\nextgroupplot[axis equal image, width=1.0\textwidth, xtick={}, ytick={}, xticklabels={}, yticklabels={}, xlabel={}, ylabel={time}, xmin=0, xmax=1, ymin=0, ymax=0.1]
\addplot []
graphics [xmin=0,xmax=1,ymin=0,ymax=0.1] { 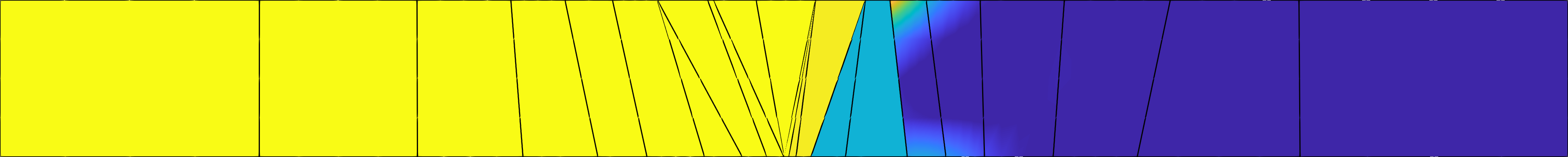};

\nextgroupplot[axis equal image, width=1.0\textwidth, xtick={0, 0.5, 1}, ytick={0, 0.1, 0.2}, xticklabels={}, yticklabels={}, xlabel={}, ylabel={time}, xmin=0, xmax=1, ymin=0, ymax=0.1]
\addplot []
graphics [xmin=0,xmax=1,ymin=0,ymax=0.1] { 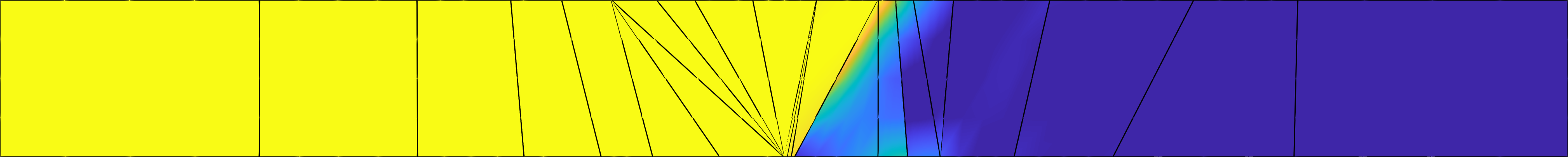};

\nextgroupplot[axis equal image, width=1.0\textwidth, xtick={}, ytick={}, xticklabels={}, yticklabels={}, xlabel={}, ylabel={time}, xmin=0, xmax=1, ymin=0, ymax=0.1]
\addplot []
graphics [xmin=0,xmax=1,ymin=0,ymax=0.1] { 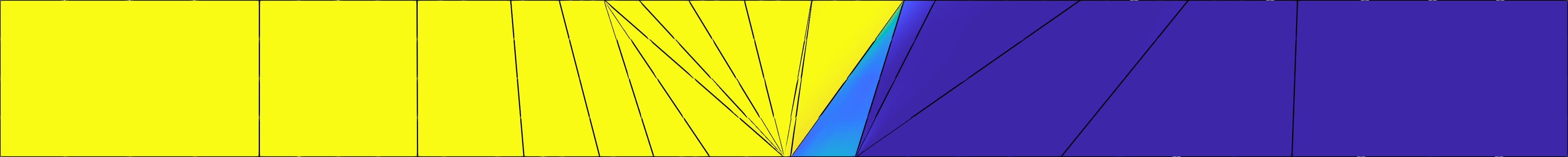};

\nextgroupplot[axis equal image, width=1.0\textwidth, xtick={0, 0.5, 1}, ytick={0, 0.1, 0.2}, xticklabels={}, yticklabels={}, xlabel={space}, ylabel={time}, xmin=0, xmax=1, ymin=0, ymax=0.1]
\addplot []
graphics [xmin=0,xmax=1,ymin=0,ymax=0.1] { 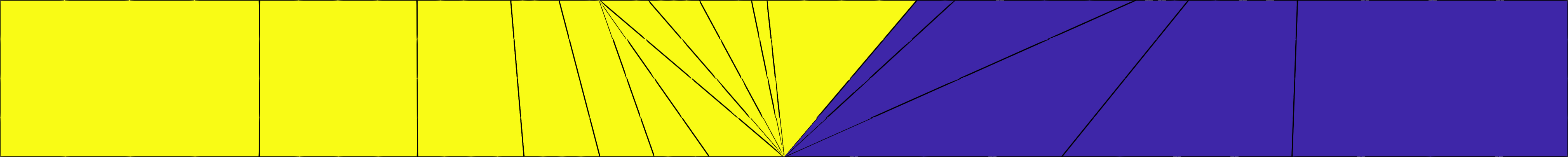};

\end{groupplot}\end{tikzpicture}}
 	\colorbarMatlabParula{0}{0.25}{0.5}{0.75}{1}        
	\caption{Two-phase HOIST SQP iterations (phase) for two-phase, ideal gas Sod shock tube problem using one slab. SQP iterations $n = 0, 25, 50, 75, 100$ (\textit{top-to-bottom}).}
 	\label{fig:sod_prob_sqp_iters_phase}
\end{figure}

\begin{figure}
 \centering
 \raisebox{-0.5\height}{\input{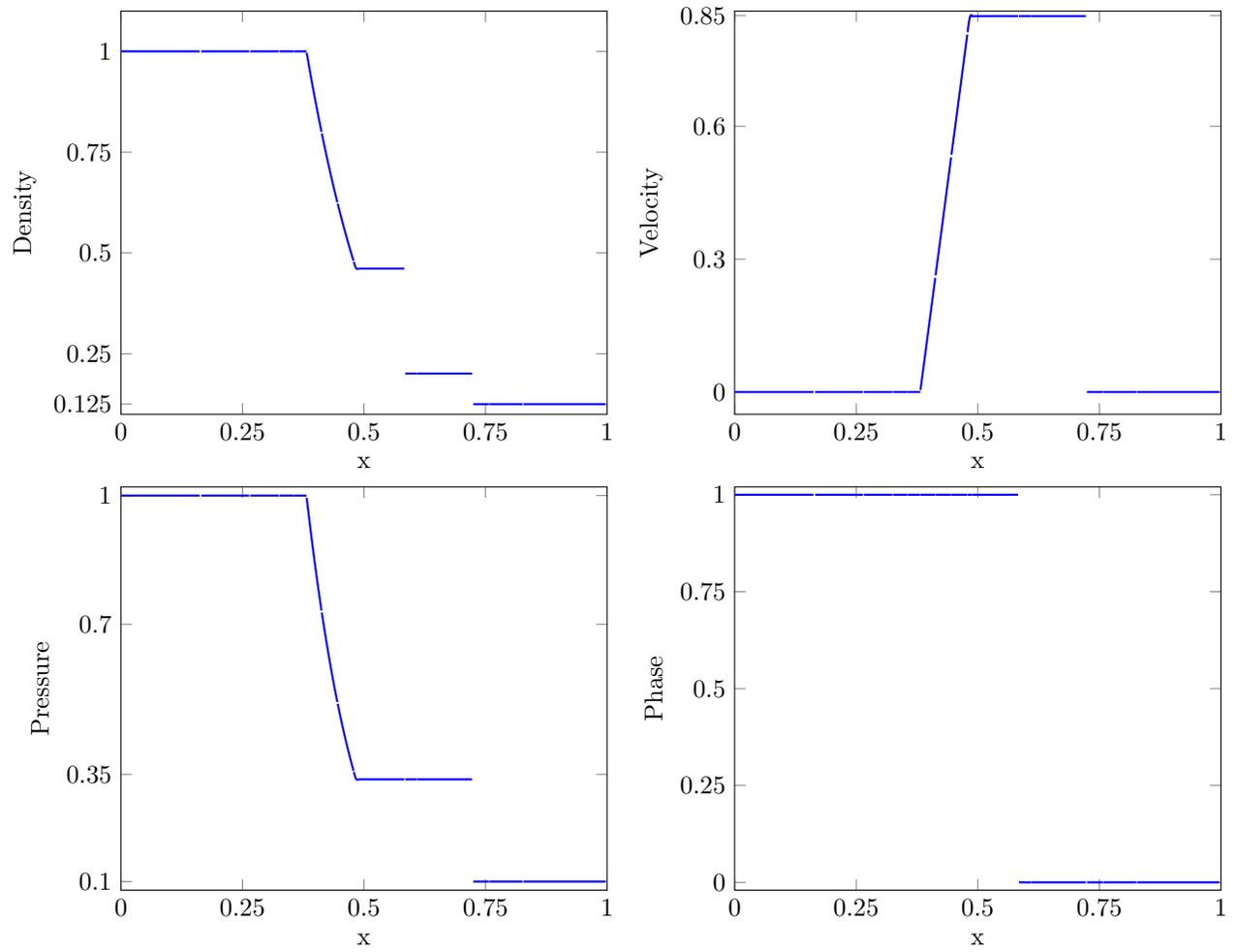}}
	\caption{Slice of two-phase HOIST solution to the two-phase, ideal gas Sod shock tube problem at the final time.}
	\label{fig:sod_prob_qoi_slices}
\end{figure}

A single slab is used to cover the entire time domain with an initial (non-aligned)
space-time mesh consisting of $22$ linear ($q=1$) quadrilateral elements with
quadratic ($p=2$) solution approximation over each. The problem is nondimensionalized
using the scalings $L^\star = t^\star = m^\star = 1$ because the spatial and temporal
scales are similar magnitudes. The HOIST solver is initialized
from the $p=0$ DG solution on the initial space-time mesh. As the SQP iterations proceed,
the faces of the mesh are driven towards alignment with the lead shock, contact
discontinuity, and head and tail of the rarefaction
(Figures~\ref{fig:sod_prob_sqp_iters_dens}-\ref{fig:sod_prob_sqp_iters_phase}).
Figure~\ref{fig:sod_prob_sqp_iters_phase} demonstrates the utility of the
phase-field formulation in converging to a sharp-interface solution. Namely,
at intermediate (non-converged) SQP iterations, there are many regions of the
domain where the materials are mixed (e.g., the phase variable takes values
in $[0,1]$), but at convergence, a sharp interface is obtained as
the phase variable is either $0$ or $1$ (separated by the contact discontinuity).
This also justifies the simplified mixture model in (\ref{eqn:mixture_eos}).
Slices of the primitive variables at the final time $t = 0.1$
(Figure~\ref{fig:sod_prob_qoi_slices}) show all variables are well-resolved
with only minimal spurious oscillations, and both the shock and contact 
re represented as perfect discontinuities (except pressure and velocity,
which are both constant across the contact).
Figure~\ref{fig:sod_prob_qoi_slices} further confirms a sharp interface
is obtained because the phase variable is $1$ to the left of
the contact discontinuity and $0$ to the right of the discontinuity.


\subsection{Ideal-stiffened gas Riemann problem}
\label{sec:numexp:euler:IGSG}
Next, we apply our method to a Riemann problem of
(\ref{eqn:gen_cons_law}) and (\ref{eqn:multiphase_gov_eqn}) involving two
different gases: an ideal gas with $\gamma_1 = 5/3$ and a stiffened gas with
$\gamma_2 = 4.4$ and $P^\star = 6 \times 10^8$. We consider the space-time
domain $\Omega_x \coloneqq (0, 1)$ and $\Tcal = (0, 5 \times 10^{-5}]$ with initial condition
\begin{equation}
 \rho(x,0) = \begin{cases} 1200 & x<0.5 \\  1000 & x \geq 0.5 \end{cases}, \quad
  v(x,0) = 0, \quad
  P(x,0) = \begin{cases} 1 \times 10^9 & x<0.5 \\ 1 \times 10^5 & x \geq 0.5 \end{cases} , \quad
  \phi(x,0) = \begin{cases} 1 & x<0.5 \\ 0 & x \geq 0.5
   \end{cases},
\end{equation}
where standard international (SI) units are used for all variables. Because of the large
pressure differential in the left and right states, there are fast moving waves that
emanate from the initial discontinuity which cause a large disparity in the spatial and
temporal scales. Therefore, we nondimensionalize the problem using the scalings
$L^\star = 1$, $t^\star = 5\times 10^{-4}$, and $m^\star = 10^3$ (SI units).

\begin{figure}
	\centering
 	\begin{tikzpicture}
\begin{groupplot} [
group style={group size = 1 by 5, horizontal sep = 0.05cm, vertical sep = 0.8cm},
title style={at={(current bounding box.north)}}]
\nextgroupplot[axis equal image, width=1.0\textwidth, xtick={0, 0.5, 1}, ytick={0, 0.1, 0.2}, xticklabels={}, yticklabels={}, xlabel={}, ylabel={time}, xmin=0, xmax=1, ymin=0, ymax=0.1]
\addplot []
graphics [xmin=0,xmax=1,ymin=0,ymax=0.1] { 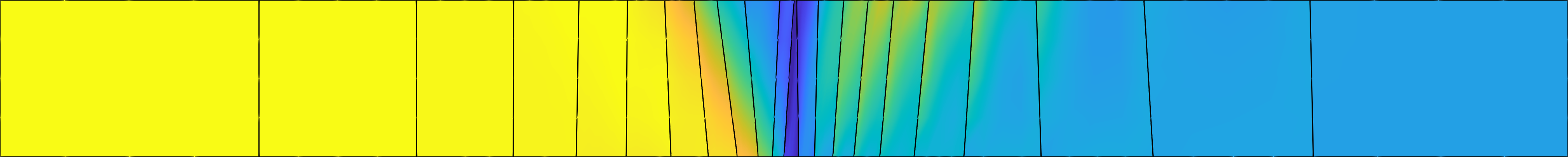};

\nextgroupplot[axis equal image, width=1.0\textwidth, xtick={}, ytick={}, xticklabels={}, yticklabels={}, xlabel={}, ylabel={time}, xmin=0, xmax=1, ymin=0, ymax=0.1]
\addplot []
graphics [xmin=0,xmax=1,ymin=0,ymax=0.1] { 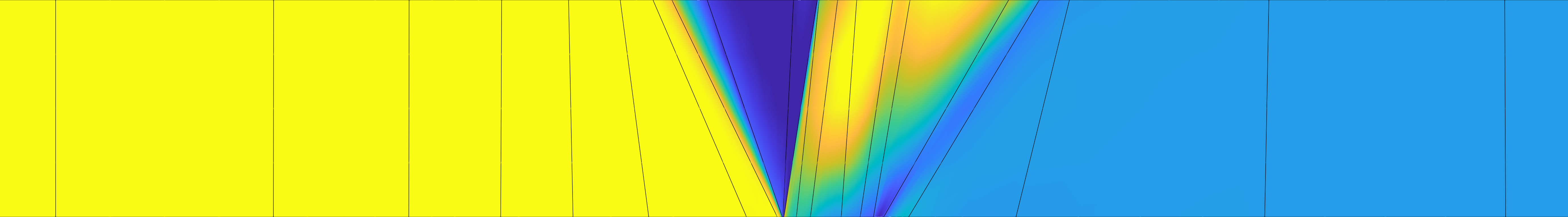};

\nextgroupplot[axis equal image, width=1.0\textwidth, xtick={0, 0.5, 1}, ytick={0, 0.1, 0.2}, xticklabels={}, yticklabels={}, xlabel={}, ylabel={time}, xmin=0, xmax=1, ymin=0, ymax=0.1]
\addplot []
graphics [xmin=0,xmax=1,ymin=0,ymax=0.1] { 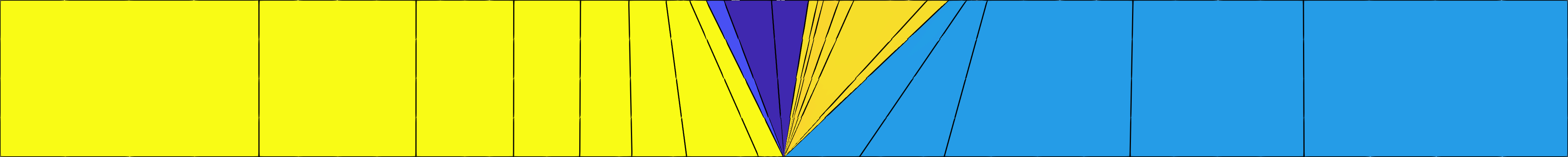};

\nextgroupplot[axis equal image, width=1.0\textwidth, xtick={}, ytick={}, xticklabels={}, yticklabels={}, xlabel={}, ylabel={time}, xmin=0, xmax=1, ymin=0, ymax=0.1]
\addplot []
graphics [xmin=0,xmax=1,ymin=0,ymax=0.1] { 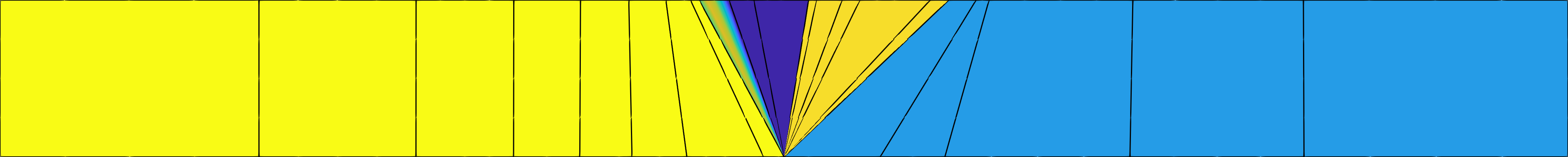};

\nextgroupplot[axis equal image, width=1.0\textwidth, xtick={0, 0.5, 1}, ytick={0, 0.1, 0.2}, xticklabels={}, yticklabels={}, xlabel={space}, ylabel={time}, xmin=0, xmax=1, ymin=0, ymax=0.1]
\addplot []
graphics [xmin=0,xmax=1,ymin=0,ymax=0.1] { 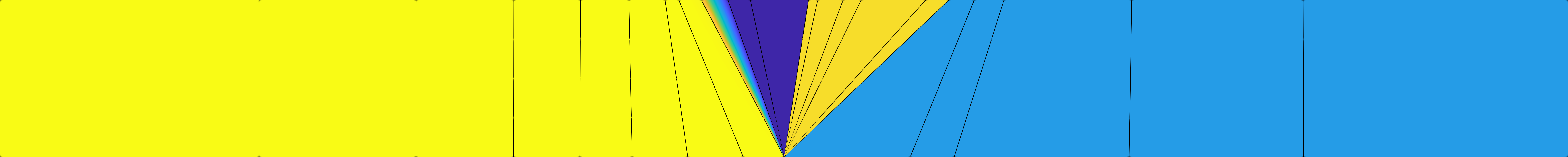};

\end{groupplot}\end{tikzpicture}
 	\colorbarMatlabParula{890}{975}{1050}{1125}{1200}             
	\caption{Two-phase HOIST SQP iterations (density) for ideal-stiffened gas Riemann problem using one slab. SQP iterations $n = 0, 10, 20, 30, 100$ (\textit{top-to-bottom}).}
 	\label{fig:ideal_stiff_blast_prob_sqp_iters_dens}
\end{figure}

\begin{figure}
	\centering
 	\begin{tikzpicture}
\begin{groupplot} [
group style={group size = 1 by 5, horizontal sep = 0.05cm, vertical sep = 0.8cm},
title style={at={(current bounding box.north)}}]
\nextgroupplot[axis equal image, width=1.0\textwidth, xtick={0, 0.5, 1}, ytick={0, 0.1, 0.2}, xticklabels={}, yticklabels={}, xlabel={}, ylabel={time}, xmin=0, xmax=1, ymin=0, ymax=0.1]
\addplot []
graphics [xmin=0,xmax=1,ymin=0,ymax=0.1] { 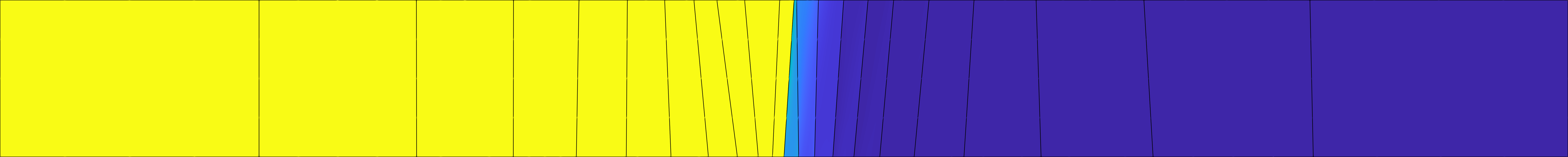};

\nextgroupplot[axis equal image, width=1.0\textwidth, xtick={}, ytick={}, xticklabels={}, yticklabels={}, xlabel={}, ylabel={time}, xmin=0, xmax=1, ymin=0, ymax=0.1]
\addplot []
graphics [xmin=0,xmax=1,ymin=0,ymax=0.1] { 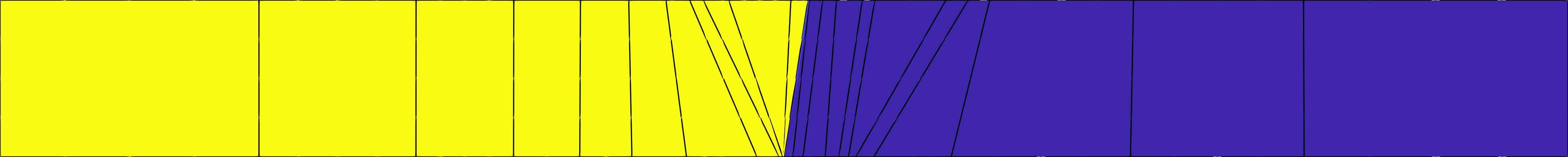};

\nextgroupplot[axis equal image, width=1.0\textwidth, xtick={0, 0.5, 1}, ytick={0, 0.1, 0.2}, xticklabels={}, yticklabels={}, xlabel={}, ylabel={time}, xmin=0, xmax=1, ymin=0, ymax=0.1]
\addplot []
graphics [xmin=0,xmax=1,ymin=0,ymax=0.1] { 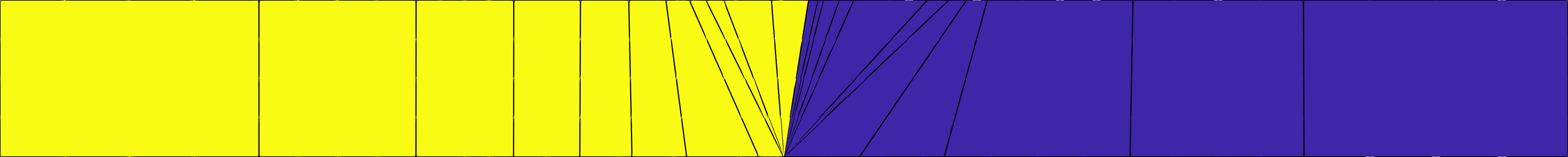};

\nextgroupplot[axis equal image, width=1.0\textwidth, xtick={}, ytick={}, xticklabels={}, yticklabels={}, xlabel={}, ylabel={time}, xmin=0, xmax=1, ymin=0, ymax=0.1]
\addplot []
graphics [xmin=0,xmax=1,ymin=0,ymax=0.1] { 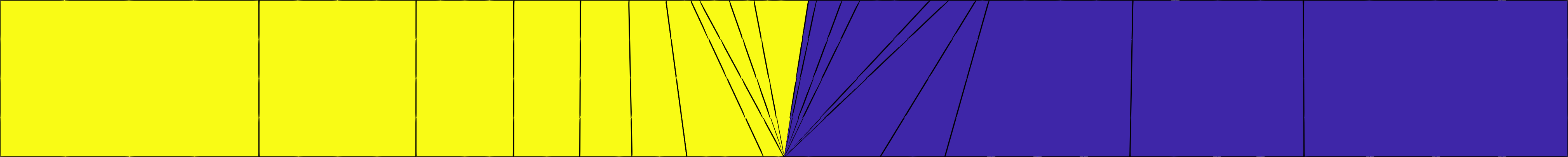};

\nextgroupplot[axis equal image, width=1.0\textwidth, xtick={0, 0.5, 1}, ytick={0, 0.1, 0.2}, xticklabels={}, yticklabels={}, xlabel={space}, ylabel={time}, xmin=0, xmax=1, ymin=0, ymax=0.1]
\addplot []
graphics [xmin=0,xmax=1,ymin=0,ymax=0.1] { 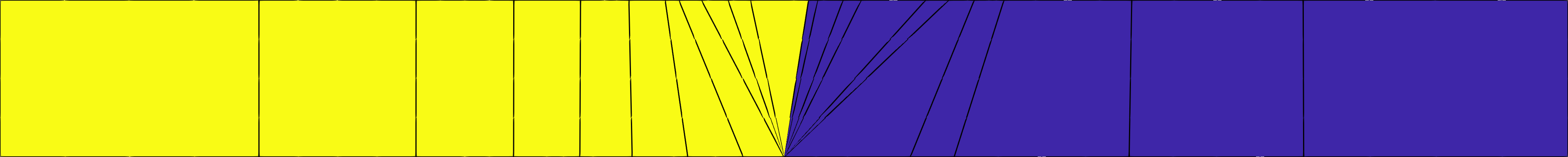};

\end{groupplot}\end{tikzpicture}
 	\colorbarMatlabParula{0}{0.25}{0.5}{0.75}{1}        
	\caption{Two-phase HOIST SQP iterations (phase) for ideal-stiffened gas Riemann problem using one slab. SQP iterations $n = 0, 10, 20, 30, 100$ (\textit{top-to-bottom}).}
 	\label{fig:ideal_stiff_blast_prob_sqp_iters_phase}
\end{figure}

\begin{figure}
	\centering
 	\begin{tikzpicture}
\begin{groupplot} [
	group style={
		group size = 1 by 2,
		horizontal sep = 0.05cm,
		vertical sep = 0.4cm
	},
	title style={at={(current bounding box.north west)}, anchor=west}
]

\nextgroupplot[axis equal image, width=1.0\textwidth, xtick={0, 0.5, 1}, ytick={0, 0.1, 0.2}, xticklabels={}, yticklabels={}, xlabel={}, ylabel={time}, xmin=0, xmax=1, ymin=0, ymax=0.1]
\addplot [] graphics [xmin=0,xmax=1,ymin=0,ymax=0.1] { 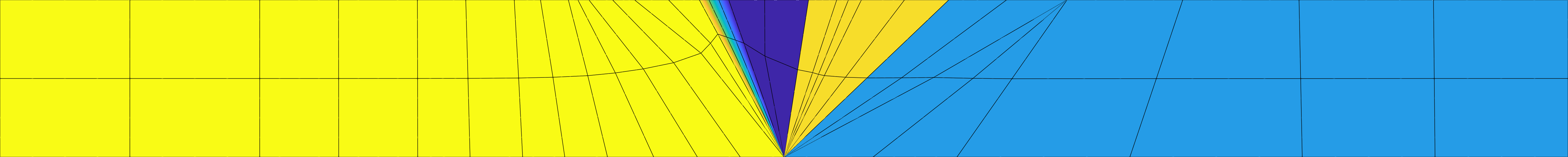};

\nextgroupplot[axis equal image, width=1.0\textwidth, xtick={0, 0.5, 1}, ytick={0, 0.1, 0.2}, xticklabels={}, yticklabels={}, xlabel={space}, ylabel={time}, xmin=0, xmax=1, ymin=0, ymax=0.1]
\addplot [] graphics [xmin=0,xmax=1,ymin=0,ymax=0.1] { 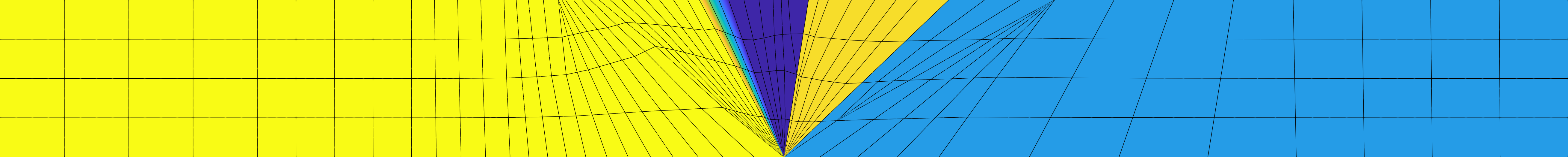};

\end{groupplot}
\end{tikzpicture}
 	\colorbarMatlabParula{890}{975}{1050}{1125}{1200}       
	\caption{Two-phase HOIST solution (density) for ideal-stiffened gas Riemann problem using one slab after one (\textit{top}) and two (\textit{bottom}) levels of refinement relative to the original simulation in (\ref{fig:ideal_stiff_blast_prob_sqp_iters_dens}).}
 	\label{fig:ideal_stiff_refstudy_meshes}
\end{figure}

\begin{figure}
 	\centering
 	\input{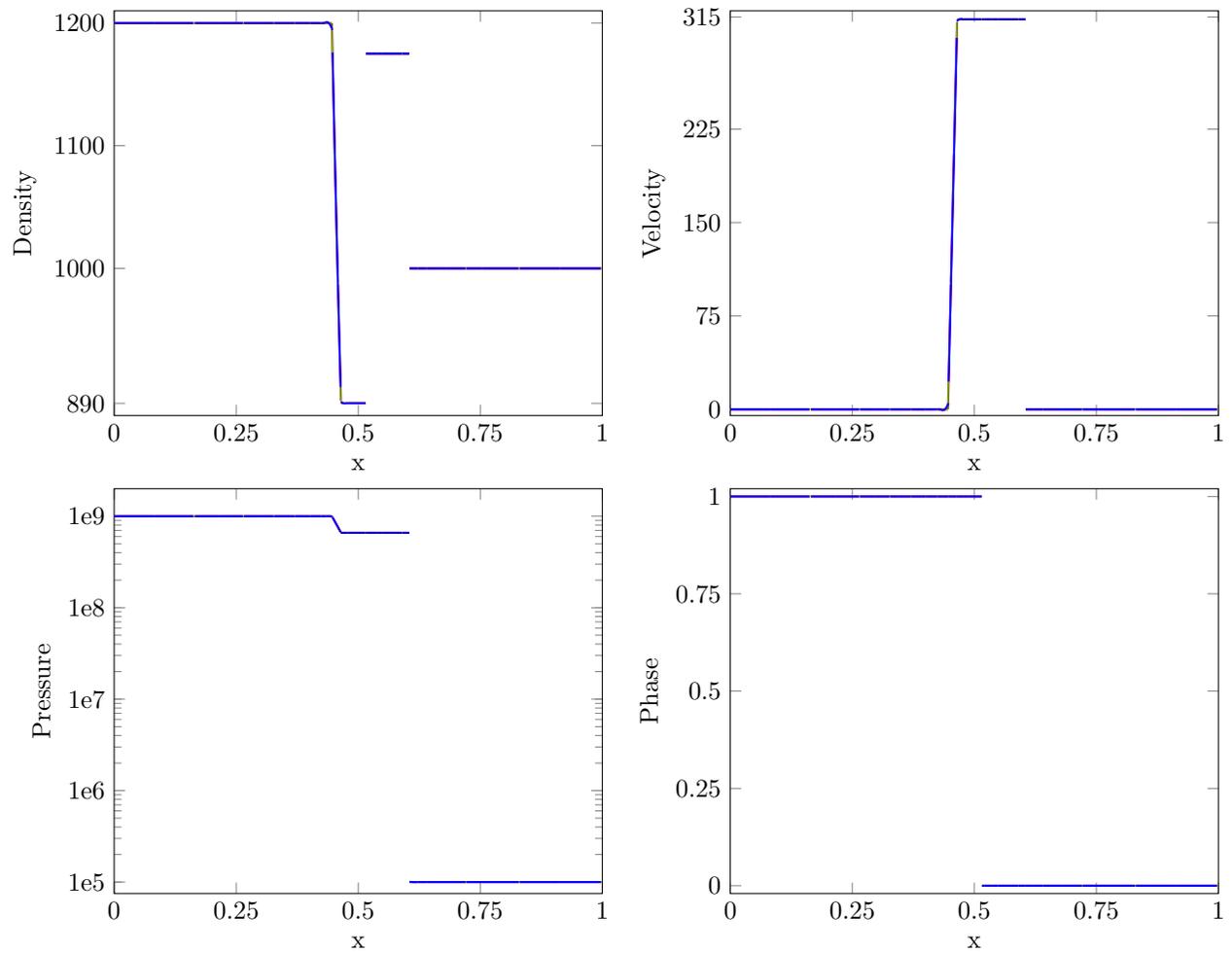}
	\caption{Slices of two-phase HOIST solution to the ideal-stiffened gas Riemann problem at the final time from the coarse (\ref{line:euler_multi_igsg_density_r0}), intermediate (\ref{line:euler_multi_igsg_density_r1}), and fine (\ref{line:euler_multi_igsg_density_r2}) meshes.}
\label{fig:ideal_stiff_blast_prob_qoi_slices}
\end{figure}

A single slab is used to cover the entire time domain with an initial (non-aligned)
space-time mesh consisting of $22$ linear ($q=1$) quadrilateral elements with
quadratic ($p=2$) solution approximation over each. The HOIST solver is initialized
from the $p=0$ DG solution on the initial space-time mesh. Similar behavior is
observed for the ideal gas problem in Section~\ref{sec:numexp:euler:IGIG}, i.e.,
as the SQP iterations proceed, the faces of the mesh are driven towards alignment
with the lead shock, contact discontinuity, and head and tail of the rarefaction
(Figures~\ref{fig:ideal_stiff_blast_prob_sqp_iters_dens}-\ref{fig:ideal_stiff_blast_prob_sqp_iters_phase}). Figure~\ref{fig:ideal_stiff_blast_prob_sqp_iters_phase} shows the
phase-field formulation converges to a sharp-interface solution despite
intermediate iterations containing non-trivial mixtures of the fluids. The
grid is refined twice, each time by splitting each quadrilateral into four
smaller ones (Figure~\ref{fig:ideal_stiff_refstudy_meshes}); the space-time
density compares well between the coarse
(Figure~\ref{fig:ideal_stiff_blast_prob_sqp_iters_dens}) 
and intermediate/fine (Figure~\ref{fig:ideal_stiff_refstudy_meshes}) grids.
Slices of the primitive variables at the final time $t = 5 \times 10^{-5}$
(Figure~\ref{fig:ideal_stiff_blast_prob_qoi_slices}) show all variables are
well resolved with only minimal spurious oscillations, both the shock and
contact are represented as perfect discontinuities (except pressure and velocity,
which are both constant across the contact), and a sharp interface is
obtain ($\phi(x,t) \in \{0,1\}$). There is a minor rarefaction shock at the head
of the rarefaction wave on the coarsest grid, most likely due to the large variation
of density at the head and tail of the rarefaction and the single element approximation
of the rarefaction at the final SQP iteration
(Figure~\ref{fig:ideal_stiff_blast_prob_sqp_iters_dens}); however, this disappears
under mesh refinement (Figure~\ref{fig:ideal_stiff_blast_prob_qoi_slices}).


\subsection{Spherically symmetric underwater explosion}
\label{sec:numexp:euler:IGBKW}
Lastly, we consider a spherically symmetric underwater blast consisting of a
high-pressure gas $\text{N}_2$ modeled using the BKW equation of state
($a = 0.0072051$ (J/mol/K$^2$), $b = 23.4866$ (J/mol/K), $c = -9545$ (J/mol), $\beta = 0.403$, $\kappa = 10.86 \times 10^{-6}$ (m$^3$ K$^\alpha$/mol), $\theta = 5441$ (K), $\alpha = 0.5$) and the
surrounding water modeled as a stiffened gas ($\gamma_2 = 4.4$,
$P_2^\star = 6 \times 10^8$). The governing equations are the spherically symmetric
two-phase Euler equations, which are a time-dependent conservation law of the form
(\ref{eqn:gen_cons_law}) with
\begin{equation} 
	U_x = \begin{bmatrix} \rho \\
	\rho v \\
	\rho E \\
	\rho \phi \end{bmatrix}, \quad \Fcal_x(U_x) = \begin{bmatrix} \rho v \\
	\rho v^2 + P(\rho, e) \\
	\left[\rho E + P(\rho, e)\right]  v \\
	\rho \phi v  \end{bmatrix}, \quad S_x(U_x) = -\frac{2}{r} \begin{bmatrix} \rho v \\
	 \rho v^2  \\
 	\left[\rho E + P(\rho, e)\right] v \\
	 \rho \phi v  \end{bmatrix},
\end{equation}
where $\rho : \Omega_x \times \Tcal \rightarrow \Rbb_{\ge 0}$ is the density
of the fluid, $v : \Omega_x \times \Tcal \rightarrow \Rbb$ is the radial
velocity of the fluid, $E : \Omega_x \times \Tcal \rightarrow \Rbb$ is the total
energy of the fluid, and $e : \Omega_x \times \Tcal \rightarrow \Rbb_{\ge 0}$ is
the specific internal energy of the fluid. The pressure is defined by the
mixture equation of state (\ref{eqn:mixture_eos}) with fluid 1 modeled as
a BKW gas (Section~\ref{sec:gov_eos:eos_BKW}) and fluid 2 modeled as water
(Section~\ref{sec:gov_eos:eos_sg}). We consider the space-time domain
$\Omega_x \coloneqq (0, 1)$ and $\Tcal = (0, 8\times 10^{-4}]$ with
initial condition
\begin{equation}
 \rho(x,0) = \begin{cases} 1600 & x<0.5 \\  1000 & x \geq 0.5 \end{cases}, \quad
  v(x,0) = 0, \quad
  P(x,0) = \begin{cases}  1.14 \times 10^{10}& x<0.5 \\ 1 \times 10^5 & x \geq 0.5 \end{cases} , \quad
  \phi(x,0) = \begin{cases} 1 & x<0.5 \\ 0 & x \geq 0.5
   \end{cases},
\end{equation}
where standard international (SI) units are used for all variables and, from these initial
conditions, the temperature in the high-pressure region is $3000^\circ$. Because of the
large pressure differential in the left and right states, there are fast moving waves
that emanate from the initial discontinuity which cause a large disparity in the spatial
and temporal scales. Therefore, we nondimensionalize the problem using the scalings
$L^\star = 1$, $t^\star = 10^{-4}$, and $m^\star = 10^2$.

\begin{figure}
	\centering
 	\raisebox{-0.5\height}{\begin{tikzpicture}
\begin{groupplot} [
group style={group size = 2 by 1, horizontal sep = 0.5cm, vertical sep = 0.8cm},
title style={at={(current bounding box.north west)}, anchor=west}]
\nextgroupplot[axis equal image, width=1.2\textwidth, xtick={0, 1, 2, 3, 4}, ytick={0, 2, 4, 6, 8}, xticklabels={0, 1, 2, 3, 4}, yticklabels={0, 2, 4, 6, 8}, xlabel={space}, ylabel={time}, xmin=0, xmax=4, ymin=0, ymax=8]
\addplot []
graphics [xmin=0,xmax=4,ymin=0,ymax=8] { 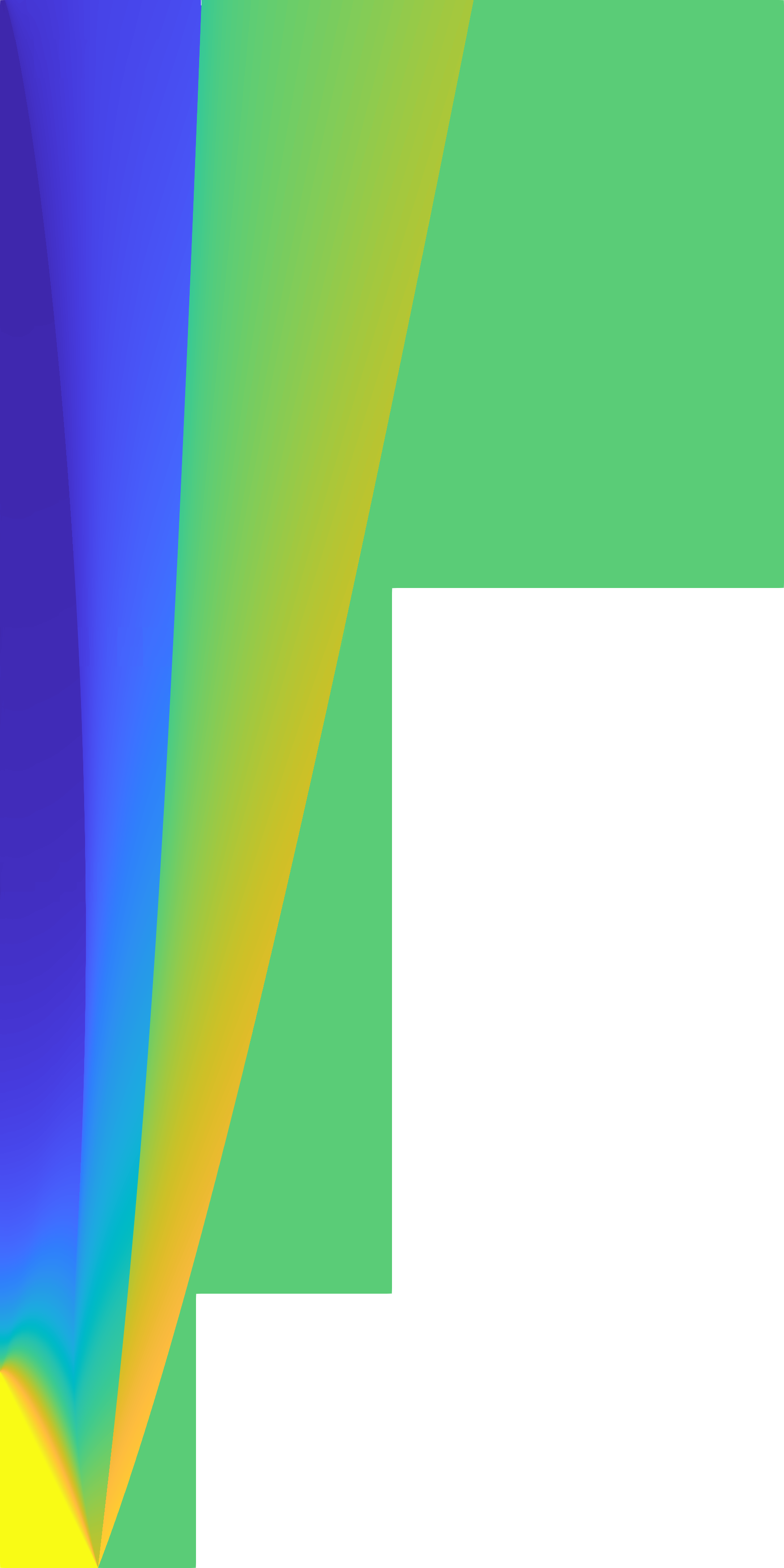};

\nextgroupplot[axis equal image, width=1.2\textwidth, xtick={0, 1, 2, 3, 4}, ytick={}, xticklabels={0, 1, 2, 3, 4}, yticklabels={}, xlabel={space}, ylabel={}, xmin=0, xmax=4, ymin=0, ymax=8]
\addplot []
graphics [xmin=0,xmax=4,ymin=0,ymax=8] { 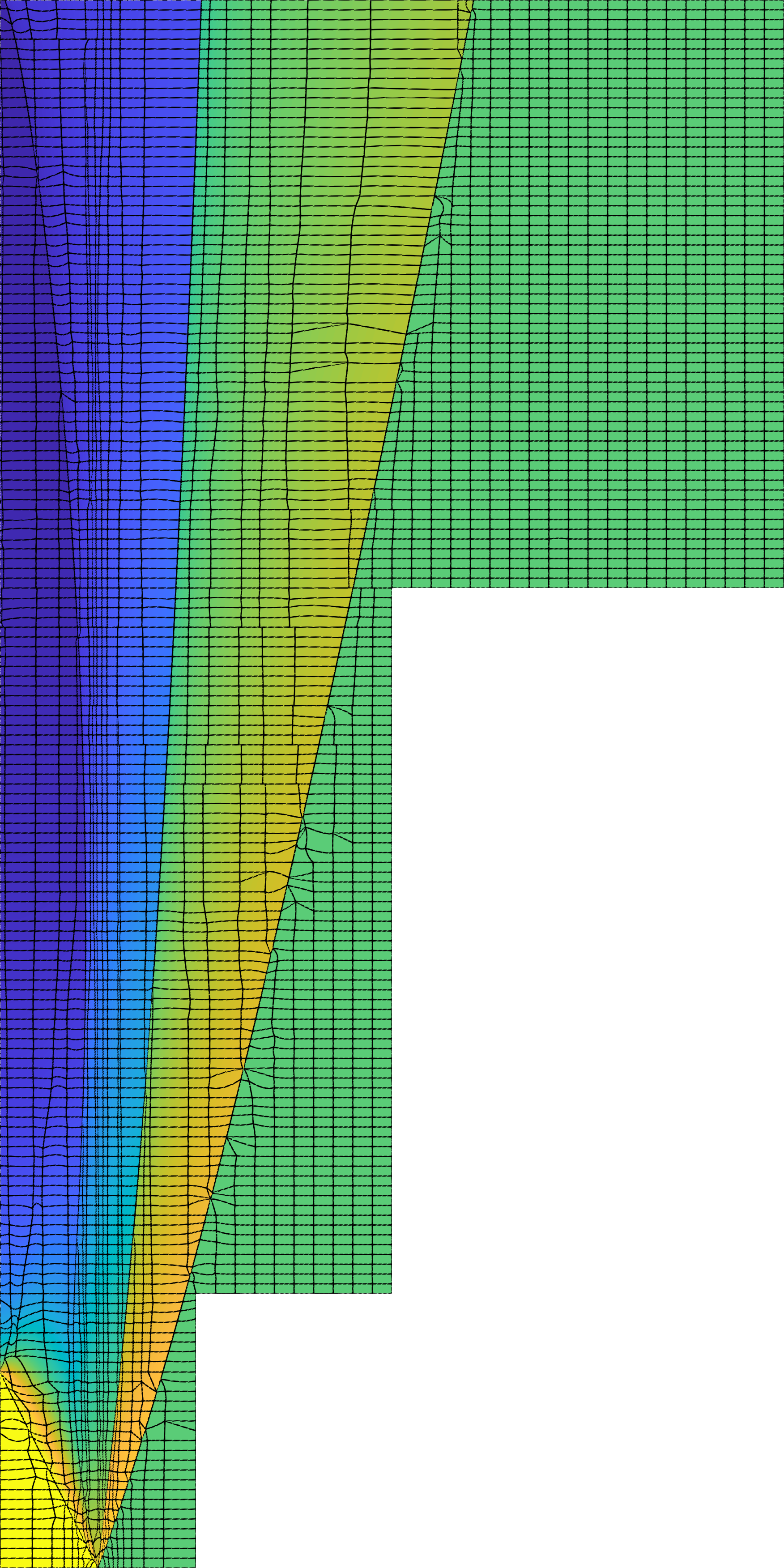};

\end{groupplot}\end{tikzpicture}}
 	\colorbarMatlabParula{1}{400}{800}{1200}{1600}        
	\caption{Two-phase HOIST solution (density) of the underwater blast problem, without (\textit{left}) and with (\textit{right}) edges shown.}
 	\label{fig:BKW_cont_dens}
\end{figure}

\begin{figure}
	\centering
 	\raisebox{-0.5\height}{\begin{tikzpicture}
\begin{groupplot} [
group style={group size = 2 by 1, horizontal sep = 0.5cm, vertical sep = 0.8cm},
title style={at={(current bounding box.north west)}, anchor=west}]
\nextgroupplot[axis equal image, width=1.2\textwidth, xtick={0, 1, 2, 3, 4}, ytick={0, 2, 4, 6, 8}, xticklabels={0, 1, 2, 3, 4}, yticklabels={0, 2, 4, 6, 8}, xlabel={space}, ylabel={time}, xmin=0, xmax=4, ymin=0, ymax=8]
\addplot []
graphics [xmin=0,xmax=4,ymin=0,ymax=8] { 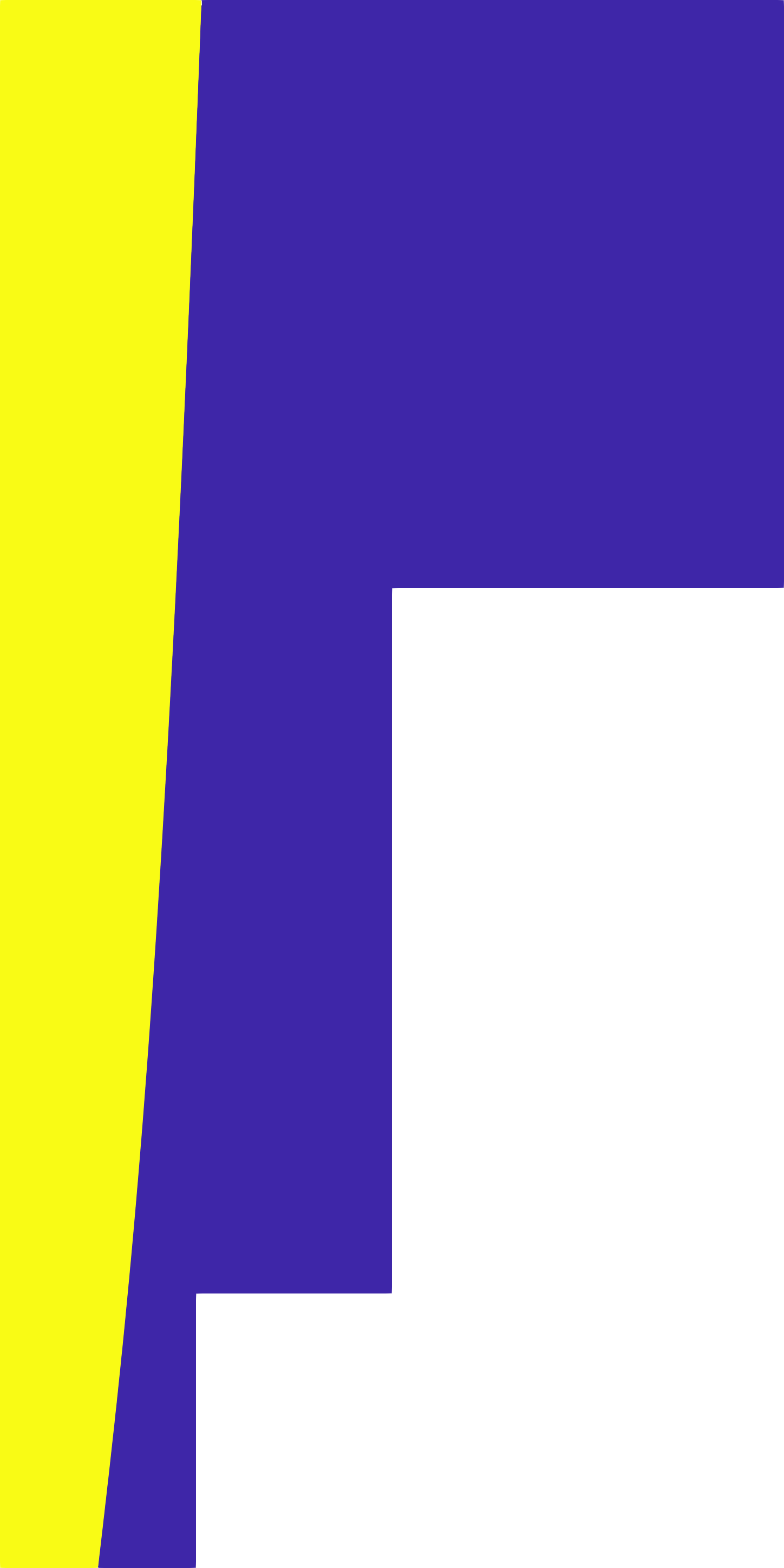};

\nextgroupplot[axis equal image, width=1.2\textwidth, xtick={0, 1, 2, 3, 4}, ytick={}, xticklabels={0, 1, 2, 3, 4}, yticklabels={}, xlabel={space}, ylabel={}, xmin=0, xmax=4, ymin=0, ymax=8]
\addplot []
graphics [xmin=0,xmax=4,ymin=0,ymax=8] { 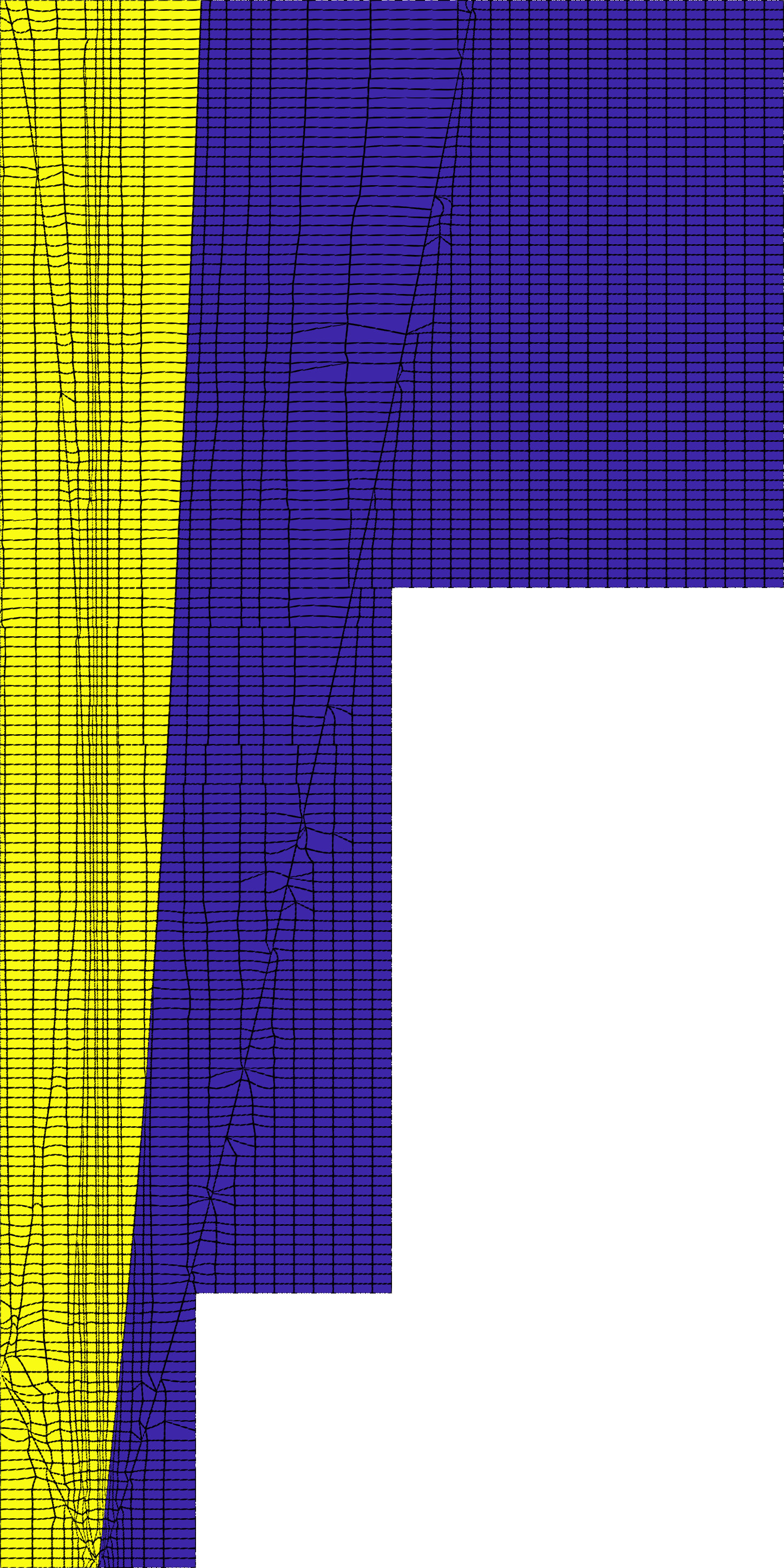};

\end{groupplot}\end{tikzpicture}}
 	\colorbarMatlabParula{0}{0.25}{0.5}{0.75}{1}        
	\caption{Two-phase HOIST solution (phase) of the underwater blast problem, without (\textit{left}) and with (\textit{right}) edges shown.}
 	\label{fig:BKW_cont_phase}
\end{figure}

Forty slabs are used to cover the time domain. The initial slab consists of
$20$ quadratic ($q=2$) quadrilateral elements with quadratic ($p=2$) solution
approximation over each. Subsequent slabs may have a different number of elements
due to element collapses \cite{2021_huang_HOIST,naudet2024space} during the HOIST
iterations of the current slab and the extract-extrude-split approach introduced
in \cite{naudet2024space} (Section~\ref{sec:ist}); however, each slab has four layers
of elements in the temporal dimension. A lead shock, rarefaction wave, and material interface
separation the BKW gas from the water emanate from $x = 0.5$ (origin of the blast).
The shock and material interface quickly travel away from the blast origin due to the
large pressure differential and at some time $t >0$, a secondary shock forms in the
BKW phase. The secondary shock eventually over expands and reverses direction. The
material interface decelerates as time evolves. The HOIST method produces a mesh
in each space-time slab that aligns with the lead shock, rarefaction wave, material
interface, and secondary shock (after its formation) (Figure \ref{fig:BKW_cont_dens}).
The motion of these features are tracked in time, including the dramatic motion of
the secondary shock that changes direction and approaches the origin at the end of
the time interval. The phase field clearly shows each slab converges to a sharp interface
(Figure \ref{fig:BKW_cont_phase}).

\section{Conclusions}
\label{sec:conclude}
In this work, we develop a space-time implicit shock tracking method to
simulate two-phase flow involving real gases. A unique feature of
our method is it utilizes a phase-field formulation of the two-phase
Euler equations to converge to a sharp-interface solution. That is,
material mixtures $0 < \phi(x,t) < 1$ are only encountered at intermediate
solver iterations; however, no mixtures $\phi(x,t)\in\{0,1\}$ are present
at the final iteration (solver convergence). As a result, the mixture
equation of state must only recover the thermodynamic properties of the
individual fluids in the limiting case where $\phi(x,t) \in \{0,1\}$,
which allows for an extremely simple mixture model (\ref{eqn:mixture_eos}).
The proposed method is built on top of the space-time implicit shock tracking
method in \cite{naudet2024space}. As such, it produces highly accurate,
sharp-interface solutions to two-phase flows without artificial stabilization,
even if the dynamics of shocks or material interface are complex.

The proposed method is validated using a single-phase and two-phase version of
Sod's shock tube with ideal gases. Finally, the method is demonstrated on Riemann
problems involving ideal and non-ideal (modeled with the Becker-Kistiakowsky-Wilson
equation of state) gases and water (modeled as a stiffened gas). The space-time
two-phase HOIST method successfully produces space-time slab meshes aligned
with all shocks (even secondary shocks that form at time $t>0$) and contacts,
delivers highly accurate solutions on coarse meshes with all shocks and
contacts represented as perfect discontinuities. For all problems considered,
the phase field $\phi(x,t)$ converges to a sharp-interface solution where
$\phi(x,t) \in \{0,1\}$ for all $(x,t) \in \Omega_x\times \Tcal$. Future
research should investigate the proposed method for more complex two-phase
flows in higher dimensions, such as the asymmetric collapse of underwater
explosion bubbles, which is extremely challenging and computationally
expensive to simulate using conventional methods.

\section*{Acknowledgments}
This work is supported by
AFOSR award numbers FA9550-20-1-0236, FA9550-22-1-0002, FA9550-22-1-0004,
ONR award number N00014-22-1-2299, and
NSF award number CBET-2338843.
The content of this publication does not necessarily reflect the position
or policy of any of these supporters, and no official endorsement should
be inferred.

\appendix

\section{Ideal gas limit of real gas quantities}
\label{sec:idealgas_lim}
In this section we show the speed of sound and projected flux Jacobian in
(\ref{eqn:proj_jac_non_ideal}) for a real gas reduce to the well-known expressions
\cite{naudet2024space} in the ideal gas limit. That is, we take the ideal gas equation of
state (\ref{eqn:P_IG}) with partial derivatives (\ref{eqn:P_derivs_IG}). The derivatives
of pressure under constant conservative variables are
\begin{equation} \label{eqn:ideal_pders}
	\begin{aligned}
		\bar{P}_\rho(\rho,\rho v,\rho E) &= (\gamma-1)(e - E + \norm{v}^2) = (\gamma-1) \norm{v}^2/2 \\
		\bar{P}_{\rho v}(\rho,\rho v,\rho E) &= -(\gamma-1) v \\
		\bar{P}_{\rho E}(\rho,\rho v,\rho E) &= (\gamma-1).
	\end{aligned}
\end{equation}
First, we substitute these expressions into the sound speed formula
(\ref{eqn:sound_speed_nonideal_single}) to obtain
\begin{equation}
		c^2 = (\gamma-1)\left(H-\frac{1}{2}\norm{v}^2\right) 
		= (\gamma-1)\left(E + \frac{P}{\rho}-\frac{1}{2}\norm{v}^2\right) 
		= (\gamma-1)\left(e + \frac{P}{\rho}\right) 
		= \gamma\frac{P}{\rho},
\end{equation}
where the first equality results from direct substitution of the ideal gas equation
of state into (\ref{eqn:sound_speed_nonideal_single}) and simplification, the next
two equalities follow from the expressions for enthalpy and internal energy
(\ref{eqn:int_ener_to_total_ener}), and the final equality follows from the
ideal gas equation of state (\ref{eqn:P_IG}) and simplification. This
agrees with the speed of sound for an ideal gas.

Next, we substitute (\ref{eqn:ideal_pders}) into (\ref{eqn:proj_jac_non_ideal}) to obtain
\begin{eqnarray} \label{eqn:proj_jac_ideal}
	B_x(U_x,\eta_x) =
	\begin{bmatrix} 0 & \eta_x^T & 0\\
	-v_n v + \frac{(\gamma - 1)}{2} ||v||^2  \eta_x & v_n I_{d'} + v \eta_x^T - (\gamma -1) v \eta_x & (\gamma -1) \eta_x \\
	(\frac{(\gamma - 1)}{2}||v||^2- H) v_n & H \eta_x^T - v_n (\gamma -1) v^T & \gamma v_n \end{bmatrix},
\end{eqnarray}
which agrees with the projected inviscid flux Jacobian from \cite{naudet2024space}. 
Repeating this process with (\ref{eqn:right_evec_non_ideal}) yields
\begin{equation}
	V_x(U_x,\eta_x) = 
	\begin{bmatrix}
                1 & \eta_x^T & 1 \\
                v - c \eta_x & (v-\eta_x)\eta_x^T  + I_{d'} & v + c \eta_x \\
                H - v_n c & v^T + (\theta - v_n) \eta_x^T & H + v_n c
        \end{bmatrix}
\end{equation}
with $\theta = \norm{v}^2/2$, which agrees with the right eigenvectors of the
projected inviscid flux Jacobian from \cite{naudet2024space}. Finally,
repeating this process with (\ref{eqn:left_evec_non_ideal}) yields
\begin{equation}
	V_x(U_x,\eta_x)^{-1} =
        \frac{\gamma-1}{2 c^2}
        \begin{bmatrix} 
		\norm{v}^2/2 + \frac{v_n c}{\gamma-1} & -v^T-\frac{c}{\gamma-1} \eta_x^T & 1\\
		\frac{2 c^2}{\gamma-1} (v_n\eta_x + \eta_x - v) -\norm{v}^2\eta_x & 2\eta_x v^T + \frac{2 c^2}{\gamma-1} (I_{d'} - \eta_x \eta_x^T) & -2 \eta_x   \\
		\norm{v}^2/2 - \frac{v_n c}{\gamma-1} & -v^T + \frac{c}{\gamma-1}\eta_x^T & 1 
        \end{bmatrix}, 
\end{equation}
where $H = \frac{c^2}{\gamma-1} + \norm{v}^2/2$ for an ideal gas was used to simplify
the $(2,1)$ component. This expression agrees with the left eigenvectors of the projected
inviscid flux Jacobian from \cite{naudet2024space}.
The matrix of eigenvalues (\ref{eqn:evals_single}) is independent of $P$
and agrees with the ideal gas eigenvalues in \cite{naudet2024space}.
Thus, all terms in Section~\ref{sec:gov_singlephase} reduce to the well-known
expressions for an ideal gas when the ideal equation of state is used
(\ref{eqn:P_IG}), which serves as a sanity check for the framework in
this limiting case.

\section{Speed of sound derivations}
\label{sec:speed_of_sound_deriv}
In this section we derive the speed of sound for a single-phase real gas
(\ref{sec:speed_of_sound_deriv_single}) and a two-phase real gas using an arbitrary mixture
equation of state (\ref{sec:speed_of_sound_deriv_multi}). Furthermore, we prove the
two-phase real gas sound speed reduces to the true material sound speeds in the
limits $\phi \rightarrow 0$ and $\phi \rightarrow 1$
(\ref{sec:multiphase_sound_speed_to_material}).

\subsection{Speed of sound derivation, single-phase flow}
\label{sec:speed_of_sound_deriv_single}
For a single phase gas with arbitrary equation of state $P(\rho, e)$, the
speed of sound, $c$, is defined as
\begin{equation} \label{eqn:sound_speed_ent}
	c^2 = \frac{\partial P}{\partial \rho} \bigg|_{s},
\end{equation}
where $s$ is the entropy. The Gibbs relation for the first law of thermodynamics
states that the change in internal energy $e$ of a system equals the heat added to
the system less the work done by the system, i.e.,
\begin{equation} \label{eqn:gibbs_rel1}
	T ds = d e + P d\nu,
\end{equation}
where $\nu = 1/\rho$ is the specific volume  of the fluid. The change in the specific
volume is related to the density change as $d\nu = -d\rho/\rho^2$, which reduces the
Gibbs relation to
\begin{equation} \label{eqn:gibbs_rel2}
	de = T ds + \frac{P}{\rho^2} d \rho.
\end{equation}
Because our equation of state is defined directly from density and internal energy,
we have
\begin{equation}
	dP = P_\rho d\rho + P_e de
	= P_\rho d\rho + P_e \left(T ds + \frac{P}{\rho^2} d \rho\right),
\end{equation}
where the Gibbs relation was used to eliminate $de$. Then, the pressure differential
at constant entropy ($ds = 0$) is
\begin{equation}
	dP\bigg|_s = \left(P_\rho + P_e\left(\frac{P}{\rho^2}\right)\right)d\rho,
\end{equation}
which gives the following expression for the derivative of pressure with respect
to density at constant entropy
\begin{equation}
	\pder{P}{\rho}\bigg|_s = P_\rho + P_e\left(\frac{P}{\rho^2}\right).
\end{equation}
Thus, the sound speed for a real gas with equation of state, $P(\rho,e)$, is
\begin{equation} \label{eqn:sound_speed_derived_nonideal}
	c = \sqrt{P_\rho + P_e \left(\frac{P}{\rho^2} \right) }.
\end{equation}
From the definition of enthalpy, we have $P/\rho^2 = (H - E)/\rho$,
which reduces (\ref{eqn:sound_speed_derived_nonideal}) to
\begin{equation}
	c = \sqrt{P_\rho - \frac{P_e}{\rho} E + \frac{P_e}{\rho} H}
	  = \sqrt{P_\rho - \frac{P_e}{\rho} (E - \norm{v}^2) + \frac{P_e}{\rho} (H - \norm{v}^2)}
\end{equation}
where the first equality comes directly from the substitution and the second equality
comes from adding and subtracting $\frac{P_e}{\rho}\norm{v}^2$ inside the radical.
Using the definition of $\bar{P}_\rho$ in (\ref{eqn:pressure_derivs2}), the expression
for the sound speed reduces to
\begin{equation} \label{eqn:sndsp_nonideal_from_prim}
	c = \sqrt{\bar{P}_\rho + \frac{P_e}{\rho} (H - \norm{v}^2)}.
\end{equation}

\subsection{Speed of sound derivation, two-phase flow}
\label{sec:speed_of_sound_deriv_multi}
Now, consider an arbitrary two-phase flow equation of state, $P(\rho, e, \rho\phi)$.
The sound speed definition (\ref{eqn:sound_speed_ent}) and Gibbs relations
(\ref{eqn:gibbs_rel1})-(\ref{eqn:gibbs_rel2}) are given in the previous section.
Because our equation of state is defined directly from density, internal energy,
and phase variable, we have
\begin{equation}
	dP = P_\rho d\rho + P_e de + P_{\rho\phi}d(\rho\phi)
	= P_\rho d\rho + P_e \left(T ds + \frac{P}{\rho^2} d \rho\right) + P_{\rho\phi}\phi d\rho,
\end{equation}
where the Gibbs relation was used to eliminate $de$ and constant material
mixture ($d\phi = 0$) is assumed to simplify $d(\rho\phi)$. Then, the pressure differential
at constant entropy ($ds = 0$) is
\begin{equation}
	dP\bigg|_s = \left(P_\rho + P_e\left(\frac{P}{\rho^2}\right) + \phi P_{\rho\phi}\right)d\rho,
\end{equation}
which gives the following expression for the derivative of pressure with respect
to density at constant entropy
\begin{equation}
	\pder{P}{\rho}\bigg|_s = P_\rho + P_e\left(\frac{P}{\rho^2}\right) + \phi P_{\rho\phi}.
\end{equation}
Thus, the sound speed for a two-phase real gas with equation of state,
$P(\rho,e,\rho\phi)$, for constant mixtures ($d\phi = 0$) is
\begin{equation} \label{eqn:sound_speed_derived_nonideal_multi}
	c = \sqrt{P_\rho + P_e \left(\frac{P}{\rho^2}\right) + \phi P_{\rho\phi}}.
\end{equation}
From the definition of enthalpy, we have $P/\rho^2 = (H - E)/\rho$,
which reduces (\ref{eqn:sound_speed_derived_nonideal_multi}) to
\begin{equation}
	c = \sqrt{P_\rho - \frac{P_e}{\rho} E + \frac{P_e}{\rho} H + \phi P_{\rho\phi}}
	  = \sqrt{P_\rho - \frac{P_e}{\rho} (E - \norm{v}^2) + \frac{P_e}{\rho} (H - \norm{v}^2) + \phi P_{\rho\phi}}
\end{equation}
where the first equality comes directly from the substitution and the second equality
comes from adding and subtracting $\frac{P_e}{\rho}\norm{v}^2$ inside the radical.
Using the definition of $\bar{P}_\rho$ in (\ref{eqn:pressure_derivs2}), the expression
for the sound speed reduces to
\begin{equation} \label{eqn:sndsp_nonideal_from_prim_multi}
	c = \sqrt{\bar{P}_\rho + \frac{P_e}{\rho} (H - \norm{v}^2) + \phi P_{\rho\phi}}.
\end{equation}

\subsection{Proof of two-phase sound speed approaching true material sound speeds}
\label{sec:multiphase_sound_speed_to_material}
In this section we prove the two-phase sound speed in
(\ref{eqn:sndsp_nonideal_from_prim_multi}) approaches the
true material sound speeds in (\ref{eqn:sndsp_nonideal_from_prim})
in the limiting cases $\phi = 0$ and $\phi = 1$ assuming the mixture
equation of state in (\ref{eqn:mixture_eos}). In particular, we will show
\begin{equation} \label{eqn:lim_cond1}
	\left.c\right|_{\phi=1} = \sqrt{\bar{P}_{1,\rho} + \frac{P_{1,e}}{\rho}(H-\norm{v}^2)}, \qquad 
	\left.c\right|_{\phi=0} = \sqrt{\bar{P}_{2,\rho} + \frac{P_{2,e}}{\rho}(H-\norm{v}^2)},
\end{equation}
where $c$ is given by (\ref{eqn:sndsp_nonideal_from_prim_multi}).
From (\ref{eqn:mixture_deriv_rho}), the limiting cases of $P_\rho$ are
\begin{equation}
	\left.P_\rho\right|_{\phi=1} = P_{1,\rho} + \frac{P_2-P_1}{\rho}, \qquad
        \left.P_\rho\right|_{\phi=0} = P_{2,\rho}.
\end{equation}
From (\ref{eqn:mixture_deriv_e}), the limiting cases of $P_e$ are
\begin{equation} \label{eqn:Pe_lim}
	\left.P_e\right|_{\phi=1} = P_{1,e}, \qquad
	\left.P_e\right|_{\phi=0} = P_{2,e}.
\end{equation}
From (\ref{eqn:mixture_deriv_rhophi}), $P_{\rho\phi}$ is independent of $\phi$
and takes the value
\begin{equation} \label{eqn:P_rhophi_lim}
	\left.P_{\rho\phi}\right|_{\phi=1} = \left.P_{\rho\phi}\right|_{\phi=1} = \frac{P_1-P_2}{\rho}.
\end{equation}
Direct substitution of these expressions into (\ref{eqn:pressure_derivs_multi})
for $\bar{P}_\rho$, we have
\begin{equation} \label{eqn:Pbar_rho_lim}
	\left.\bar{P}_\rho\right|_{\phi=1} =  P_{1,\rho} + \frac{P_2-P_1}{\rho} - \frac{P_{1,e}}{\rho}(E-\norm{v}^2), \qquad
	\left.\bar{P}_\rho\right|_{\phi=0} = P_{2,\rho} - \frac{P_{2,e}}{\rho}(E-\norm{v}^2).
\end{equation}
Combining (\ref{eqn:P_rhophi_lim}) and (\ref{eqn:Pbar_rho_lim}), the limiting cases
of $\bar{P}_\rho + \phi P_{\rho\phi}$ are
\begin{equation} \label{eqn:lim_last}
	\begin{aligned}
		\left(\bar{P}_\rho + \phi P_{\rho\phi}\right)_{\phi=1} &= P_{1,\rho} + \frac{P_2-P_1}{\rho} - \frac{P_{1,e}}{\rho}(E-\norm{v}^2) + \frac{P_1-P_2}{\rho} = P_{1,\rho} - \frac{P_{1,e}}{\rho}(E-\norm{v}^2) \\
		\left(\bar{P}_\rho + \phi P_{\rho\phi}\right)_{\phi=0} &= P_{2,\rho} - \frac{P_{2,e}}{\rho}(E-\norm{v}^2).
	\end{aligned}
\end{equation}
Finally, we obtain (\ref{eqn:lim_cond1}) by direct substitution of (\ref{eqn:Pe_lim})
and (\ref{eqn:lim_last}) into (\ref{eqn:sndsp_nonideal_from_prim_multi}).

\section{Roe averages}
\label{sec:roeavg}
In this section we present the Roe averages for Roe's approximate Riemann solver
\cite{roe1981approximate} for the real gas, single- and two-phase Euler equations
using the Roe-Pike method \cite{roe1985efficient} derived in \cite{glaister1988approximate}.
The Roe averages are constructed such that the numerical flux function is conservative.
In the ideal gas case, the Roe averages are defined for the density ($\hat\rho$),
velocity ($\hat{v}$), and enthalpy ($\hat{H}$) as
\begin{equation} \label{eqn:ideal_gas_roe_avg}
        \hat{\rho} = \sqrt{\rho_L \rho_R}, \quad \hat{v} = \frac{\sqrt{\rho_L} v_L + \sqrt{\rho_R} v_R}{\sqrt{\rho_L} + \sqrt{\rho_R}}, \quad \hat{H} = \frac{\sqrt{\rho_L} H_L + \sqrt{\rho_R} H_R}{\sqrt{\rho_L} + \sqrt{\rho_R}}.
\end{equation}
In this case, the partial derivatives of $P(\rho,e)$ can easily be expressed in term of
these variables and the corresponding Roe averages are obtained by applying those
expressions to (\ref{eqn:ideal_gas_roe_avg}).

On the other hand, the partial derivatives of $P(\rho, e)$ cannot be written in terms
of the variables $\rho$, $v$, and $H$, which requires Roe averages for these partial
derivatives to ensure the Roe flux is conservative. The Roe averages for the density,
velocity, and enthalpy are given in (\ref{eqn:ideal_gas_roe_avg}) and the Roe averages
for the partial derivatives are \cite{glaister1988approximate}
\begin{equation}
	\begin{aligned}
		\hat{P}_\rho &=
		\begin{dcases}
			\frac{1}{2\Delta\rho}(P_{RR} + P_{RL} - P_{LR} - P_{LL}) & \text{ if } \Delta\rho\neq 0 \\
			\frac{1}{2}(P_\rho(\rho, e_L)+P_\rho(\rho,e_R)) & \textit{ otherwise}
		\end{dcases} \\
		\hat{P}_e &=
                \begin{dcases}
                        \frac{1}{2\Delta e}(P_{RR} + P_{LR} - P_{RL} - P_{LL}) & \text{ if } \Delta e\neq 0 \\
			\frac{1}{2}(P_e(\rho_L, e)+P_e(\rho_R,e)) & \textit{ otherwise}
                \end{dcases}
	\end{aligned}
\end{equation}
where $\Delta(\cdot) = \cdot_L - \cdot_R$ and $P_{ab} = P(\rho_a, e_b)$ for $a,b\in\{L,R\}$,
$\rho = \rho_L = \rho_R$ in the case $\Delta\rho = 0$, and $e = e_L = e_R$ in the case
$\Delta e = 0$. In finite precision, the check for $\Delta(\cdot) = 0$ is replaced with
\begin{equation}\label{eqn:arith_diff_norm}
	\frac{ \Delta(\cdot) } { \sqrt{(\cdot)_L^2 + (\cdot)_R^2}} \leq \tau, 
\end{equation}
where $\tau$ is a relative tolerance ($\tau = 10^{-4}$ in this work). This is
particularly important in this work where the left and right states can vary
by nine orders of magnitude.

Finally, the Roe averages for the two-phase, real-gas Euler equations follows directly
from the single-phase expressions. The Roe averages for density, velocity, and enthalpy
agree with the ideal gas case (\ref{eqn:ideal_gas_roe_avg}) and the Roe average for the
phase ($\phi$) is defined consistently
\begin{equation}
	\hat{\phi} = \frac{\sqrt{\rho_L} \phi_L + \sqrt{\rho_R} \phi_R}{\sqrt{\rho_L} + \sqrt{\rho_R}}.
\end{equation}
The Roe averages for the partial derivatives of pressure $P(\rho,e,\rho\phi)$ are
\begin{equation}
	\begin{aligned}
		\hat{P}_\rho &=
		\begin{dcases}
			\frac{1}{4\Delta\rho}(P_{RRR} + P_{RLL} + P_{RLR} + P_{RRL} - P_{LRR} - P_{LRL} - P_{LLR} - P_{LLL}) & \text{ if } \Delta\rho\neq 0 \\
			\frac{1}{4}(P_\rho(\rho, e_R, \rho\phi_R)+P_\rho(\rho,e_L,\rho\phi_R)+P_\rho(\rho,e_R,\rho\phi_L) + P_\rho(\rho, e_L, \rho\phi_L)) & \textit{ otherwise}
		\end{dcases} \\
		\hat{P}_e &=
                \begin{dcases}
			\frac{1}{4\Delta e}(P_{RRR} + P_{LRR} + P_{RRL} + P_{LRL} - P_{RLR} - P_{LLR} - P_{RLL} - P_{LLL}) & \text{ if } \Delta e\neq 0 \\
			\frac{1}{4}(P_e(\rho_R, e, \rho\phi_R)+P_e(\rho_L,e,\rho\phi_R)+P_e(\rho_R,e,\rho\phi_L) + P_e(\rho_L, e, \rho\phi_L)) & \textit{ otherwise}
                \end{dcases} \\
		\hat{P}_{\rho\phi} &=
                \begin{dcases}
                        \frac{1}{2\Delta\rho\phi}(P_{RRR} + P_{LLR} - P_{RRL} - P_{LLL}) & \text{ if } \Delta\rho\phi\neq 0 \\
			\frac{1}{4}(P_{\rho\phi}(\rho_R,e_R,\rho\phi)+P_{\rho\phi}(\rho_R,e_L,\rho\phi)+P_{\rho\phi}(\rho_L,e_R,\rho\phi)+P_{\rho\phi}(\rho_L,e_L,\rho\phi)) & \textit{ otherwise,}
                \end{dcases}
	\end{aligned}
\end{equation}
where $\Delta(\cdot) = \cdot_L - \cdot_R$ and $P_{abc} = P(\rho_a, e_b, \rho\phi_c)$
for $a,b,c\in\{L,R\}$, $\rho = \rho_L = \rho_R$ in the case $\Delta\rho = 0$,
$e = e_L = e_R$ in the case $\Delta e = 0$, and $\rho\phi = \rho\phi_L = \rho\phi_R$
in the case $\Delta\rho\phi = 0$. In practice, (\ref{eqn:arith_diff_norm}) is used
to numerically check for zero jumps.

\bibliographystyle{plain}
\bibliography{biblio}

\end{document}